\documentclass[a4paper,fleqn,usenatbib]{mnras}

\usepackage[T1]{fontenc}
\usepackage{ae,aecompl}
\usepackage{subfig}

\usepackage{graphicx}	
\usepackage{amsmath}	
\usepackage{array}
\usepackage{mathabx}
\usepackage{threeparttable}

\title[TTV Analyses of Ten Systems]{Homogeneous Transit Timing Analyses of Ten Exoplanet Systems}

\author[Ba\c{s}t\"urk et al.]{
\"O. Ba\c{s}t\"urk$^{1}$,\thanks{E-mail: obasturk@ankara.edu.tr}
E. M. Esmer$^{1}$,
S. Yal\c{c}{\i}nkaya$^{1}$,
\c{S}. Torun$^{1}$,
L. Mancini$^{2,3,4}$
\newauthor
F. Helweh$^{5}$,
E. Karamanl{\i}$^{5}$,
J. Southworth$^{6}$,
S. Ali\c{s}$^{8,9}$,
A. W\"unsche$^{7}$,
F. Tezcan$^{10}$,
\newauthor
Y. Alada\u{g}$^{11}$,
N. Aksaker$^{11,12}$,
E. Tun\c{c}$^{5}$,
F. Davoudi$^{13}$,
S. Fi\c{s}ek$^{8,9}$,
M. Bretton$^{7}$,
\newauthor
D. F. Evans$^{6}$,
C. Ye\c{s}ilyaprak$^{10,14}$,
M. Y{\i}lmaz$^{1}$,
C. T. Tezcan$^{10,14}$,
K. Yelkenci$^{8,9}$
\\
\\
$^{1}$Ankara University, Faculty of Science, Astronomy \& Space Sciences Department, Tandogan, TR-06100, Ankara, Turkey\\
$^{2}$\,Department of Physics, University of Rome ``Tor Vergata'', Via
della Ricerca Scientifica 1, I-00133, Rome, Italy \\
$^{3}$\,Max Planck Institute for Astronomy, K\"{o}nigstuhl 17,
D-69117, Heidelberg, Germany \\
$^{4}$\,INAF -- Osservatorio Astrofisico di Torino, via Osservatorio
20, I-10025, Pino Torinese, Italy \\
$^{5}$ Bilkent University, Science Faculty, Physics Department, TR-06800 Ankara, Turkey \\
$^{6}$ Astrophysics Group, Keele University, Staffordshire ST5 5BG, UK \\
$^{7}$ Baronnies Proven\c{c}ales Observatory, Hautes Alpes - Parc Naturel
Regional des Baronnies Proven\c{c}ales, F-05150 Moydans, France \\
$^{8}$Department of Astronomy and Space Sciences, Faculty of Science, Istanbul University, 34119 Istanbul, Turkey \\
$^{9}$ Istanbul University Observatory Research and Application Center, 34119 Istanbul, Turkey \\
$^{10}$ Atat\"urk University, Science Faculty, Department of Astronomy and Space Sciences, 25240, Erzurum, Turkey \\
$^{11}$ Space Science and Solar Energy Research and Application Center (UZAYMER), University of \c{C}ukurova, 01330, Adana, Turkey\\
$^{12}$ Adana Organised Industrial Zones Vocational School of Technical Science, University of \c{C}ukurova, 01410, Adana, Turkey\\
$^{13}$ Department of Physics, University of Zanjan, P.O. Box 45195-313, Zanjan, Iran\\
$^{14}$ Atat\"urk University Astrophysics Research and Application Center (ATASAM), Yakutiye, 25240, Erzurum, Turkey 
}

\date{Accepted 2022-02-25. Received 2022-02-25; in original form 2021-03-09}

\pubyear{2020}

\begin{document}
\label{firstpage}
\pagerange{\pageref{firstpage}--\pageref{lastpage}}
\maketitle

\begin{abstract}
  We study the transit timings of 10 exoplanets in order to investigate potential Transit Timing Variations (TTVs) in them. We model their available ground-based light curves, some presented here and others taken from the literature, and homogeneously measure the mid-transit times. We statistically compare our results with published values and find that the measurement errors agree. However, in terms of recovering the possible frequencies, homogeneous sets can be found to be more useful, of which no statistically relevant example has been found for the planets in our study. We corrected the ephemeris information of all ten planets we studied and provide these most precise light elements as references for future transit observations with space-borne and ground-based instruments. We found no evidence for secular or periodic changes in the orbital periods of the planets in our sample, including the ultra-short period WASP-103\,b, whose orbit is expected to decay on an observable timescale. Therefore, we derive the lower limits for the reduced tidal quality factors (Q$^{\prime}_{\star}$) for the host stars based on best fitting quadratic functions to their timing data. We also present a global model of all available data for WASP-74\,b, which has a Gaia parallax-based distance value $\sim$25\% larger than the published value.
\end{abstract}

\begin{keywords}
planetary systems - methods: observational - techniques: photometric - stars: individual: HAT-P-23, WASP-37, WASP-69, WASP-74, HAT-P-56, WASP-2, WASP-14, HAT-P-32, WASP-103, HAT-P-37.
\end{keywords}



\section{Introduction}
\label{sec:introduction}
Analysis of exoplanet transit timing variations (TTVs) is important for several reasons. First of all, follow-up observations heavily rely on accurate and precise transit timings, especially those from space for which the observing window should be carefully planned. When ephemeris information is not up-to-date, at least a part of out-of-transit flux, or in worse cases, ingress/egress times may be missed due to the accumulation of uncertainties on their timings. Moreover, unseen additional bodies can be recovered from the periodic changes they cause in transit timings due to orbital perturbations and/or the so-called Light Time Effect (LiTE) \citep{2011ApJ...743..200B, 2012ApJ...750..113F, 2018haex.bookE...7A}. These multi-planet systems should be common, and it is also possible to determine the individual masses on the basis of the transit times of one or more transiting bodies in these systems without further and more challenging observations \citep{2018A&A...613A..68G}. This enables a full characterization of the systems and their planets provided that they are also transiting \citep{2021PSJ.....2....1A}. Most of the planets discovered by ground-based transit surveys are the so-called hot Jupiters because of their proximity to the host star. In these short-period orbits, they tidally interact with their host star causing them to lose angular momentum and even mass \citep{2018Sci...362.1388N, 2020AJ....159..278V}. The long-term evolution of planets orbiting F- and G-type stars can be significantly different owing to the combined effect of magnetic braking and tidal dissipation. It has been argued by \citet{2015A&A...574A..39D} that the existence of a quasi-stationary state, in the case of short-period planets, magnetic braking can significantly delay their tidal evolution that would otherwise bring the planet to fall into its host star. Angular momentum transfer between the orbital and rotational motion of the planet due to a potential spin-orbit coupling can induce observable deviations from a constant transit period \citep{2020MNRAS.497.3911L}.  In systems with a transiting planet on a highly eccentric orbit, even apsidal motion can be observed \citep{2017AJ....154....4P, 2019AJ....157..217B, 2019MNRAS.490.4230S, 2020ApJ...893L..29B}. In order to test such explanations of the observed orbital period modulations, the sample size of planets with in-depth analysis of their transit timings should be large enough to provide insight into the processes that lead to TTVs. Consequently, precise follow-up transit observations and their repeated analyses are needed to encompass the largest possible data sets of the systems, from which an observable TTV signal is possible for one reason or another.

We aim at a homogeneous analysis of the transit light curves of ten exoplanets (HAT-P-23\,b, WASP-37\,b, WASP-69\,b, WASP-74\,b, HAT-P-56\,b, WASP-2\,b, WASP-14\,b, HAT-P-32\,b, WASP-103\,b, HAT-P-37\,b) accumulated so far in the literature (176 in total for the planets in our sample) and various open databases\footnote{http://var2.astro.cz/ETD/index.php}$^,$\footnote{https://www.exoclock.space/}$^,$\footnote{http://brucegary.net/AXA/x.htm} (339 in total) as well as our own light curves (35 in total) from different observatories that are being published for the first time, adding up to 550 light curves in total (see Table-\ref{tab:planetdata} for details). The Transiting Exoplanet Survey Satellite (TESS) has observed only two of our targets (HAT-P-32 and HAT-P-37), one of which (HAT-P-32) has only long-cadence data, and the other (HAT-P-37) was blended by a bright source not resolved by TESS, making both inadequate for precise measurements of the mid-transit times. We are therefore limited to ground-based observations. We also note that HAT-P-56\,b is being observed by TESS at the time of our study during sectors 43, 44, and 45.

We describe the instruments that we used for transit observations of the planets in our sample, the data reduction techniques that we employed, and the quantitative criteria for light curve selection procedure in Section~\ref{sec:observations_datared}. Information on light curve modelling, mid-transit time measurements, the TTVs, and the updated ephemeris based on the underlying datasets are presented in Section~\ref{sec:analysis}. We then analyze the mid-transit timings, search for potential secular and/or periodic changes in the orbital periods of the planets we study, compare the data sets composed of homogeneous measurements that are based on our own measurements of mid-transit times with those that consist of the measurements by the observers from the same light curves. We also refine the parameters of the transiting exoplanet WASP-74\,b based on the global model of the target's most precise light curves acquired so far, radial velocity (RV) measurements including our own measurements from unpublished HARPS archival data, broadband magnitudes that we gathered from different databases, reliable stellar atmospheric parameters from the literature, and the ultra-precise distance value based on its Gaia parallax, which is $\sim$25\% higher than the value referred to in the literature (Section~\ref{sec:ttv_WASP74}). Finally, we discuss our results in Section~\ref{sec:conclusions}.

\begin{table*}
\centering
	\caption{Basic properties and the number of light curves analyzed for each planet in our sample}
	\label{tab:planetdata}
        \begin{threeparttable}
	\begin{tabular}{lcccccccc} 
		\hline
		Planet & P$_{\rm orb}$ & M$_{\rm p}$ / M$_{\star}$ & a / R$_{\star}$  & T$_{\rm eff,\star}$ & Database & Literature & Our & Total \\
                Name & [days] & $\times 10^{-3}$ & - & [K] & LC Number & LC Number & LC Number & \\ 
		\hline
		HAT-P-23\,b$^1$ & 1.212884(2) & 1.77 $\pm$ 0.11 & 4.140 $\pm$ 0.230 & 5905 $\pm$ 80 & 41 & 39 & 5 & 85 \\
                WASP-37\,b$^2$  & 3.577469(11) & 1.86 $\pm$ 0.30 & 9.017 $\pm$ 0.626 & 5800 $\pm$ 150 & 10 & 2 & 3 & 15 \\
                WASP-69\,b$^3$  & 3.8681382(17) & 0.30 $\pm$ 0.02 & 11.968 $\pm$ 0.435 & 4715 $\pm$ 50  & 5 & 2 & 7 & 14 \\
                WASP-74\,b$^4$  & 2.137750(1) & 0.64 $\pm$ 0.04 & 4.849 $\pm$ 0.038 & 5970 $\pm$ 110 & 3 & 35 & 4 & 42 \\
                HAT-P-56\,b$^5$  & 2.7908327(47) & 1.61 $\pm$ 0.19 & 6.370 $\pm$ 0.110 & 6566 $\pm$ 50  & 18 & 0 & 2 & 20 \\
                WASP-2\,b$^6$  & 2.152226(4) & 1.06 $\pm$ 0.22 & 8.463 $\pm$ 0.718  & 5200 $\pm$ 200 & 50 & 4 & 3 & 57 \\
                WASP-14\,b$^7$  & 2.243752(10) & 5.79 $\pm$ 0.73 & 5.927 $\pm$ 0.370  & 6475 $\pm$ 100 & 24 & 13 & 2 & 39 \\
                HAT-P-32\,b$^{8}$  & 2.150008(1) & 0.71 $\pm$ 0.01 & 6.050 $\pm$ 0.040  & 6207 $\pm$ 88 & 81 & 22 & 1 & 104 \\
                WASPP-103\,b$^{9}$  & 0.925542(19) & 1.17 $\pm$ 0.08 & 2.978 $\pm$ 0.096  & 6110 $\pm$ 160 & 22 & 55 & 6 & 83 \\
                HAT-P-37\,b$^{10}$ & 2.797436(7) & 1.20 $\pm$ 0.12 & 9.320 $\pm$ 0.570 & 5500 $\pm$ 100 & 85 & 4 & 2 & 91 \\
                \hline
                Total & & & & & 339 & 176 & 35 & 550 \\ 
		\hline
	\end{tabular}
        \begin{tablenotes}
          \footnotesize{
        \item[1] \citet{2011ApJ...742..116B}, $^2$\citet{2011AJ....141....8S},$^3$\citet{2014MNRAS.445.1114A}, $^4$\citet{2015AJ....150...18H}, $^5$\citet{2015AJ....150...85H},  $^6$\citet{2007MNRAS.375..951C}, $^7$\citet{2009MNRAS.392.1532J}, $^{8}$\citet{2011ApJ...742...59H}, $^{9}$\citet{2014A&A...562L...3G}, $^{10}$\citet{2012AJ....144...19B}
          }
        \end{tablenotes}
        \end{threeparttable}
\end{table*}

\section{Observations and Data Reduction}
\label{sec:observations_datared} 
\subsection{Observations}
\label{subsec:observations}
\subsubsection{TUG T100 Observations}
We observed 14 transits in total with the 1-metre Turkish telescope T100, located in T\"UB\.{I}TAK National Observatory of Turkey's (TUG) Bak{\i}rl{\i}tepe Campus at 2500~m altitude. With the well-established defocusing technique \citep{2009MNRAS.396.1023S, 2015ASPC..496..370B}, we achieved high precision in photometry of our stars using a cryo-cooler SI 1100 CCD with $4096\times4096$ pixels, which gives an effective field of view (FoV) of $20^{\prime}\times20^{\prime}$. We employed a Bessell $R$ filter in our exposures, the times of which were determined according to the brightness of the target, the duration of the observed transit, and the apparent proximity of nearby stars on the CCD to obtain the best signal-to-noise ratio (SNR).


\subsubsection{CAHA Observations}
We observed three transits of WASP-69\,b, two transits of WASP-103\,b, and one transit of HAT-P-23\,b using the 1.23\,m telescope at the Calar Alto Observatory (CAHA), Spain, equipped with the DLR-MKIII CCD camera. Two transit light curves of WASP-69\,b were obtained through a Cousins $I_c$ filter, while one light curve for the same target was acquired in the Sloan $i^{\prime}$ passband on 8 September 2016. One WASP-103\,b light curve was obtained in the Cousins $R_c$ filter, and another in the Sloan-$r^{\prime}$ filter. The FoV was $21\arcmin.5\times21\arcmin.5$ at a plate scale of $0\arcsec.32$ per pixel. We used the defocusing technique in the observations to increase photometric precision.

\subsubsection{IST60 Observations}
We observed a transit of WASP-74\,b on 14 August 2020 with the 60 cm telescope (IST60) of Istanbul University Observatory (IUGUAM), located at the \c{C}anakkale Onsekiz Mart University Ulup{\i}nar Observatory, Turkey, at 410 m altitude above sea level. IST60 is a Ritchey-Chr\'etien (RC) telescope on an NTM-500, German equatorial mount. An Andor iXon Ultra 888 model CCD is attached at its focal plane. This setup gives a pixel scale of $0\arcsec.56$ per pixel and a FoV of 9.6$\times$9.6 arcmin$^2$.

\subsubsection{UT50 Observations}
The transits of WASP-37\,b on 5 May 2020 and HAT-P-23\,b on 20 July 2021 were observed with the 50 cm RC telescope, equipped with an Apogee Aspen CG type CCD at \c{C}ukurova University UZAYMER Observatory's campus in Adana, Turkey. The CCD has $1024\times1024$ pixels with a pixel scale of $2\arcsec.4$ per pixel, which gives a similar effective FoV of T100. The details of the observatory and observing conditions are provided in \cite{2020arXiv200600528P}. We made use of the standard Johnson $R$ filter during the observation of HAT-P-56\,b transiting its star, while the observation of WASP-37 was acquired without a filter (clear) to increase the SNR.

\subsubsection{ATA50 Observations}
We observed a transit of WASP-14\,b on 14 May 2020 and a transit of WASP-2\,b on 12 October 2020 using the 50~cm aperture, RC telescope, located at Atat\"urk University's ATASAM Observatory in Erzurum, Turkey at an altitude of 1824\,m above sea level. An Apogee Alta U230 CCD with $2048\times2048$ pixels of 15\,$\mu$m size was employed, giving a pixel-scale of $0\arcsec.77$ per pixel. Unfortunately, we had a dome failure at the end of this observation, and we were not able to use it again throughout our study. We made use of a Cousins $R_c$ filter in the observation and defocused the target slightly to be able to expose for a longer duration and thus increase the SNR. 

\subsubsection{T35 Observations}
We used the 35 cm T35 telescope in Ankara University Kreiken Observatory (AUKR) primarily to check the reliability of the ephemeris information for the transit observations of our targets. When the weather conditions are optimal, it is also possible to achieve the required precision in photometry to derive mid-transit times having uncertainties smaller than $\sim$2 minutes. We observed 10 transits in total with this telescope and the 1024$\times$1024 Apogee Alta U47 CCD attached at its focal plane. Only the transit light curve of WASP-103\,b on 1 July 2018 was obtained without a filter, while the rest of the light curves were acquired through the Bessel $R$ filter.\\

We summarize the basic information about the telescopes that we employed for data acquisition in Table~\ref{tab:instruments}.

\begin{table}
  \centering
  \caption{Basic information on the instruments that we employed in our observations}
  \label{tab:instruments}
  \footnotesize
  \begin{tabular}{lcccc} 
    \hline
    Telescope & Aperture & Camera & FoV & Plate Scale \\
    Label & cm & px $\times$ px & $arcmin^{2}$ & $^{\prime\prime}$ / px \\
    \hline
    T100 & 100 & $4096 \times 4096$ & $20.0 \times 20.0$ & 0.11 \\
    CAHA & 123 & $2048 \times 2048$ & $21.5 \times 21.5$ & 0.32 \\
    IST60 & 60 & $1024 \times 1024$ & $9.6 \times 9.6$ & 0.56 \\
    UT50 & 50 & $1024 \times 1024$ & $11.2 \times 11.2$ & 2.40 \\
    ATA50 & 50 & $2048 \times 2048$ & $12.4 \times 12.4$ & 0.77 \\
    T35 & 35 & $1024 \times 1024$ & $13.2 \times 13.2$ & 0.75 \\
    \hline
	\end{tabular}
\end{table}

\subsection{Data Reduction}
\label{subsec:data_reduction}
We corrected our images acquired with T100, IST60, UT50, ATA50, and T35 with the AstroImageJ (hereafter AIJ) software package \citep{2017AJ....153...77C} for the instrumental effects (bias-dark-flat corrections) in the standard manner by making use of calibration images obtained in the same nights with the observations. We employed 2$\times$2 pixel-binning when needed to decrease the exposure and readout times. We converted all the observation timings to Dynamical Barycentric Julian Days (BJD-TDB) before an ensemble photometry \citep{1992PASP..104..435H} with the AIJ relative to a number of comparison stars, selected according to their brightness, colour, and the unchanging behaviour during an observing run in the passband of observations. We determined the centres of the apertures visually, to avoid incorrect flux-centre positions by centroid methods when applied to ``ring-like'' shapes of the stellar signal on CCD images due to defocusing. We then corrected for the airmass, and then normalized the relative fluxes determined by the AIJ with the line fit to the out-of-transit relative fluxes. 

CAHA raw images were reduced by using the {\sc defot} code (\cite{2014MNRAS.444..776S} and references therein), which performs aperture photometry using the {\sc aper} algorithm. Differential photometry was performed with respect to an ensemble of comparison stars, the number of which depends on the availability of adequate stars with constant flux within the limits of the observation setup during an observing night.

Finally, we detrended the light curves from red-noise by making use of Gaussian Processes (GP) since instrumental effects such as the drifts in the x-y positions of targets and comparison stars on the CCD images, and non-linearity issues as well as astrophysical noise sources such as stellar variability and spot-induced asymmetries affect transit profiles differently for each of the stars in our sample. We used a quasi-periodic kernel for the stochastic part of a light curve model, the deterministic part of which was composed of a dilution factor, mean out-of-transit flux and the transit model based on the system parameters from the studies announcing their discoveries. These parameters were also used during the light curve modelling  to derive the mid-transit timings for consistency since they are the default values provided by the NASA Exoplanet Archive for a given planet. We checked the level of scatter in the light curve in the out-of-transit flux through the standard deviation and tried to preserve it to avoid decreasing the level of white noise artificially in the data during the process.

We present a log of all our observations in Table~\ref{tab:observations} to provide statistical information summarizing the quality of our observations. We also provide all the acquired light curves in the Appendix~\ref{sec:appendixa} and the data in the online materials as well as their noise statistics separately in relevant files for each of the planets together with that from the light curves of others.

\begin{table*}
  \scriptsize
\centering
	\caption{A log of photometric observations performed for this study. The dates of the light curves that are eliminated and hence not used in the TTV diagrams are marked and the reasons for the elimination are given in the footnotes.}
	\label{tab:observations}
        \begin{threeparttable}
	\begin{tabular}{cccccccccccc} 
		\hline
		System & Telescope & Date & Start  & End & Filter & Exp. Time & Images & PNR & $\beta$ & Mid-Transit & Error \\
                Name & & UTC & UTC & UTC & & [s] & Number &  &  & BJD-TDB & [days] \\ 
		\hline
		HAT-P-23 & T100 & 2014-09-25 & 17:32:33 & 20:24:20 & R & 135 & 65 & 0.357 & 0.529 & 2456926.300832 & 0.000181 \\
                HAT-P-23 & CAHA & 2015-08-28 & 19:43:29 & 01:54:53 & R$_{\rm c}$ & 40 & 511 & 2.288 & 0.344 & 2457263.483520 & 0.000259 \\
                HAT-P-23 & T35 & 2019-08-08 & 20:05:30 & 01:00:02 & R & 60 & 285 & 10.263 & 0.546 & 2458704.391921 & 0.001378 \\
                HAT-P-23 & T35 & 2019-09-28 $^1$ & 17:38:08 & 22:45:35 & R & 90 & 188  & 5.911 & 0.401 & 2458755.329150 & 0.001652 \\
                HAT-P-23 & UT50 & 2021-07-20 $^2$ & 19:06:30 & 22:39:37 & R & 150 & 78  & 1.340 & 1.542 & 2459416.355578 & 0.001028 \\
                WASP-37 & T100 & 2017-04-27 & 20:45:50 & 01:28:57 & R & 135 & 105 & 0.974 & 0.624 & 2457871.472553 & 0.000692 \\
                WASP-37 & UT50 & 2020-05-28 $^{1,2}$ & 18:19:19 & 22:56:44 & Clear & 60 & 236 & 10.118 & 2.018 & 2458998.368941 & 0.000917 \\
                WASP-37 & T100 & 2020-06-22 $^{5}$ & 20:55:51 & 23:59:47 & Clear & 125 & 70 & 1.933 & 0.853 & 2459023.423285 & 0.001028 \\
                WASP-69 & CAHA & 2015-08-22 & 19:47:03 & 01:05:04 & I$_{\rm c}$ & 50 & 316 & 0.971 & 0.238 & 2457257.407761 & 0.000155 \\
                WASP-69 & CAHA & 2015-09-18 & 19:44:12 & 01:04:23 & I$_{\rm c}$ & 75 & 254 & 0.962 & 0.454 & 2457284.484812 & 0.000162 \\
                WASP-69 & CAHA & 2016-09-08 & 18:38:41 & 22:48:28 & i$^{\prime}$ & 45 & 320 & 2.203 & 0.216 & 2457640.353797 & 0.000241 \\
                WASP-69 & T100 & 2016-10-09 & 17:30:55 & 21:45:34 & R & 90 & 105 & 0.810 & 0.497 & 2457671.298629 & 0.000301 \\
                WASP-69 & T100 & 2017-08-26 & 18:23:04 & 22:57:27 & I & 100 & 103 & 0.484 & 0.826 & 2457992.354849 & 0.000199 \\
                WASP-69 & T35 & 2019-07-27 & 20:27:08 & 01:26:27 & R & 30 & 511 & 5.495 & 0.437 & 2458692.486662 & 0.000519 \\
                WASP-69 & T35 & 2019-08-27 & 19:20:50 & 00:25:27 & R & 20 & 732 & 6.326 & 0.110 & 2458723.431989 & 0.000381 \\
                WASP-74 & T35 & 2019-07-07 $^5$ & 21:09:03 &0 0:34:41 & R & 90 & 131 & 1.869 & 0.348 & 2458672.437582 & 0.000834 \\
                WASP-74 & T100 & 2020-08-12 & 18:39:19 & 22:14:41 & R & 120 & 75 & 6.341 & 0.428 & 2459074.330869 & 0.000479 \\
                WASP-74 & IST60 & 2020-08-14 $^1$ & 21:07:20 & 01:19:39 & R & 150 & 95 & 0.677 & 0.953 & 2459076.465929 & 0.000999 \\
                WASP-74 & T100 & 2020-08-27 & 17:42:37 & 21:15:55 & R & 120 & 74 & 2.697 & 0.519 & 2459089.297844 & 0.000815 \\
                HAT-P-56 & T35 & 2019-02-17 & 17:06:25 & 22:08:35 & R & 45 & 359 & 4.334 & 0.450 & 2458532.312128 &  0.001577 \\
                HAT-P-56 & T100 & 2019-11-09 & 20:49:28 & 01:02:04 & R & 90 & 103 & 0.458 & 0.470 & 2458797.442744 & 0.000556 \\
                WASP-2 & T35 & 2019-07-20 & 19:54:21 & 01:38:04 & R & 40 & 492 & 10.709 & 0.475 & 2458685.509228 & 0.001260 \\
                WASP-2 & ATA50 & 2020-10-12 & 17:00:41 & 21:11:19 & R & 120 & 109 & 1.573 & 0.412 & 2459135.323994 & 0.000610 \\
                WASP-2 & T100 & 2020-10-25 & 15:58:10 & 19:01:51 & R & 120 & 63 & 1.202 & 1.228 & 2459148.237230 & 0.000464 \\
                WASP-14 & T35 & 2019-07-03 & 19:14:08 & 23:26:15 & R & 45 & 267 & 6.594 & 0.188 & 2458668.396037 & 0.001491 \\
                WASP-14 & ATA50 & 2020-05-14 $^5$ & 17:24:45 & 20:30:34 & R & 20 & 307 & 1.635 & 0.548 & 2458980.286843 & 0.000612 \\
                HAT-P-32 & T35 & 2019-11-03 & 16:45:39 & 02:57:18 & R & 100 & 297 & 1.240 & 0.504 & 2458791.414706 & 0.000351 \\
                WASP-103 & T100 & 2014-05-30 & 22:48:14 & 01:48:47 & R & 145 & 53 & 0.542 & 0.863 & 2456808.530535 & 0.000296 \\
                WASP-103 & CAHA & 2015-04-30 & 23:56:16 & 04:16:09 & R$_{c}$ & 145 & 106 & 0.380 & 0.650 & 2457143.577475 & 0.000183 \\
                WASP-103 & CAHA & 2015-06-07 & 20:24:31 & 01:58:08 & I$_{c}$ & 90 & 183 & 0.694 & 0.898 & 2457181.524650 & 0.000241 \\
                WASP-103 & T100 & 2017-06-11 & 19:42:54 & 01:37:55 & R & 130 & 117 & 1.077 & 0.553 & 2457916.407767 & 0.000683 \\
                WASP-103 & T35 & 2018-07-01 & 19:40:25 & 00:04:00 & Clear & 60 & 187 & 1.053 & 2.313 & 2458301.434320 & 0.000447 \\
                WASP-103 & T100 & 2020-07-29 & 19:02:37 & 22:38:46 & R & 120 & 75 & 0.983 & 0.318 & 2459060.382613 & 0.000595 \\
                HAT-P-37 & T100 & 2015-08-04 & 21:30:11 & 01:56:51 & R & 150 & 79 & 0.712 & 0.818 & 2458532.312128 & 0.001577 \\
                HAT-P-37 & T100 & 2019-05-17 & 19:45:14 & 00:53:57 & R & 115 & 139 & 0.832 & 0.526 & 2458621.418178 & 0.000250 \\
		\hline
        \end{tabular}
        \begin{tablenotes}
          \scriptsize{
          \item[1] Eliminated because it is an outlier on TTV-diagram.
          \item[2] Eliminated because its depth is out of 3$\sigma$ of the average.
          \item[3] Eliminated because its PNR value is larger than its depth.
          
          \item[4] Eliminated because its $\beta$-factor is larger than 2.5.
          \item[5] Eliminated because it is incomplete.
          }
        \end{tablenotes}
        \end{threeparttable}
\end{table*}

\subsection{Light Curve Selection Criteria}
\label{subsec:criteria}
Quality light curves are needed to infer mid-transit timings, precise enough to study potential transit timings in a system. On the other hand, the number and the frequency  of observations are also crucial to reveal the potential trends in the transit timings and understand their causes. Therefore, not all available transit light curves should be used prior to careful and quantitative inspection. Otherwise, it might lead to incorrect interpretations due to the lack of required data quality. Two well-defined metrics are widely used to quantify the quality of transit light curves in the literature. Photometric Noise Rate (PNR) \citep{2011AJ....142...84F} is a primary indicator of the white noise although it is also affected by the red-noise component but to a lesser extent and the $\beta$-factor quantifies the red (or correlated) noise \citep{2008ApJ...683.1076W}. We computed both metrics for each of the light curves we made use of, after checking the timings of observations carefully and converting the light curves given in the timing frames other than BJD-TDB. We employed a simple Python code for measurements based on the definitions of these two metrics given in respective publications, and previously used in our study of the HAT-P-19 system \citep{2020MNRAS.496.4174B}. However, we have found out that low-quality light curves leading to incorrect mid-transit timings can survive the thresholds based on these two metrics, especially the one that is based on the PNR. Therefore we decided to follow the procedure we outline in the selection of light curves for analysis in this section.

First of all, we visually inspected all the available light curves in open databases, in the literature, as well as our own. On the basis of this visual inspection, we eliminated the obviously problematic cases as a result, including all the incomplete ones since the timing of a mid-transit heavily depends on the timings of the ingress and the egress. We then checked if the noise level is comparable to the transit depth, making it difficult to achieve a model consistent with the properties of the system and to measure an accurate and precise mid-transit time as a result. We made use of PNR for the purpose and if a light curve has a PNR value larger than or equal to the transit depth, we eliminated it. This ensured we eliminated the lowest-quality light curves but kept those that would help secure the sampling, especially in time intervals where there are not many observations although their error bars will also be large.

We also empirically determined a $\beta$-factor of 2.5 to exclude the red noise-dominated light curves from our analysis. Moreover, the depths of the light curves have been taken into consideration making use of the model output value $\delta$. We discarded the light curves with transit depths outside the 3$\sigma$ of the weighted average transit depth of the light curves of a given system in a given passband. Finally, we discarded the outliers in the final TTV diagrams based on two-tailed 3$\sigma$ deviations from a simple linear fit with least-squares minimization. At the end of the entire selection procedure, 150 light curves have been eliminated out of 550, leaving 399 light curves behind for TTV analysis. We relaxed the final criterion for WASP-69 because otherwise only one data point, its first-ever transit observation published by \cite{2014MNRAS.445.1114A}, would be eliminated, despite the fact that it is very precise, leaving a large gap of almost 4 years ($\sim1412$ days) without a transit observation after it. The other transit observation published in the same study was also eliminated because it doesn't meet the depth criterion whereas the transit parameters derived from this first light curve are in very good agreement with that from all the other light curves analyzed in this study. As a result, if we hadn't relaxed this criterion, the linear model's ability to correct its ephemeris would have been heavily influenced by more recent observations of lower quality.

Each of the light curves of a given system that survived the visual inspection was assigned a 4-digit binary code composed of 1's corresponding to the elimination due to the selection criterion in the order and 0's to that survived from the same criterion. The first digit is for the $\beta$-factor, the second for the PNR to transit depth ratio, the third for the transit depth ($\delta$) as a sanity check for the transit, and the fourth for the outliers in the TTV diagram. If a light curve meets the selection criterion then the corresponding digit to that specific criterion is assigned a value 0, or else 1 is assigned to indicate that it has been eliminated by our light curve selection procedure. These 4-digit binary codes for each of the light curves in our sample have been given in the data files corresponding to each of the systems. These digital binary codes and all the statistics employed in light curve selection are provided in data files as the online supplementary material. We also provide the mid-transit times reported by observers as well as our own measurements of them from the same light curves together with their errors and the passband of the observations in those data files.

\section{Data Analysis and Results}
\label{sec:analysis} 

\subsection{Light Curve Modelling and Measurements of Mid-Transit Times}
\label{subsec:measurements}
In order to measure the timing of the transit centre and to ensure the reliability of a selected light curve, we modelled it in {\sc exofast} \citep{2013PASP..125...83E}. We preferred the first version of the code ({\sc exofast}-v1) running on a browser through the NASA Exoplanet Archive service to have homogeneous measurements of mid-transit times based on the same planetary and stellar parameters from the NASA Exoplanet Archive as well as speed and consistency in terms of CPU hours. This also makes the measurements more consistent for larger groups of researchers since all the collaborators work on the same parameter sets. We provided the priors for the linear and quadratic limb darkening coefficients ourselves based on the passband of the observation and the fundamental parameters of the host stars after interpolations in the tables from \citet{2011A&A...529A..75C} using the online tool provided by the {\sc exofast} website\footnote{http://astroutils.astronomy.ohio-state.edu/exofast/limbdark.shtml} \citep{2013PASP..125...83E}. We selected the \emph{CoRoT} passband for light curves obtained without the use of a filter, and the $R$ band for the ones obtained with an $R_{c}$ filter, because the transmission curves are similar.

Prior to the analysis of the literature and database light curves obtained by professional and amateur observers, we converted the timings of all observations from the timing reference frames they were recorded in, to BJD-TDB by using our scripts that we developed based on the relevant modules and functions of the \emph{astropy} package \citep{2013A&A...558A..33A, 2018AJ....156..123A}. We also calculated the airmass values, using the same code to normalize and detrend the light curves that were not detrended for the effect. We made use of the AIJ software package for detrending, removing the obvious outliers, and reducing the raw data when necessary. Since all our light curves were detrended and normalized before the analysis as a result, we adjusted all the parameters, ran the {\sc exofast}-v1 code, and recorded all the results of the least squares fit that it performed. 

We used the output values of the models and their uncertainties for the transit depth ($\delta$) and the mid-transit time error as well as the goodness of fit included in the computation of the PNR and the $\beta$-factor  which are based on the {\sc exofast}-v1 model for the second level of light curve elimination. Mid-transit times and their uncertainties of the light curves that survived also in this second phase of light curve selection were used in our TTV studies. Since we had also taken the mid-transit timings and their uncertainties as reported by observers in open databases and literature, we were able to compare our results at the end. This comparison allowed us to find the inconsistencies, such as incorrectly reported values by the observers as a result of modelling errors, references to incorrect timing frames (e.g. heliocentric Julian days instead of geocentric Julian days, in which the original measurements were made), and conversion errors between the timing frames and units. Where such inconsistencies were found, we contacted the observers, made sure of their source, and asked for the raw data in critical cases, and repeated the measurements. Nevertheless, discrepancies between our and the observers' measurements and their uncertainties were expected although they were made on the same light curves because different software packages and methods have been allocated to derive the reported mid-transit times by the observers. Our measurements for all available light curves provide homogeneous sets of timing data. We discuss the discrepancies in more detail in Section~\ref{subsec:comparisons}.

\subsection{Ephemeris Corrections}
\label{subsec:ephemeris}
In order to update the reference mid-transit times (T$_0$) and the orbital periods (P$_{\rm orb}$) we used {\sc emcee} \citep{2013PASP..125..306F}, a widely used Markov Chain Monte Carlo (MCMC) sampler, and created posterior probability distributions of the linear coefficients; slope for the correction term of the orbital period ($\Delta$P$_{orb}$), and y-intercept for the correction term of the reference mid-transit time ($\Delta$T$_0$). We assumed Gaussian priors centered at zero for an unbiased fit. The posterior distributions are then calculated by making use of the Gaussian priors and the likelihoods of the random normal samples, where the noise of the measurements are assumed to be independent and identically distributed as random normal variables with zero mean and constant variance. The correction terms for the reference mid-transit time and the orbital period were obtained from the median values of the posterior probability distributions, which we provide in Appendix-\ref{sec:appendixb} for all the planets in our sample. In all computations, the first 500 steps (the so-called burn-in period) were discarded in each of the random walkers until an equilibrium is settled. As a result, we corrected the ephemeris information for all the planets in our sample and listed reference values in Table-\ref{tab:ephemeris} together with their uncertainties. Results have shown that the maximum uncertainty on a reference mid-transit time is 16.85 seconds and that on the orbital period is 0.10 seconds (for HAT-P-56 b). We should point out here that we followed the same probabilistic approach in fitting quadratic functions to the TTV diagrams as well.

\begin{table}
\centering
\caption{Reference ephemeris information (T$_0$ and P$_{\rm orb}$)}
\label{tab:ephemeris}
\begin{tabular}{lcc}  
\hline
\hline
Planet & T$_0$ & P$_{\rm orb}$ \\
 & BJD-TDB  & [days]  \\
\hline
HAT-P-23\,b & 2456539.390497(040) & 1.21288643(004) \\
WASP-37\,b & 2458225.646784(161) & 3.57747771(067) \\
WASP-69\,b & 2455845.538739(142) & 3.86813577(032) \\
WASP-74\,b & 2457218.764573(056) & 2.13775148(015) \\
HAT-P-56\,b & 2458822.558285(195) & 2.79082643(116) \\
WASP-2\,b & 2456823.839482(062) & 2.15222214(008) \\
WASP-14\,b & 2455632.578654(103) & 2.24376644(022) \\
HAT-P-32\,b & 2454424.747289(055)  & 2.15000831(005) \\
WASP-103\,b & 2457511.944555(016) & 0.92554549(003) \\
HAT-P-37\,b & 2457625.530538(056)  & 2.79744121(015) \\
\hline
\end{tabular}
\end{table}

\subsection{Transit Timing Analyses}
\label{subsec:ttv}
\begin{figure*}
\includegraphics[width=0.75\paperwidth]{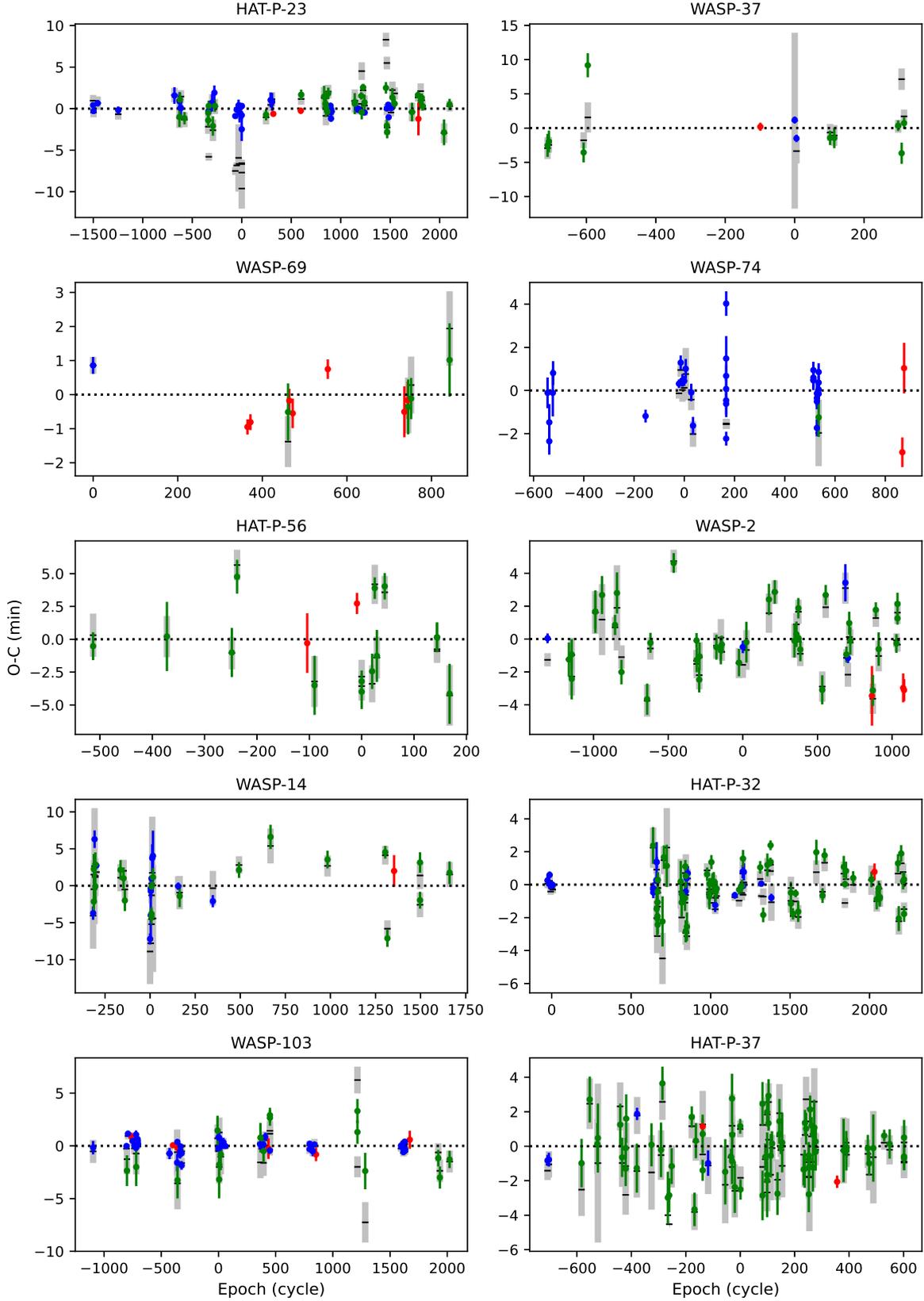}
\caption{TTV diagrams for all the planets in our sample based on observations from open databases (green), our observations (red), and light curves published in the literature (blue). The mid-transit times reported by the observers are shown in gray for comparison for the same light curves except our own.}
    \label{fig:ttv_plots_all}
\end{figure*}

We prepared TTV diagrams of all the planets in our sample by plotting the residuals of the observed mid-transit times from that calculated based on the corrected ephemeris information with respect to the epoch of observation (Fig.~\ref{fig:ttv_plots_all}). We provided both our own measurements from all the selected light curves, colour-coded according to the type of observation and the reported values of these timings by the observers (in light gray) in order to show how homogeneous measurements make a difference. We detail the TTV analysis we performed for each of these systems in the forthcoming subsections.

\subsubsection{HAT-P-23 System}
\label{sec:ttv_HATP23}
HAT-P-23\,b is a massive exoplanet (M$_{\rm p} = 2.090 \pm 0.111 M_{\rm jup}$) with an inflated radius (R$_{\rm p} = 1.368 \pm 0.090 R_{\rm jup}$) discovered by \cite{2011ApJ...742..116B} at a short period (P$_{\rm orb} = 1.212884 \pm 0.000002$ days). \citet{2011A&A...533A.113M} determined that the orbit is aligned ($\lambda = +15^{\circ} \pm 22^{\circ}$) from Rossiter-McLaughlin observations, and \cite{2014ApJ...781..109O} found that it is also circular from Spitzer observations of an occultation. Because an orbital decay is expected on the basis of its parameters, \citet{2018AcA....68..371M} and later \citet{2020AJ....159..150P} studied its transit timings and updated the lower limit of the reduced tidal quality factor for the star to Q$_{\star}^{\prime} > 5.6 \times 10^{5}$ and Q$_{\star}^{\prime} > 6.4 \pm 1.9 \times 10^5$, respectively, both in 95\% confidence levels. 

Our light curve selection procedure ended up with 73 mid-transit times spanning 11.96 years (3601 orbits in total), of which 33 are from amateur observers, 37 are from published light curves in the literature, and three from our own light curves. 12 light curves that we selected after our visual inspection of all the available light curves, have been eliminated due to not being able to meet our quantitative selection criteria. 

Relatively smaller discrepancies between the mid-transit times that we measure from a light curve and that reported by the observer should be expected. However, for a total of 10 points, the discrepancies are large enough to look for their sources. Therefore we contacted the observers and found that all 10 cases have been caused by reporting in a different timing reference than the original measurements were made in. The quality light curves published by \citet{2015A&A...577A..54C} constitute an especially interesting group which can be seen at and around BJD-TDB 2456500. We contacted the authors of this study and found out that the reported timings have been converted into BJD-TDB twice, once for the entire light curves, and the second time for the mid-transit times measured from them. When we corrected the timings, we had similar values, differing only by a few seconds at most. Nevertheless, we gave these timings as they were reported in order to point at potential problems that should be taken into account in a TTV analysis, and for consistency.

The residuals of the linear fit to the TTV diagram has a full range less than $\sim$10 minutes, which is significant considering the error bars of the measurements. Therefore we performed a frequency analysis with a Python code based on the \emph{astropy} function that computes Lomb-Scargle periodograms for time-series data. Since the frequency with the maximum power in the periodogram has a False Alarm Probability (FAP) value of 61.4\%, we concluded that within the range and frequency of our observations there is no statistically significant periodicity in the TTV diagram. We then fitted the timing data with a quadratic function and compared the results with that from the linear fit with the differences in Bayesian Information Criterion ($\Delta$BIC) and Akaike Information Criterion ($\Delta$AIC). The linear model turned out to be more successful in representing the data in both statistics. Even if the second-degree polynomial was found to be superior, its quadratic coefficient would be positive within almost $2\sigma$ (1.92 to be exact). Therefore the 5th percentile of the posterior probability distribution of the quadratic coefficient cannot be used to constrain a lower limit for Q$^{\prime}_{\star}$. Instead we provide the lower limit as Q$^{\prime}_{\star} > 3.8 \times 10^{6}$ in 99\% confidence level. As a result, it can be asserted that there is no periodic or secular change in the transit timings of HAT-P-23\,b within the limits of the observing window, the frequency of high-quality photometric observations, and observational uncertainties.

\subsubsection{Comparison Between Heterogeneous and Homogeneous Measurements}
\label{subsec:comparisons}
There are multiple sources of errors in a TTV analysis due to the usage of inhomogeneous datasets, from measurements made by different techniques and software packages giving different results and error estimates to inadequacies in timing of observations and reporting the results in different timing frames. In order to understand their consequences, we compared two data sets composed of the mid-transit times and their uncertainties as reported by the observers who published these timings either in open databases or in refereed publications, and that as measured by ourselves from the same light curves after validating the timing of observations.

\begin{figure}
\includegraphics[width=\columnwidth]{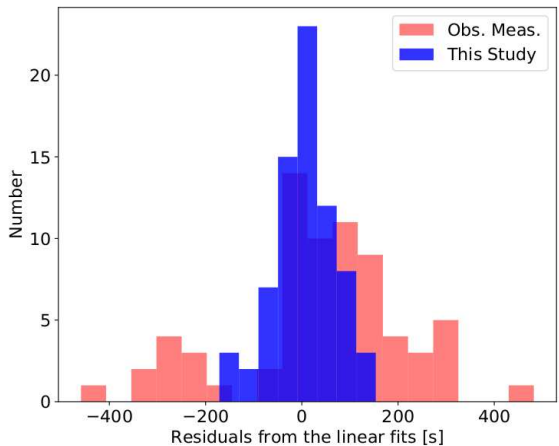}
\caption{Histograms of the residuals from the linear fits to our measurements of the mid-transit times (This Study, in blue) and the reported values of mid-transits (Observers' Measurements, in red) for HAT-P-23.}
    \label{fig:histogram_measurements_HATP23}
\end{figure}

First, we plotted the differences between mid-transit times, measured within this study and measured by the observers themselves who reported the values, with respect to the averages of both mid-transit time measurements. We selected HAT-P-23\,b as a sample case to explain our methodology because we have sufficient number of points for such an analysis.  The median difference for the transit measurements of HAT-P-23\,b indicated a positive bias between the two measurement sets, meaning that our measurements of the mid-transit times are $\sim11$ seconds larger compared to the reported values of the mid-transit times by the observers. The value of the median absolute deviation (MAD) of the differences (19 seconds); however, shows that the positive bias is not statistically significant. Since the mid-transit times reported by \citet{2015A&A...577A..54C} are found to be consistently smaller due to an error in their conversion to BJD-TDB, we excluded these measurements and repeated the analysis, and found a better agreement between the measurements with a median difference of $\sim8$ seconds, and a MAD of 28 seconds. This time three mid-transit times reported to the Exoplanet Transit Database (ETD) affected the magnitude of the bias considerably. We corrected the timings from \citet{2015A&A...577A..54C}, however, we were unable to find the sources of the deviations of three mid-transit times reported to the ETD. Since they were not eliminated by our light curve selection algorithm, we analyzed them based on the measurements from our model. We followed the same procedure for all the planets in comparison of the observer measurements and our own measurements of the mid-transit times. We provide the median of the differences and the corresponding MAD statistics in Table~\ref{tab:comparison_statistics} as well as the range and standard deviation of TTVs formed by two datasets for comparison, which are smaller for our measurements than the observer-reported values owing to our homogenesous analysis approach. This reduction of the scatter in data is illustrated in Fig.~\ref{fig:histogram_measurements_HATP23} for HAT-P-23\,b. Although median differences and associated MAD values do not indicate any disagreement between two sets of measurements, homogeneous measurements may help to recover potential periodicities with a better SNR since biases between measurement and data analysis techniques are avoided. Moreover, the advantage of working on a homogeneous set of measurements will be clearer for larger sets, and the comparisons with the original measurements will be fair. Consequently, we based our TTV analysis on our own measurements for the rest of this study.

\begin{figure}
\includegraphics[width=\columnwidth]{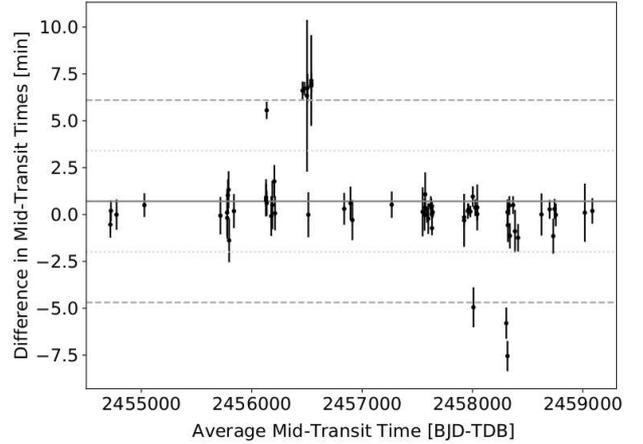}
\caption{Differences between the mid-transit times measured from {\sc exofast}-v1 models for this study and that reported by the observers with respect to the average of both measurements, for HAT-P-23. Sources of both the positive and negative outliers are provided in the text.}
    \label{fig:agreement_HATP23}
\end{figure}

\begin{table}
  \footnotesize
	\centering
	\caption{Comparison of the observers and our own measurements of the mid-transit times. Details of the comparison statistics are provided in the text.}
	\label{tab:comparison_statistics}
	\begin{tabular}{cccc} 
		\hline
		System & Range of O-C & $\sigma_{\rm O-C}$ & Median Diff. \\
                Name & our \& obs. & our \& obs. & $\pm$ MAD  \\
                 & [min] & [min] & [s]  \\ 
		\hline
		HAT-P-23 & 5.42 \& 15.69 & 1.11 \& 2.92 & 8 $\pm$ 28  \\
                WASP-37 & 12.84 \& 6.38 & 3.40 \& 1.87 & 7 $\pm$ 78  \\
                WASP-69 & 2.82 \& 3.55 & 0.93 \& 1.42 & -3 $\pm$ 38 \\
                WASP-74 & 6.90 \& 15.36 & 1.29 \& 3.50 & 27 $\pm$ 24  \\
                HAT-P-56 & 8.92 \& 8.11 &  2.89 \& 3.27  & 2 $\pm$ 37 \\
                WASP-2 & 8.29 \& 8.88 &  1.98 \& 1.78  & 12 $\pm$ 26  \\
                WASP-14 & 13.82 \& 14.15 &  3.39 \& 3.60   & 44 $\pm$ 72 \\
                HAT-P-32 & 5.28 \& 5.38 &  1.22 \& 1.06 & 17 $\pm$ 18 \\
                WASP-103 & 6.63 \& 13.49 &  1.17 \& 1.43 & 9 $\pm$ 11 \\
                HAT-P-37 & 7.31 \& 7.28 &  1.53 \& 1.66 & 14 $\pm$ 47 \\  
                \hline
	\end{tabular}
\end{table}

\subsubsection{WASP-37 System}
\citet{2011AJ....141....8S} discovered the transiting planet WASP-37\,b orbiting an old ($11^{+3}_{-4}$ Gyrs), metal-poor star. They noted that if the reduced tidal quality factor is $Q_{\star}^{\prime} > 10^7$, then the planetary spiral-in timescale should be longer than the main-sequence lifetime. They also argued that the planet has a larger radius than it should have, based on the extent of the expected photo-evaporation from its age. \citet{2019A&A...622A..81M} analyzed the transit timings of the planet accumulated until the time of their study and corrected the linear ephemeris as a result.

We analyzed 11 transit light curves of WASP-37\,b downloaded from open databases, two light curves from the literature \citep{2019A&A...622A..81M}, one UT50, and two T100 light curves (one of which is incomplete), adding up to a total of 15 light curves. Two more light curves were excluded after the second stage of the light curve selection procedure. As a result, 12 light curves formed the TTV diagram with a baseline of 10.06 years of observations. Although the observations form a sufficiently long baseline for a TTV analysis, the overall number of light curves is small. Therefore we only attempted at linear and linear+quadratic models of our measurements, the former of which was found to be superior to the latter in terms of $\Delta$BIC and $\Delta$AIC values. Consequently, we corrected the ephemeris information based on the parameters of the linear fit. The residuals from the linear fit are scattered within a range of $\sim$5 minutes if the literature minimum deviating from the linear trend by more than 10 minutes is ignored. More observations will be needed to clarify if the scatter is indicative of a real variation. 

\subsubsection{WASP-69 System}
WASP-69\,b is a Saturn-mass planet with an inflated atmosphere, orbiting a young ($\sim$1 Gyr) and magnetically active K-star in 3.868 days. It was discovered by \citet{2014MNRAS.445.1114A}, who obtained a global model of the transit and RV measurements after removing the spot-induced modulations on the light curves. They found that the eccentricity term in the RVs did not improve the results, so adopted a zero-eccentricity for the orbit. This was later confirmed by \citet{2019AJ....158..217W} based on Spitzer occultation observations. Since WASP-69\,b has an inflated atmosphere and is orbiting an active star, mass loss has been predicted due to evaporation driven by high-energy particles \citep{2014MNRAS.445.1114A}. Finally, \citet{2018Sci...362.1388N} detected excess absorption in the helium triplet at 1083~nm during a transit of WASP-69\,b, which they interpreted as the escape of part of the atmosphere trailing behind the planet in comet-like form. \citet{2020AJ....159..278V} confirmed their finding and constrained the mass loss rate to be $5.25^{+0.65}_{-0.46} \times 10^{-0.4} M_{\rm jup}$ per Gyr from their ultra-narrow band observations of a transit of WASP-69\,b.

WASP-69\,b is an overlooked planet in terms of ground-based transit observations. Only five transit light curves were found in the open databases of sufficient quality. The two published light curves are from the discovery study by \cite{2014MNRAS.445.1114A}. We observed the target twice with AUKR's T35 in order to make sure about the ephemeris information for the transit observations. In addition, we have three precise light curves from CAHA 1.23 m telescope, and two from the T100, which enabled us to form a TTV diagram covering 8.93 years. One of the light curves from \citep{2014MNRAS.445.1114A} was eliminated from our study since it does not meet the transit depth criterion. Consequently, the TTV diagram for WASP-69\,b is based on 13 light curves. The linear + quadratic model to our measurements performs better than the linear model based on the $\Delta$BIC and $\Delta$AIC values. However, there is a large gap between the first transit observation of the target, which is above the best fitting line, and the second observation. It is difficult to detect a secular change in the orbital period based on a dataset of only 13 points, with a gap of more than 350 epochs. As a result, we only corrected the ephemeris of WASP-69\,b, which will be targeted for further analysis of its inflated atmosphere in future ground-based and space-borne observations. The residuals of the linear fit have a scatter of less than 4 minutes, which is statistically significant when the measurement uncertainties are taken into account. In order to understand this potential variation, more observations are needed.  

\subsubsection{WASP-74 System}
\label{sec:ttv_WASP74}
WASP-74\,b was discovered by the SuperWASP team \citep{2015AJ....150...18H} around a late F-type (F9), bright (m$_{\rm V} = 9^m.7$), southern star. \cite{2019MNRAS.485.5168M} analyzed 18 light curves from 11 multi-colour planetary transits of WASP-74\,b, eight of which were obtained simultaneously in different filters in the NIR with the GROND instrument on MPG 2.2 m telescope in La Silla. They derived mid-transit times from the eight complete transits in their data set. They found the orbital period to be smaller by 0.475~s than the value in the discovery paper.

WASP-74\,b was classified as an inflated hot-Jupiter because its radius (R$_p$ = 1.56 $\pm$ 0.06 R$_{\rm jup}$) is relatively large for its mass (M$_p$ = 0.95 $\pm$ 0.06 M$_{\rm jup}$) \citep{2015AJ....150...18H}. However, the distance of the star was found to be d $= 120 \pm 20$ pc from its global modelling by \cite{2015AJ....150...18H}, whereas the Gaia parallax \citep{2016A&A...595A...1G, 2018A&A...616A...1G} gives a distance of $d = 148.029 \pm 1.037$ pc (after correction for the systematic offset noticed by \citet{2018ApJ...862...61S}. The small uncertainty in the Gaia parallax prompted us to refine the parameters of WASP-74 based on a global model of the best light curves in our data sets,  published and unpublished RVs from 29 archival HARPS spectra acquired in two different nights, and the directly measured distance from Gaia. \citet{2020AJ....159..137G} confirmed that the orbit is circular through Spitzer occultation observations so we fixed the eccentricity to zero.

We collected broadband photometric magnitudes from 2MASS \citep{2003yCat.2246....0C}, WISE \citep{2012yCat.2311....0C}, Tycho-2 \citep{2000AA...355L..27H}, SDSS \citep{2015ApJS..219...12A}, APASS \citep{2015AAS...22533616H} catalogues (listed in Table-\ref{tab:passband_brightness}), and fitted the Spectral Energy Distribution (SED) of WASP-74 using the second version of {\sc exofast} \citep{2017ascl.soft10003E, 2019arXiv190709480E}. We employed the atmospheric parameters (T$_{\rm eff}$, log~$g$, and [Fe/H]) obtained by \citet{2018A&A...620A..58S} as Gaussian priors with widths equal to their uncertainties. The effective wavelengths of the broadband photometry passbands were gathered from The Spanish Virtual Observatory (SVO) Filter Service \citep{2012ivoa.rept.1015R, 2013hsa7.conf..953R}. Finally, we limited the extinction value using the dust maps published by \citet{1998ApJ...500..525S}. We calculated the radius of the star R$_{\star}$ from our SED fit and provided its value as a Gaussian prior for the global modelling during which we fit the RV measurements, transit light curves, and stellar-isochrone tracks. Instead of using the uncertainty on R$_{\star}$ derived from the SED fitting, which is below the systematics in their calibration, we made use of 3.5\% of the radius value as a Gaussian width for the global modelling. This systematic error floor comes from the precision of the interferometric angular diameters \citep{2018MNRAS.477.4403W} and the disagreements in the measurements of bolometric fluxes between various methods \citep{2019ApJ...885..166Z}.

\begin{table}
\centering
\caption{Passband Brightnesses of WASP-74.}
\begin{tabular}{lcc}  
\hline
\hline
Passsband & $\lambda_{eff} \AA$ & Magnitude \\
\hline
\multicolumn{3}{l}{Tycho-2 \citep{2000AA...355L..27H}}\\
\hline
B$_T$ & 4280.0 & $10.533\pm0.042$ \\
V$_T$ & 5340.0 & $9.823\pm0.033$ \\
\hline
\multicolumn{3}{l}{APASS-DR9 \citep{2015AAS...22533616H}}\\
\hline
Johnson B & 4378.1 & $10.388\pm0.037$ \\
Johnson V & 5466.1 & $9.731\pm0.037$ \\
SDSS g' & 4640.4 & $10.102\pm0.066$ \\
\hline
\multicolumn{3}{l}{SDSS \citep{2015ApJS..219...12A}}\\
\hline
SDSS i' & 7439.5 & $9.432\pm0.002$ \\
\hline
\multicolumn{3}{l}{2MASS \citep{2003yCat.2246....0C}}\\
\hline
J & 12350.0 & $8.548\pm0.037$ \\
H & 16620.0 & $8.286\pm0.018$ \\
K$_S$ & 21590.0 & $8.221\pm0.023$ \\
\hline
\multicolumn{3}{l}{All WISE \citep{2012yCat.2311....0C}}\\
\hline
WISE1 & 33526.0 & $8.119\pm0.023$ \\
WISE2 & 46028.0 & $8.178\pm0.019$ \\
WISE3 & 115608.0 & $8.187\pm0.021$ \\
WISE4 & 220883.0 & $8.160\pm0.267$ \\
\hline
\label{tab:passband_brightness}
\end{tabular}
\end{table}

We added six new data points to the RV curve of WASP-74 from archival HARPS observations. There are 29 spectra recorded in two consecutive nights (17 and 18 August 2017) in the HARPS public archive lacking RV measurements. We downloaded the 1D-spectra reduced and corrected for the Earth's orbital motion by the HARPS pipeline. We normalized them to the continuum level with the latest version of the iSpec software package for spectral analysis \citep{2014ascl.soft09006B, 2019MNRAS.486.2075B}, making use of a synthetic spectrum as a visual template, that we created with the relevant tools of the same package based on the fundamental parameters of the star from \citet{2018A&A...620A..58S}. Then we measured the RVs by cross-correlating the observed spectra corrected for the orbital motion of the Earth with the spectral mask from the HARPS pipeline created for G2 stars. We binned the resultant RVs to form three data points for each of the nights in order to avoid the intra-night jitter after a few attempts with different amounts of binning. We provide the individual measurements and the RV values after binning within the online materials. There is an offset between the CORALIE and HARPS RVs either due to an instrumental offset or a change in a longer timescale than the baseline of each of the data sets. {\sc exofast}-v2 computes the RV of the barycentre of the system (V$_{\gamma}$) for both sets of measurements, and corrects for the offset in the RV fit, which is only $9.3~^{+8.2}_{-8.1}$ m/s. 

We selected only very precise and complete light curves from the literature and our own observations for global modelling. However, we used an incomplete light curve if it is the only one acquired in a certain passband for a wider wavelength coverage. That is why we included the RISE light curve observed in Johnson-V band (2014-08-19) given by \cite{2015AJ....150...18H}, and the GROND light curve in the K-band \citep{2019MNRAS.485.5168M} although they are affected by the white noise to a larger extent (PNR > 3.0). We observed the target on five nights with different telescopes. However, the light curves we acquired with the 43~cm telescope in Observatoire des Baronnies Proven\c{c}ales (OBP43), the 60 cm telescope IST60 in our network, and two light curves obtained with T100 in T\"UB\.{I}TAK National Observatory (T100) were either incomplete or affected by red noise. Hence, we decided to use the light curve acquired through a Bessell $R$ filter with T100 on 20/08/12 in the global modelling. Consequently, we were able to model 15 precise light curves of the target simultaneously with 20 CORALIE and six HARPS RV measurements, which form the largest and most precise dataset analyzed so far in the literature for WASP-74.  

We used {\sc exofast}-v2 \citep{2017ascl.soft10003E, 2019arXiv190709480E} for the global modelling of the light and RV curve data, fundamental atmospheric parameters derived from a detailed spectroscopic analysis by \citet{2018A&A...620A..58S},  and the brightness of the host star in different passbands. We enforced a Gaussian prior only on the orbital period ($P_{\rm orb} = 2.1377445 \pm 0.0000018$ days) determined by \cite{2019MNRAS.485.5168M} with a width of its uncertainty. We assumed a circular orbit for the planet because the eccentricity was found to be smaller than 0.07 in 3$\sigma$, and the circularization timescale is smaller than the age of the system \citep{2015AJ....150...18H}. The fundamental parameters of the stellar atmosphere and their uncertainties were taken from \citep{2018A&A...620A..58S}, and were assigned as Gaussion priors. The surface gravity was adjusted because light curve modelling of transits constrains the parameter better than spectroscopy \citep{2007MNRAS.379L..11S}. The coefficients of the quadratic limb darkening law are interpolated from the tables by \cite{2011A&A...529A..75C} during the light curve fitting process. Orbital inclination (cos~$i$) was assigned a uniform prior.

\begin{figure}
\includegraphics[width=\columnwidth]{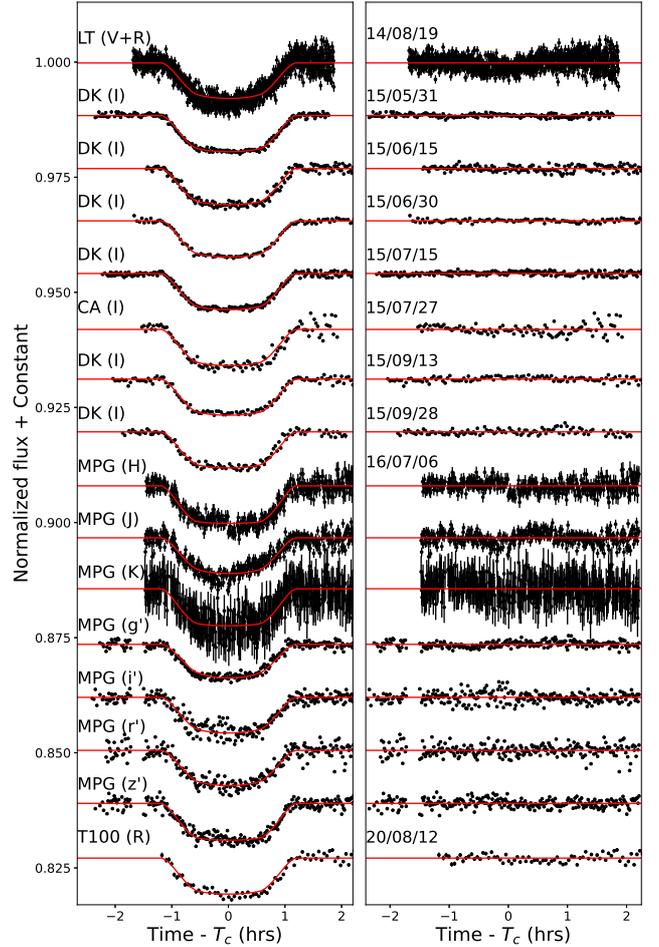}
\caption{Individual transit observations of WASP-74 (black data points) and their {\sc exofast}-v2 models (red continuous curve) to each of these light curves based on the parameters in Table-\ref{tab:global_parameters} (on left), and the residuals (on right). The telescope and the passband of observation are given on the upper left corner for each light curve. DK (1.54 m Danish Telescope at La Silla Observatory), CA (1.23 m telescope at CAHA), and MPG (2.2 m Max Planck Gesellschaft Telescope at La Silla Observator on 2016-07-06) light curves were obtained by \citet{2019MNRAS.485.5168M}, LT (2 m Liverpool Telescope at Roque de los Muchachos Observatory) light curve was published by \citet{2015AJ....150...18H}, T100 denotes 1m Telescope at TUG (this study). Observation dates are given in YY/MM/DD format next to the residuals.}

    \label{fig:wasp74_mcmc_transits}
\end{figure}

\begin{figure}
\includegraphics[width=\columnwidth]{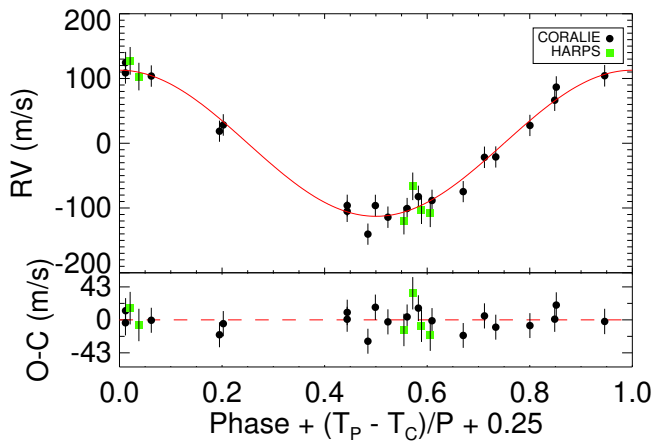}
\caption{Radial velocity observations from CORALIE \citep{2015AJ....150...18H} (black data points), and HARPS derived in this study (green data points) phased with the ephemeris information given in Table-\ref{tab:ephemeris}. The red curve represents our Keplerian model based on the parameters in Table-\ref{tab:global_parameters}. Residuals of the model are given in the lower panel.}
    \label{fig:wasp74_mcmc_rv}
\end{figure}

As a result, we achieved a global model that satisfactorily fits all the light curves in different passbands acquired with different telescopes, and the RV curves from two spectrographs. We list the parameters of our global model in Table~\ref{tab:global_parameters}, and provide the light curves in Fig.~\ref{fig:wasp74_mcmc_transits}, and the RV curve in Fig.~\ref{fig:wasp74_mcmc_rv} together with models based on those parameters.

\providecommand{\bjdtdb}{\ensuremath{\rm {BJD-TDB}}}
\providecommand{\feh}{\ensuremath{\left[{\rm Fe}/{\rm H}\right]}}
\providecommand{\teff}{\ensuremath{T_{\rm eff}}}
\providecommand{\ecosw}{\ensuremath{e\cos{\omega_*}}}
\providecommand{\esinw}{\ensuremath{e\sin{\omega_*}}}
\providecommand{\msun}{\ensuremath{\,M_\odot}}
\providecommand{\rsun}{\ensuremath{\,R_\odot}}
\providecommand{\lsun}{\ensuremath{\,L_\odot}}
\providecommand{\mj}{\ensuremath{\,M_{\rm J}}}
\providecommand{\rj}{\ensuremath{\,R_{\rm J}}}
\providecommand{\me}{\ensuremath{\,M_{\rm E}}}
\providecommand{\re}{\ensuremath{\,R_{\rm E}}}
\providecommand{\fave}{\langle F \rangle}
\providecommand{\fluxcgs}{10$^9$ erg s$^{-1}$ cm$^{-2}$}
\renewcommand*{\arraystretch}{1.3}
\begin{table}
   \scriptsize
\caption{Parameters of the global model for WASP-74 system and their 1$\sigma$ errorbars from {\sc exofast}-v2 compared to those from \citet{2015AJ....150...18H} (H2015).}
\begin{tabular}{lcc}
\hline \hline
Parameter (Unit) & Value & H2015\\
\hline
\multicolumn{2}{l}{Stellar Parameters:} \\
\hline
$M_{\star}$ (\msun) & $1.316^{+0.052}_{-0.053}$ & $1.48\pm0.12$\\
$R_{\star}$ (\rsun) & $1.578^{+0.024}_{-0.025}$ &  $1.64\pm0.05$\\
$L_{\star}$ (\lsun) & $3.03\pm0.14$ & -\\
$\rho_{\star}$ (cgs) & $0.472\pm0.011$ & $0.338\pm0.018$\\
$\log{g}$ (cgs) & $4.1608^{+0.0088}_{-0.0091}$ & $4.180\pm0.018$\\
$T_{\rm eff}$ (K) & $6064\pm42$ & $5990\pm110$ \\
$[{\rm Fe/H}]$ (dex) & $+0.465^{+0.026}_{-0.027}$ & $+0.39\pm0.13$\\
$Age$ (Gyr) & $3.43^{+1.00}_{-0.82}$ & - \\
$A_V$ (mag) & $0.215^{+0.052}_{-0.051}$ & -\\
$d$ (pc) & $148.2^{+1.9}_{-1.8}$ & $120\pm20$\\
\hline
\multicolumn{2}{l}{Planetary Parameters:}\\
\hline
$P_{\rm orb}$ (days) &$2.13775132^{+0.00000053}_{-0.00000055}$ & $2.137750 \pm 0.000001$\\
$T_C$ (\bjdtdb)&$2457173.871443\pm0.000080$ & -\\
$R_P$ (\rj)& $1.407^{+0.023}_{-0.024}$ & $1.56\pm0.06$ \\
$M_P$ (\mj) & $0.877^{+0.037}_{-0.038}$ & $0.95\pm0.06$\\
$a$ (AU) & $0.03559^{+0.00046}_{-0.00049}$ & $0.037\pm0.001$\\
$i$ (Degrees) & $79.95^{+0.12}_{-0.11}$ & $79.81\pm0.24$\\
$e$ & $0.00 {\rm (fixed)}$ & $0.00 {\rm (adopted)}$\\
$K$ (m/s) & $113.5\pm3.7$ & $114.1\pm1.4$\\
$T_{eq}$ (K) & $1947\pm16$ & $1910\pm40$\\
\hline
\multicolumn{2}{l}{Transit Parameters:}\\
\hline
$b$  & $0.8460^{+0.0031}_{-0.0033}$ & $0.860\pm0.006$\\
$\delta$ (fraction) & $0.008393\pm0.000051$ & $0.00961\pm0.00014$\\
$a/R_*$ & $4.849^{+0.038}_{-0.037}$ & -\\
$\tau$ (days) & $0.02574\pm0.00055$ & $0.0288\pm0.0014$\\
$T_{14}$ (days) & $0.09865^{+0.00038}_{-0.00037}$ & $0.0955\pm0.0008$\\
\hline
\multicolumn{3}{l}{RV Parameters:}\\
\hline
& CORALIE & HARPS \\
\hline
RV Offset (m/s) & $-15764.9^{+3.2}_{-3.1}$&$-15755.6^{+7.6}_{-7.5}$\\
RV Jitter (m/s) & $13.0^{+3.0}_{-2.3}$&$19.1^{+2.3}_{-3.1}$\\
\hline
\label{tab:global_parameters}
\end{tabular}
\end{table}

\subparagraph{TTV Analysis of WASP-74: } After obtaining its global parameters, we analyzed 42 transit light curves in total, including 3 light curves from open databases, 23 from the literature (7 from \cite{2015AJ....150...18H}, and 16 from \cite{2019MNRAS.485.5168M}), and 2 T100, 1 IST60, and 1 T35 light curves. 36 of these light curves survived the quantitive light curve elimination and formed the TTV diagram with an observation baseline of 8.305 years. We corrected the reference mid-transit time and the orbital period from a linear fit to this TTV diagram composed of our measurements of the mid-transit times. We not find any evidence for either a quadratic or a periodic variation in the analysis of the TTV diagram of WASP-74\,b.

\subsubsection{HAT-P-56 System}
\cite{2015AJ....150...85H} detected the inflated and massive hot-Jupiter type planet HAT-P-56\,b with the HATNet telescopes, later confirmed by \textit{Kepler}/K2 observations. Since the transit is grazing, the uncertainty on the radius was found to be large because the transit depth is highly dependent on the limb darkening parameters. The advantage of the large impact parameter is that the timings of the transit will be measured within more precision thanks to the V-shape of the light curve, and hence the system has more potential to be studied in terms of its TTV and TDV. An interesting finding is that the rotation period of HAT-P-56 (P$_{\rm rot} = 1.98 \pm 0.08$ days) turned out to be shorter than the orbital period of the planet HAT-P-56\,b (P$_{\rm orb} = 2.79$ days) \citep{2016MNRAS.460.3376Z}, which is in agreement with the period derived from a strong peak at $1.744 \pm 0.023$ days in the periodogram of the long cadence K2-light curve of the target and the spectroscopic projected rotational velocity (${\rm v~sini} = 40.1$ km/s) found by \citet{2015AJ....150...85H}. This can be indicative of a weak tidal interaction between the star and the planet or the planet might have been pushed away by the tides from the star from an even closer orbit, which might also be detected in its TTVs.

Only 18 transit light curves were found in open databases for HAT-P-56\,b. We observed the target with T35 and T100 in 2019. Unfortunately, the light curves published in the literature were not adequate in deriving reliable mid-transit times, and four of the 20 light curves we had were eliminated during our second stage of light curve selection based on quantitative criteria. Therefore, the resultant TTV diagram was formed by only 16 data points in total, which can be best described with a linear model. We corrected the reference light elements as a result and listed them in Table~\ref{tab:ephemeris}. The residuals of this linear fit scatter within a full range of $\sim$10 minutes, which should be tracked with future transit observations of the target. Although the orbit was also found to be aligned ($\lambda = 7^{+2}_{-2}$ deg) by \citet{2016MNRAS.460.3376Z}, the RVs of the system obtained to verify the planetary nature of the companion with TrES spectrograph can be modelled with a slight eccentricity (e $ = 0.130 \pm 0.058$) \citep{2015AJ....150...85H}, which strengthens its importance for future TTV studies.

\subsubsection{WASP-2 System}
WASP-2\,b is the second planet discovered during the SuperWASP survey for transiting planets \citep{2007MNRAS.375..951C}, orbiting its star in a retrograde orbit (projected angle is $\lambda = -153^{+15}_{-11}$ degrees) \citep{2010A&A...524A..25T}. \cite{2009A&A...498..567D} identified a nearby contaminating source in their Lucky Imaging survey. This potential stellar companion was studied several times \citep{2013MNRAS.428..182B, 2015ApJ...800..138N,2015ApJ...814..148P, 2015A&A...575A..23W, 2016A&A...589A..58E}. VLT/SPHERE observations by \cite{2020A&A...635A..73B} confirmed that this visual companion is comoving with the WASP-2 system. \cite{2020A&A...635A..74S} corrected the RVs for this contamination and reanalyzed the system, and refined its parameters as a result. \cite{2019PASP..131k5003A} observed a transit of WASP-2\,b with a high precision with MINERVA telescopes in the Sloan $r^{\prime}$ filter. Although they found a significantly smaller period, they did not attempt an explanation. Despite an expectation of an orbital decay from the system based on its parameters, \cite{2019MNRAS.490.1294B} studied the TTV of the planet WASP-2\,b and were able to fit the transit timings with an upward parabola implying an orbital period increase with $\frac{dP}{dn} = 3 - 10 \times 10^{-10}$ days per cycle. From their fit to previous RV observation, they found a deceleration of $-3.9$ m s$^{-1}$ yr$^{-1}$. Their global fit to RV and TTV resulted in an slightly eccentric orbit  ($e = 0.0134 \pm 0.0056$) with an orbital decay time scale (linear time scale of the orbital period to be zero) of T$_{d} = \frac{1}{q} = \frac{P}{\dot P} = -12 \pm 11$ Myr. Non-zero eccentricity, which was assumed to be zero in all previous analyses, was also confirmed by Spitzer occultation observations \citep{2010arXiv1004.0836W}.

We assembled 57 complete transit light curves of WASP-2\,b, with sufficient quality for a TTV analysis. The open databases have 50 transits, there is one from  \cite{2019PASP..131k5003A}, one from \cite{2017MNRAS.472.3871T}, one from \cite{2007ApJ...658.1322C}, one from \cite{2020AcAT....1...13I}, and three from our own observations (T35, ATA50 and T100). 47 of these light curves survived our second stage of the light-curve elimination procedure, forming the TTV diagram with a 14.07 year-baseline. Having corrected the TTV diagram with an updated ephemeris, we performed a frequency analysis using a Lomb-Scargle periodogram. We found a peak at 268.31 days, with a False Alarm Probability of 38\%, which is not significant. We then attempted to fit the timings with a linear + quadratic model, which turned out to be inferior to the linear lit. However, a $\sim$10-minute range scatter, larger than the observational error bars, is observed in the residuals of the linear fit that we present in Fig.\ref{fig:ttv_plots_all}.

\subsubsection{WASP-14 System}
WASP-14\,b is a dense planet ($\rho_{\rm p}$ = 3.501 g~cm$^{\rm -3}$) in a short-period (P$_{\rm orb} \sim 2.24$ days), slightly eccentric ($e \sim 0.09$), and misaligned ($\lambda = +33^{\circ}.1 \pm 7^{\circ}.4$) orbit \citep{2009MNRAS.392.1532J}. The eccentricity of the orbit was later confirmed based on the Spitzer measurements as $e = 0.087 \pm 0.002$ by \cite{2013ApJ...779....5B}. On the basis of its parameters and the fiducial values of tidal dissipation parameters for both the star and the planet from \citet{2008ApJ...678.1396J}, \citet{2009MNRAS.392.1532J} found that the timescales of both circularization and decay of the orbital semi-major axis should be smaller than 1 Gyr. On the other hand, the age of the star was found to be between 0.5 and 1 Gyr, both from its lithium abundance and from stellar evolution models. Therefore, the WASP-14 system may be very close to the lower limit for the point at which orbital eccentricity can survive tidal evolution for a sufficiently long time. If this is not the case, and the reduced tidal quality factor is not large enough (Q$^{\prime}_{\star} > 6.6 \times 10^6$) for the planet to survive on an eccentric orbit, then an unseen potential third body can explain the small but statistically significant eccentricity of the orbit. \cite{2015ApJ...800..138N} detected a $0.33 \pm 0.04 M_{\odot}$ stellar companion to WASP-14 at a separation of $300 \pm 20$ au, which was later found to be an early-type background source from flux ratios by \cite{2015A&A...575A..23W}. \cite{2015MNRAS.451.4139R} provided the first TTV analysis of the system, but found no significant periodic signal in their frequency analysis based on a 7 year-long baseline of observations including their own 19 light curves of 13 transit events. Nevertheless, they reported an orbital period 1.2 seconds longer and 10 times more precise than the previous value.

We analyzed 39 transit light curves of WASP-14\,b, 24 of which have been reported to the ETD, 13 of which appeared in the literature, and two of which are our own observations with ATA50 and T35. 31 of these light curves were found to be of sufficient quality by our selection criteria. We attempted linear a linear + quadratic fits to the data. Positive $\Delta$BIC and $\Delta$AIC values showed that the latter model was more successful in describing the data set formed by our own measurements. The case was the same for the data set composed of the reported values of the mid-transit times. The orbital period of WASP-14\,b is not expected to increase because the timescale for an orbital decay was found to be less than 1 Gyr, while the lithium abundance and the location of the host star on isochrones both give an age value between 0.5 and 1.0 Gyr \citep{2009MNRAS.392.1532J}. However, the only other TTV analysis of the system by \citet{2015MNRAS.451.4139R} also resulted in a larger orbital period than the previous value. On the other hand, the residuals from both linear and the linear + quadratic models have scatters larger than 15 minutes. Non-zero eccentricity increases the probability of the existence of an unseen third body on a larger orbit, that would explain both the eccentricity and the statistically significant TTV variation. Our analysis is based on only 31 precise and complete transit light curves spanning a baseline of 12.15 years, which is an improvement on the data set that \citet{2015MNRAS.451.4139R} used in their study. We searched for periodicity in the residuals of the linear fit with a Lomb-Scargle periodogram, finding the peak of maximum power to be at 59.06 days. However, its FAP is only 5\% so is not sufficient to claim any periodicity.

\subsubsection{HAT-P-32 System}
HAT-P-32\,b is orbiting a relatively young star ($2.7 \pm 0.08$ Gyr), which is a magnetically active star manifested by its high RV jitter of over 80 m/s \citep{2011ApJ...742...59H}. There has been dispute on the orbital eccentricity of the system in the past based on the RV models \citep{2011ApJ...742...59H, 2013MNRAS.436.2974G, 2014ApJ...785..126K}; however, the orbital phase of the occultation observations by Spitzer revealed the circular nature of the orbit \cite{2014ApJ...796..115Z}. The linear trend found by \citet{2014ApJ...785..126K} in the RV observations with a slope of $-33 \pm 10$ m s$^{\rm -1}$ yr$^{\rm -1}$ was not found to be caused by the nearby companion detected by \citet{2013AJ....146....9A} and later \cite{2014ApJ...796..115Z} from flux ratios in multiple visual and IR passbands. \cite{2014MNRAS.441..304S} analyzed all the precise transit timings of the system available at the time, and ruled out TTV amplitudes larger than $\sim1.5$ minutes. \citet{2019AJ....157...82W} analyzed more mid-transit times accumulated until the time of their study and reached the same conclusion. 

We found 104 complete transit light curves of HAT-P-32\,b in the open databases (81) and the literature (22 in total from \cite{2011ApJ...742...59H}, \cite{2012PASP..124..212S}, \cite{2018MNRAS.474.5485T}, and \cite{2019AJ....157...82W}), and observed the target once with T35 to check the ephemeris information. Although HAT-P-32 was observed by TESS in Sector-18 (November 2019), the observations were in the long-cadence (30 minutes), which is not adequate to acquire precise mid-transit timings. On the basis of our quantitative criteria, we selected 82 of the light curves in our sample. These are evenly distributed apart from a gap of 3.175~yr after the first few data points from the discovery paper \citep{2011ApJ...742...59H}. A Lomb-Scargle periodogram has no peaks with a FAP smaller than 12\%, which is for an unrealistically small period of 23.71 day. We thus ruled out a periodicity in TTV within the limits of the data set, and only corrected the ephemeris information for the system (Table~\ref{tab:ephemeris}). The residuals of the linear fit have a range of $\sim$5 minutes.

\subsubsection{WASP-103 System}
\citet{2014A&A...562L...3G} announced the discovery of the ``ultra'' hot-Jupiter WASP-103\,b on a very short-period (P$_{\rm orb}< 1$ day) orbit, larger than its Roche radius by only 20\%, making it a candidate for mass loss and eventual tidal disruption, hence a decrease in its orbital period. \citet{2015A&A...579A.129W} detected a nearby source at an angular distance of $0\arcsec.242 \pm 0\arcsec.0.016$ from WASP-103. \citet{2016MNRAS.463...37S} reanalyzed their light curves they presented in \citet{2015MNRAS.447..711S}, characterized this visual companion, and found that its effect on the measured physical properties of the host star will be small. \citet{2017AJ....153...34C} derived the effective temperature of this visual companion as T$_{\rm eff} = 4400 \pm 200$ K from their SED fitting for the visual companion, which is consistent with the results of \cite{2015ApJ...800..138N}, and that of \citet{2015MNRAS.447..711S}, who reported a mass value of $0.72 \pm 0.08 M_{\odot}$ for this star. The system is not resolved by Gaia in either the DR2 and EDR3 data releases. \citet{2017AJ....153...34C} corrected their secondary eclipse observation for this companion, and found that its timing was consistent with zero eccentricity. \citet{2017MNRAS.472.3871T} presented the first near-UV transits (Bessell $U$) of WASP-103\,b, and constructed a TTV plot with the data accumulated until the time of their study. However, they found no variation in its transit timings. \citet{2018AcA....68..371M} also studied the TTV diagram but did not find a variation other than the linear ephemeris suggested. They constrained the reduced tidal quality factor ($Q^{\prime}$ to be larger than $10^6$ at $3.5\sigma$ confidence. \citet{2018MNRAS.474.2334D} analyzed new and precise transit and occultation light curves and RV data, and found consistent results with the previous findings. \citet{2020AJ....159..150P} studied the TTV of the system based on the most precise light curves of the target and four light curves they acquired with FLWO. Their best-fit orbital decay model resulted in a positive period change. They constrained the reduced tidal quality factor to be $Q^{\prime} > (1.1 \pm 0.1) \times 10^5$ with 95\% confidence as a result, contradicting the previous findings and potential mass-loss expected from the nature of the planet. Most recently \citet{2022A&A...658C...1B} also found an increase in the orbital period based on their data set extended by their own observations with CHEOPS as well as re-reduced HST and Spitzer data. However, their quadratic model is not statistically superior to the linear model.

WASP-103\,b is a frequently studied planet owing to its very short period (P < 1 day), leading to an expectation of a decay in its orbit. We observed the transits of the planet in high precision with 1 m T100 three times, and 1.23 m telescope of CAHA twice in addition to an observation with T35 with inferior quality. We collected 55 precise light curves from the literature, including 14 from \cite{2015MNRAS.447..711S}, the TRAPPIST light curve from \citep{2014A&A...562L...3G}, three from \cite{2017A&A...606A..18L}, one light curve from \citep{2017MNRAS.472.3871T}, two from \citep{2018MNRAS.474.2334D}, eight in total from \cite{2018AcA....68..371M}, four from \cite{2020AJ....159..150P}, 10 very precise light curves obtained with different instruments to form a transmission spectrum of WASP-103\,b from \cite{2021AJ....162...34K}, and most recently twelve light curves from \cite{2022A&A...658C...1B}, four of which we had to eliminate due to insufficient sampling for timing analyis. We found 22 complete light curves in the ETD, five of which were eliminated by our selection criteria. We continued our analysis with our measurements of the conjunction times of 75 light curves selected after our second stage of the selection procedure spanning 7.90 years. The linear model turned out to be superior to the linear + quadratic model, the quadratic term of which is found to be negative. However, until the time of the recent publication of very precise transit observations of WASP-103\,b by \cite{2021AJ....162...34K}, the trend that was observed in the TTV was an upward parabola as found by \cite{2020AJ....159..150P} and \cite{2022A&A...658C...1B}, which lack the data from \cite{2021AJ....162...34K} in their analysis,  in contrast of the expectation of an orbital decay from the ultra-short period. These observations seem to have changed the picture and showed the need for further precise observations of the target to reveal its orbital period behaviour. Nevertheless, we determined a lower limit for Q$^{\prime}_{\star}$ as $> 1.4 \times 10^6$. Although it is difficult to argue for a period change in its TTVs with only 75 data points, a peak at frequency corresponding to a periodicity of 76.15 days was found with a FAP value of 1.63\% in its Lomb-Scargle periodogram, which we note for future observations of the target with a reservation that it probably arises from seasonal sampling.  

\subsubsection{HAT-P-37 System}
HAT-P-37 b is a hot-Jupiter discovered in the HAT-Net survey by \citet{2012AJ....144...19B}. \citet{2016AcA....66...55M} published two complete and two incomplete transit light curves of the target. They analyzed them and obtained similar results to \citet{2012AJ....144...19B}. They computed the mid-transit timings of their own light curves together with that of \citet{2012AJ....144...19B}, constructed the TTV diagram, and corrected the ephemeris information from a linear fit for the uncertainties on reference light elements. Then \citet{2017MNRAS.472.3871T} observed the system in Harris B and R filters on July 1, 2015. They found a difference in R${_P} / R_{\star}$ between the values they derived from the transit depths in $B$ and $R$ bands by 2.6 $\sigma$, which could be indicative of a TiO / VO absorption in its atmosphere. 

We analyzed 91 transit light curves of HAT-P-37\,b. Four of these light curves are from the literature \citep{2011ApJ...742..116B, 2018AcA....68..371M}, 85 are from ETD, and two are from our own high-precision observations with T100. TESS has also observed HAT-P-37 in both long and short cadence, but these data are afected by a star 29\arcsec.62 with large intrinsic variability. The two sources cannot be resolved by TESS and have different distances from Gaia: 377.3 $\pm$ 6.1 pc for HAT-P-37 and 667.7 $\pm$ 54.2 pc for the background star. Although we were able to recover the transit signal when the light curve is folded in orbital phase, it is impossible to measure reliable times of midpoint for the individual transits due to the contaminating light. 76 transit light curves were selected in the second stage of light curve elimination, which resulted in a rejection of 15 light curves from open databases. The linear + quadratic model is slightly better than the linear model in terms of $\Delta$BIC and $\Delta$AIC, this is not statistically significant (4.26 and 6.59, respectively). Therefore we preferred the simpler linear model and corrected the ephemeris based on its parameters. The residuals from this model have a scatter less than 10 minutes. Since the number of data points is sufficient, and they are evenly distributed over a 10.033 year-baseline, we carried out a frequency search. We found that the peak with the maximum power (at a period of 61.87 days) has only 11.3\% FAP so is not significant.

\section{Summary and Discussion}
\label{sec:conclusions}
\subsection{Discussion on the Parameters of WASP-74\,b}
\label{subsec:discssion_parameters}

We attempted at a global model of WASP-74 which we found necessary before a TTV analysis of the system because the reference distance value in the literature (d $ = 120 \pm 20$ pc; \citealt{2015AJ....150...18H}) of WASP-74 is smaller by more than $1\sigma$ from that measured by Gaia (d $= 148.029 \pm 1.037$ pc). Our results agree with that found by \citet{2019MNRAS.485.5168M} in terms of both stellar and planetary parameters; hence we confirm their findings, while masses and radii of both the host star and the planet are smaller compared to what had been found by \citet{2015AJ....150...18H}. Our analysis was based on a confirmed value of zero eccentricity by the orbital phase of the occultation observation from Spitzer \citep{2020AJ....159..137G}, which ended the discussion on a potentially eccentric orbit. The evolutionary stage of the host star was found from MESA-isochrones (MIST) based on a global model of all available data and a semi-empirically derived stellar radius. As a result, we found an age value of $3.43^{+1.00}_{-0.82}$ Gyrs, consistent with and  in between that given by \citet{2015AJ....150...18H} and \citet{2019MNRAS.485.5168M}.

\subsection{Discussion on Transit Timing Variations}
\label{subsec:discussion_ttv}
We updated the ephemeris information of all 10 transiting exoplanets in our sample based on a homogeneously compiled set of mid-transit time measurements, resulting in the most precise light elements (T$_0$, P$_{\rm orb}$) so far,  which will be useful for future observations.  The ephemeris information listed in Table~\ref{tab:ephemeris} is based on the most reliable light curves of each target owing to the selection criteria we employed and homogeneous measurements of the mid-transit times from models constructed with the same parameter sets. This approach also helped to decrease the scatter observed in the TTV diagrams for most of the cases significantly. 

In each case, we compared the linear + quadratic model with the best linear model in terms of Bayesian and Akaike Information Criteria. On this basis, the linear + quadratic model was found superior only for WASP-14\,b and WASP-69\,b. However, there is a large gap of almost 4 years in WASP-69\,b's TTV data, which is composed only of 13 light curves. WASP-14\,b has only 31 reliable light curves, which is also not sufficient to argue for a secular change in its TTVs. In such cases, addition of new precise data can change the situation dramatically as we experienced in the TTV analysis of WASP-103\,b. Before the inclusion of 10 precise light curves from \cite{2021AJ....162...34K}, our results were in agreement with the finding of \cite{2018AcA....68..371M}, \citet{2020AJ....159..150P}, and most recently \cite{2022A&A...658C...1B}, which also lacks these data points, of an increase in the orbital period against the expectation of an orbital decay from this ultra short-period hot-Jupiter (P$_{\rm orb} \sim 0.926$ days). When the recent data were added, a negative quadratic coefficient was found. However, the quadratic model is not statistically superior to the linear model and the change in the orbital period is consistent with zero within less than 1$\sigma$.

We also searched for periodicity in the residuals from the linear model with Lomb-Scargle periodograms. However, we have not found any reliable periods in the transit timings of our targets smaller than a False Alarm Probability (FAP) value of 1\% other than that found in the TTV diagrams of WASP-69\,b and HAT-P-56\,b which have only 13 and 16 points, respectively. We found a minimum FAP value of $\sim5$\% for WASP-14\,b at 59.08 days, which also has the largest scatter in our sample with a range larger than 15 minutes. This planet has a non-zero eccentricity orbit (e = 0.087 $\pm$ 0.02), confirmed by Spitzer occultation observations. Therefore, its TTV analysis should be refined in the future, with more data than the 31 reliable light curves that we selected and analyzed.

We provide 35 new transit observations of the planets in our sample. They are shown in Appendix~\ref{sec:appendixa} and their data are included in the online materials. In addition, the mid-transit time measurements from our light curves and those of other observers, together with all the related statistics and information, are listed in a file for each of the planets in the online materials. 

\subsection{Discussion on Tidal Quality Factors}
\label{subsec:tidal_quality}
We did not find any statistical evidence for either periodic or secular changes in the O-C diagrams of our targets, some of which are especially promising candidates (HAT-P-23\,b and WASP-103\,b in particular) to detect orbital decay. Since the tidal quality factor is an informative parameter for the efficiency of the host star (Q$_{\star}^{\prime}$) to dissipate the energy arising from the tidal interactions with the close-in giant planet, we calculated lower limits for the reduced form of the parameter by making use of the formulae provided by \cite{2014ARA&A..52..171O} and \cite{2018AcA....68..371M}, the fifth percentile of the posterior probability distribution for the quadratic coefficient of the second degree polynomial fits, and the system parameters we gathered from the references in Table-\ref{tab:planetdata} except for WASP-74\,b for which we made use of our model values in Table-\ref{tab:global_parameters} for consistency. We list these lower limits in 95\% confidence based on the results of our analyses of their respective O-C diagrams together with $\Delta$AIC and $\Delta$BIC values, showing why the quadratic models were not found to be statistically significant, and the rates of change in the orbital periods implied by the best fitting quadratic functions in Table-\ref{tab:tidal_quality_factors}. The exceptions are HAT-P-23 (denoted with an asterisk in Table-\ref{tab:global_parameters}) for which we only computed the lower limit of Q$^{\prime}_{\star}$ within 99\% confidence, and WASP-14 \& WASP-69, for which the positive quadratic coefficient is statistically significant. Nevertheless, we cannot argue an increase in the orbital period for the planets in these latter two systems because we do not have enough data points sufficiently sampling the baseline.

\begin{table*}
\centering
\caption{Lower limits for the Reduced Tidal Quality Factors (Q$_{\star}^{\prime}$) for the host stars in our sample in 95\% confidence level. The lower limit for HAT-P-23\,b has 99\% confidence. The positive signs of the orbital period change in WASP-14\,b and WASP-69\,b are statistically significant. $\Delta$AIC and  $\Delta$BIC values were used for comparisons between linear and quadratic models in this order. A negative value, therefore, means that the linear model should be favored.}
\label{tab:tidal_quality_factors}
\begin{tabular}{lccrrc}  
\hline
\hline
System & $\Delta$AIC & $\Delta$BIC & dP/dE (days/cycle) & $\dot{P}$ (ms/yr) & Q$_{\star}^{\prime}$ \\
\hline
HAT-P-23 & -0.558  & 1.732  & $(1.44 \pm 0.75) \times 10^{-10}$ &  $3.7 \pm 2.0 $ & $> 3.8 \times 10^{6 *}$ \\
WASP-37 & -2.316  & -1.831 & $(-3.64 \pm 3.48) \times 10^{-9}$ & $-32.1 \pm 30.7$ & $> 5.0 \times 10^2$  \\
WASP-69 & 17.987  & 18.472 & $(9.96 \pm 2.32) \times 10^{-9}$ & $81.2 \pm 19.0$ &  -  \\
WASP-74 & 7.813  & 9.368 & $(-2.86 \pm 0.83) \times 10^{-9}$ & $-42.1 \pm 12.4$ & $> 5.0 \times 10^{3}$  \\
HAT-P-56 & -3.060  & -2.288  & $(0.13 \pm 1.09) \times 10^{-8}$ & $149.3 \pm 124.0$  & $> 1.2 \times 10^3$ \\
WASP-2 & -3.856  & -2.005 & $(0.09 \pm 2.43) \times 10^{-10}$ & $0.1 \pm  3.5$ & $> 5.7 \times 10^3$  \\
WASP-14 & 20.422  & 21.856 & $(4.84 \pm 0.83) \times 10^{-9}$ & $68.3 \pm  11.7$ & -  \\
HAT-P-32 & -1.902  & 0.346 & $(1.91 \pm 1.25) \times 10^{-10}$ & $2.8 \pm 1.8$ & $> 5.9 \times 10^5$   \\
WASP-103 & -4.062 & -1.757 & $(-3.23 \pm 6.64) \times 10^{-11}$ & $-1.1 \pm 2.3$ & $> 1.4 \times 10^6$ \\
HAT-P-37 & 4.257 & 6.588 & $(-2.41 \pm 0.77) \times 10^{-9}$ & $-27.2 \pm 8.8$ & $> 5.5 \times 10^2$  \\
\hline
\end{tabular}
\end{table*}

HAT-P-23 and WASP-103 are two systems that deserve a mention for their tidal quality factors due to their very short orbital periods. Although a linear model was found to be superior to a quadratic model of the TTVs observed in HAT-P-23, we were able to constrain HAT-P-23's reduced tidal quality factor as Q$_{\star}^{\prime} > 3.8 \times 10^6$ only in 99\% confidence. This value is in agreement with some of the theoretical expectations \citep{2011ApJ...731...67P,2012ApJ...751...96P,2014ARA&A..52..171O, 2019AJ....158..190H} while it exceeds some others \citep{2008ApJ...678.1396J,2016APS..APR.L1022E}. The value of the quadratic coefficient of the best fitting parabola to the TTV diagram of WASP-103\,b suggests an orbital period decrease with a rate of $\dot{P} = -1.1 \pm 2.3$ miliseconds per year. Although it is statistically inferior to the linear model and it is consistent with zero given the magnitude of its uncertainty, a lower limit on the reduced tidal quality factor can be determined. Based on the stellar and planetary parameters from \cite{2014A&A...562L...3G}, we computed this limit as  Q$^{\prime}_{\star}$ > $1.4 \times 10^6$ for WASP-103 in 95\% confidence.

More observations, spanning longer baselines are needed to reveal the nature of the TTVs expected from WASP-14\,b \citep{2019MNRAS.485.5168M} based on its statistically significant non-zero eccentricity, and any other target in our sample. The lower limit we derived for Q$^{\prime}_{\star}$ is smaller than the $6.6 \times 10^6$ needed for it to survive on its orbit. There is an expectation for TTVs from WASP-2\,b as well from the linear trend observed in the RVs of its host star. \cite{2019MNRAS.490.1294B} found a positive derivative for the orbital period, which is supported by our finding ($\frac{\rm dP}{\rm dt} = (0.04 \pm 1.13) \times 10^{-10}$ seconds per year), although the quadratic fit was found to be inferior to a linear-only fit. All these systems need further and more precise follow-up observations to reveal the nature of potential TTVs in them. Therefore, we urge observers to obtain more light curves with good photometric precision of these targets. Future observations from the TESS and CHEOPS satellites will help tremendously in terms of characterizations of these planets as well.

\section*{Acknowledgments}
We gratefully acknowledge the support by The Scientific and Technological Research Council of Turkey (T\"UB\.{I}TAK) with the project 118F042. We thank T\"UB\.{I}TAK for the partial support in using T100 telescope with the project numbers 12CT100-378 and 16CT100-1096. We thank the observers in AUKR, UZAYMER, ATASAM, TUG, OBP, IUGUAM observatories, and all the observers around the world who kindly shared their data with us. This research has made use of data obtained using the ATA50 telescope and CCD attached to it, operated by Atat\"urk University Astrophysics Research and Application Center (ATASAM). Funding for the ATA50 telescope and the attached CCD have been provided by Atat\"urk University (P.No. BAP-2010/40) and Erciyes University (P.No. FBA-11-3283) through Scientific Research Projects Coordination Units (BAP), respectively. Authors from Bilkent University thank the internship program of Ankara University Kreiken Observatory (AUKR). IST60 telescope and its equipment are supported by the Republic of Turkey Ministry of Development (2016K12137) and Istanbul University with the project numbers BAP-3685, FBG-2017-23943. We thank Jason Eastman for his help in using both versions of the {\sc exofast} code. New RV measurements from HARPS spectra are based on data obtained from the ESO Science Archive Facility under request number 556296. NA thanks to \c{C}ukurova University Scientific Research Center with the project number of FYL-2019-11726. NA also thanks T\"UB\.{I}TAK National Observatory for its support in providing the CCD to UZAYMER. We thank all the observers who report their observations to ETD, ExoClock, and AXA open databases.

\section{Data Availability}
Some of the light curves to derive mid-transit times were downloaded from Exoplanet Transit Database at http://var2.astro.cz/ETD/. All other light curves appearing for the first time in this article are presented as online material. Mid-transit times derived from our own light curves as well as that of other observers' are presented as online data sets too. Individual radial velocity values derived from the HARPS instrument for WASP-74\,b which are being published for the first time within this study are in the online data as well.  



\bibliographystyle{mnras}
\bibliography{ttv_10_exoplanets_final}


\newpage
\appendix
\section{Light Curves}
\label{sec:appendixa}

We provide our own light curves (black data points) and {\sc exofast}-v1 models (red continuous curves) in Figs.{\ref{fig:hatp23_lightcurves} - {\ref{fig:hatp37_lightcurves}.

  
\begin{figure}
  \centering
  \subfloat[T100 2014/09/25]{\includegraphics[width=4cm]{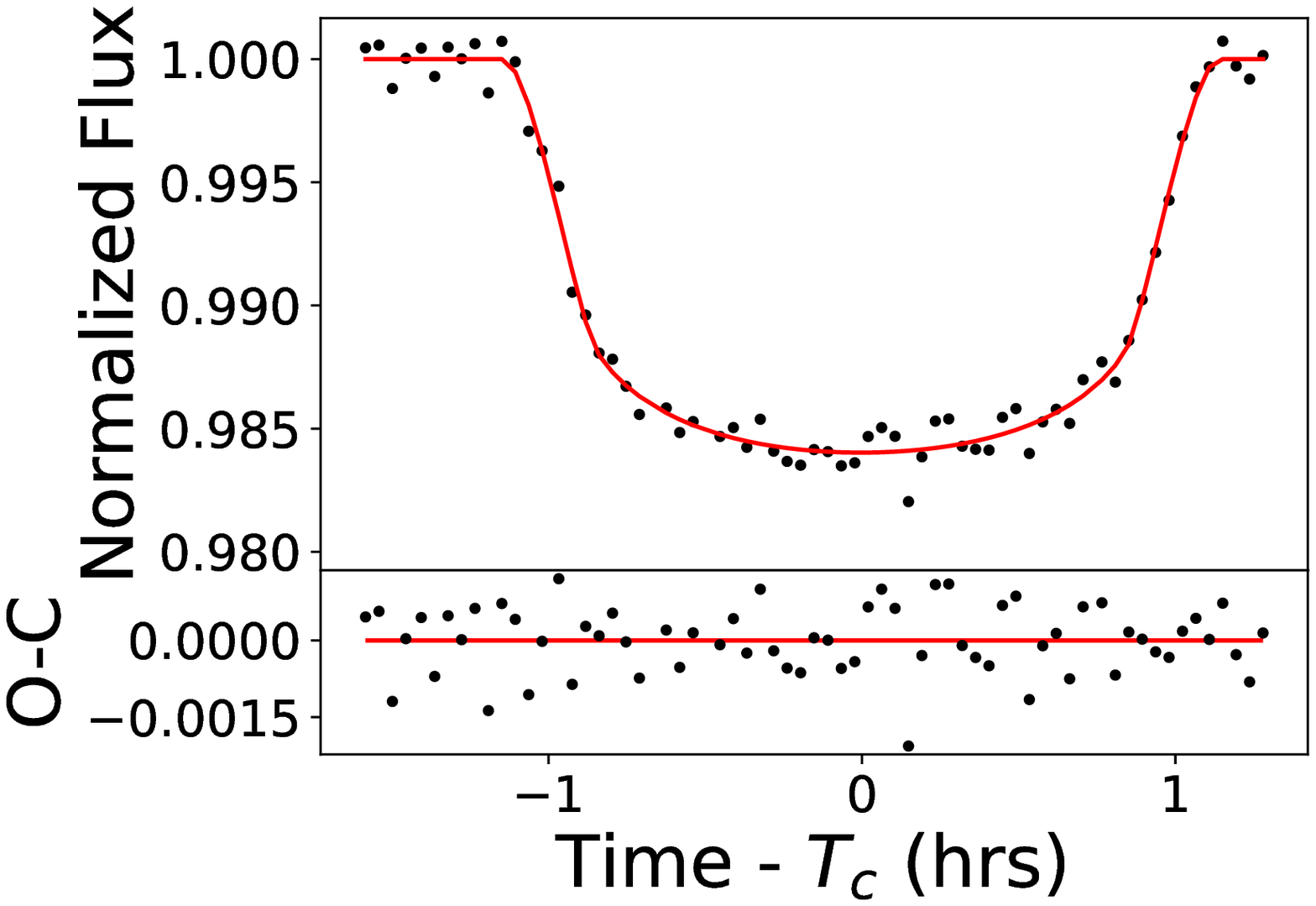}}
  \hfill
  \subfloat[CAHA 2015/08/28]{\includegraphics[width=4cm]{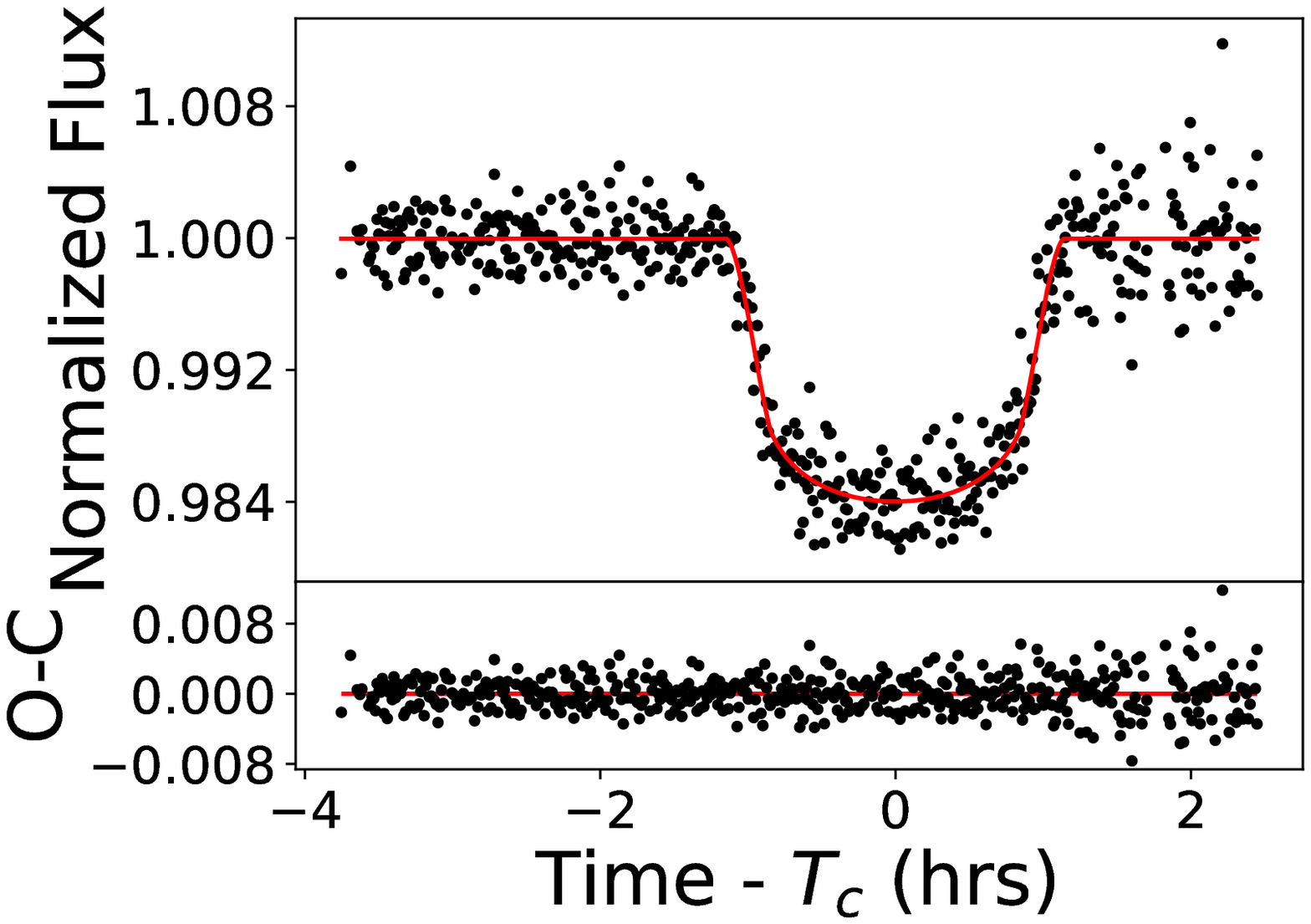}}
  \qquad
  \subfloat[T35 2019/08/08]{\includegraphics[width=4cm]{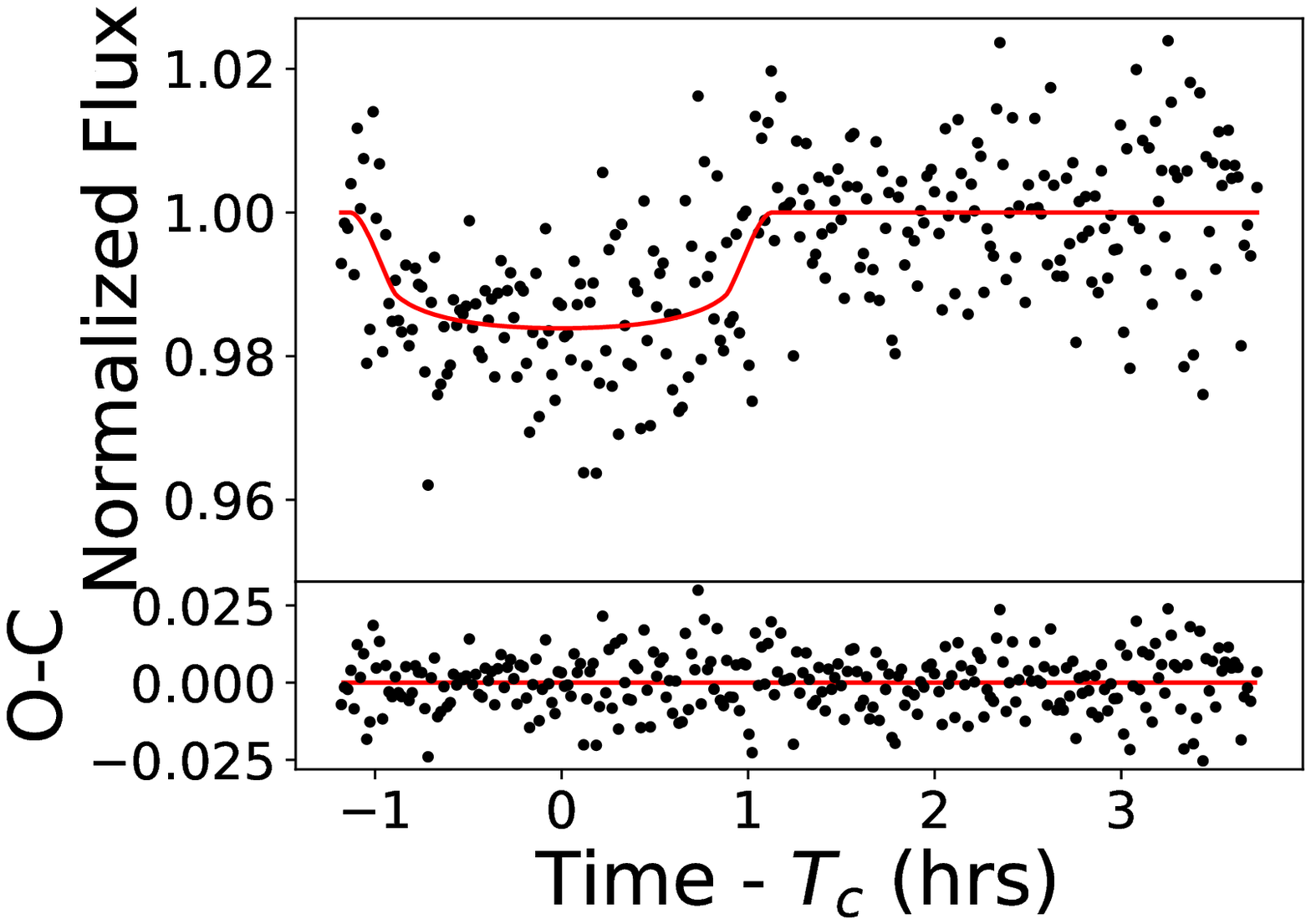}}
  \hfill
  \subfloat[T35 2019/09/28*]{\includegraphics[width=4cm]{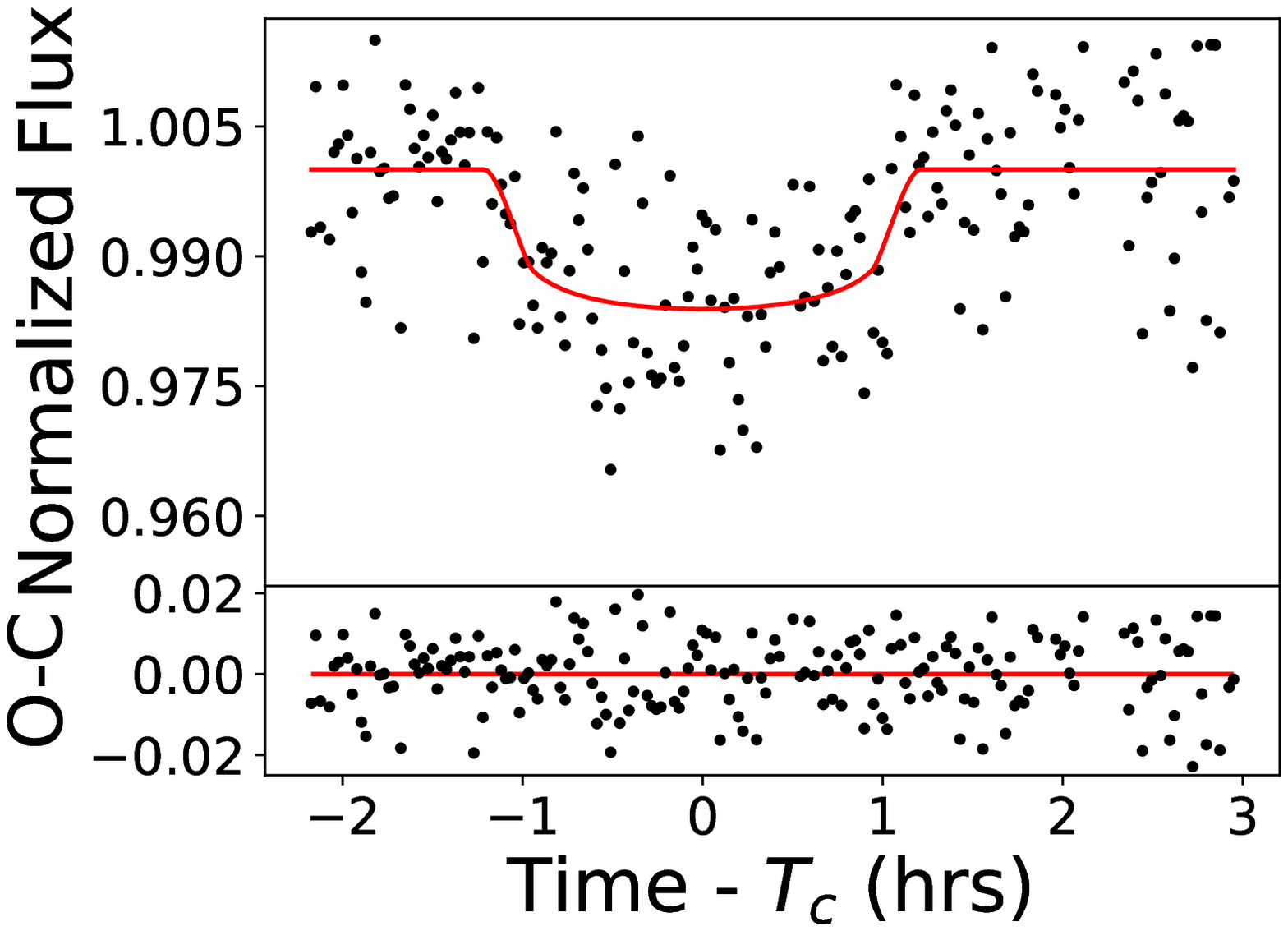}}
    \qquad
  \subfloat[UT50 2021/07/20*]{\includegraphics[width=4cm]{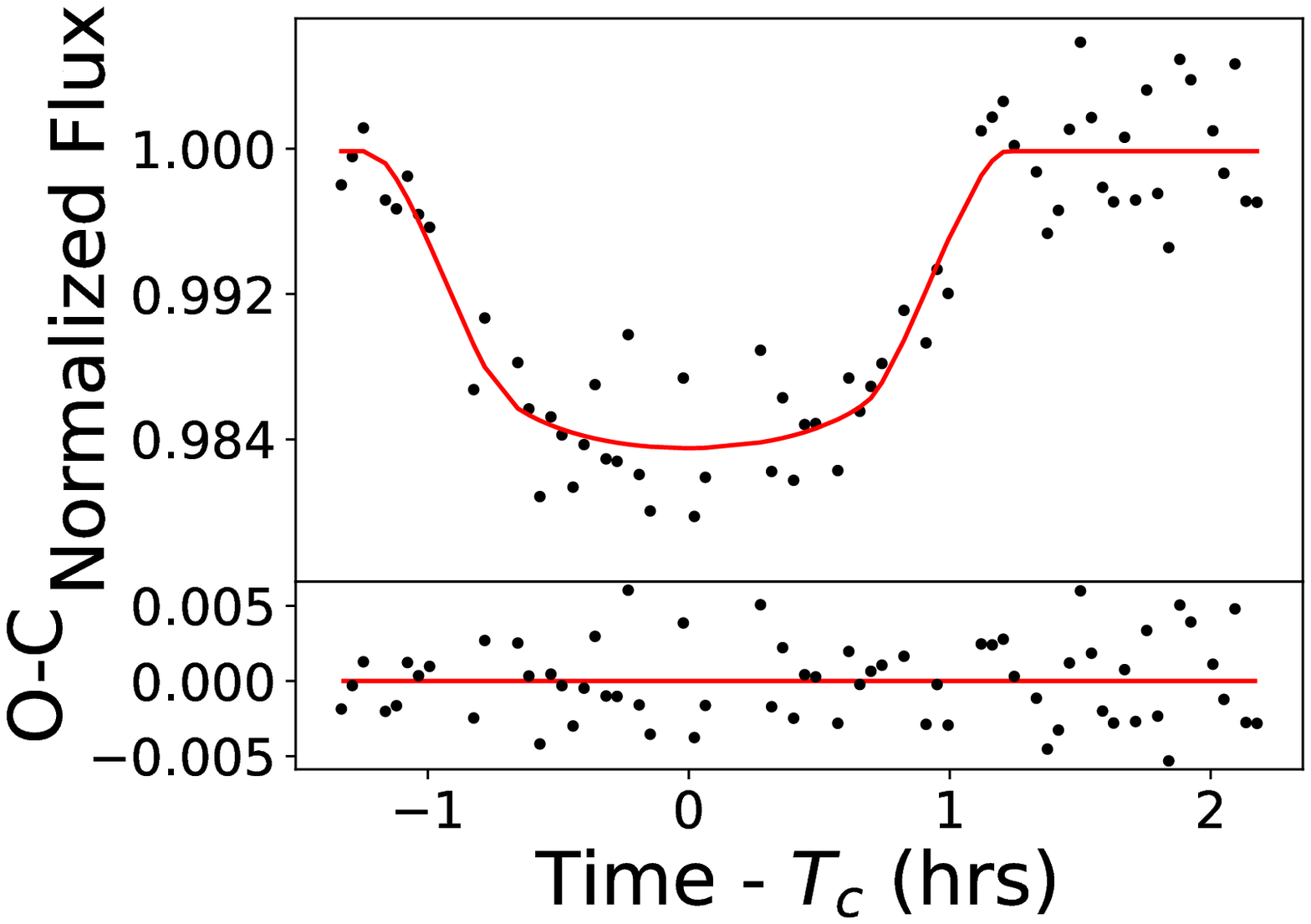}}
    \caption{HAT-P-23\,b Light Curves. Black dots are data points while the red continuous curve is for the {\sc exofast} model in all the light curves presented in this section. The ones that were eliminated based on our quantitative light curve selection criteria, therefore not used in timing analyses are marked with asterisks.}
    \label{fig:hatp23_lightcurves}
\end{figure}
\begin{figure}
  \centering
  \subfloat[T100 2017/04/27]{\includegraphics[width=4cm]{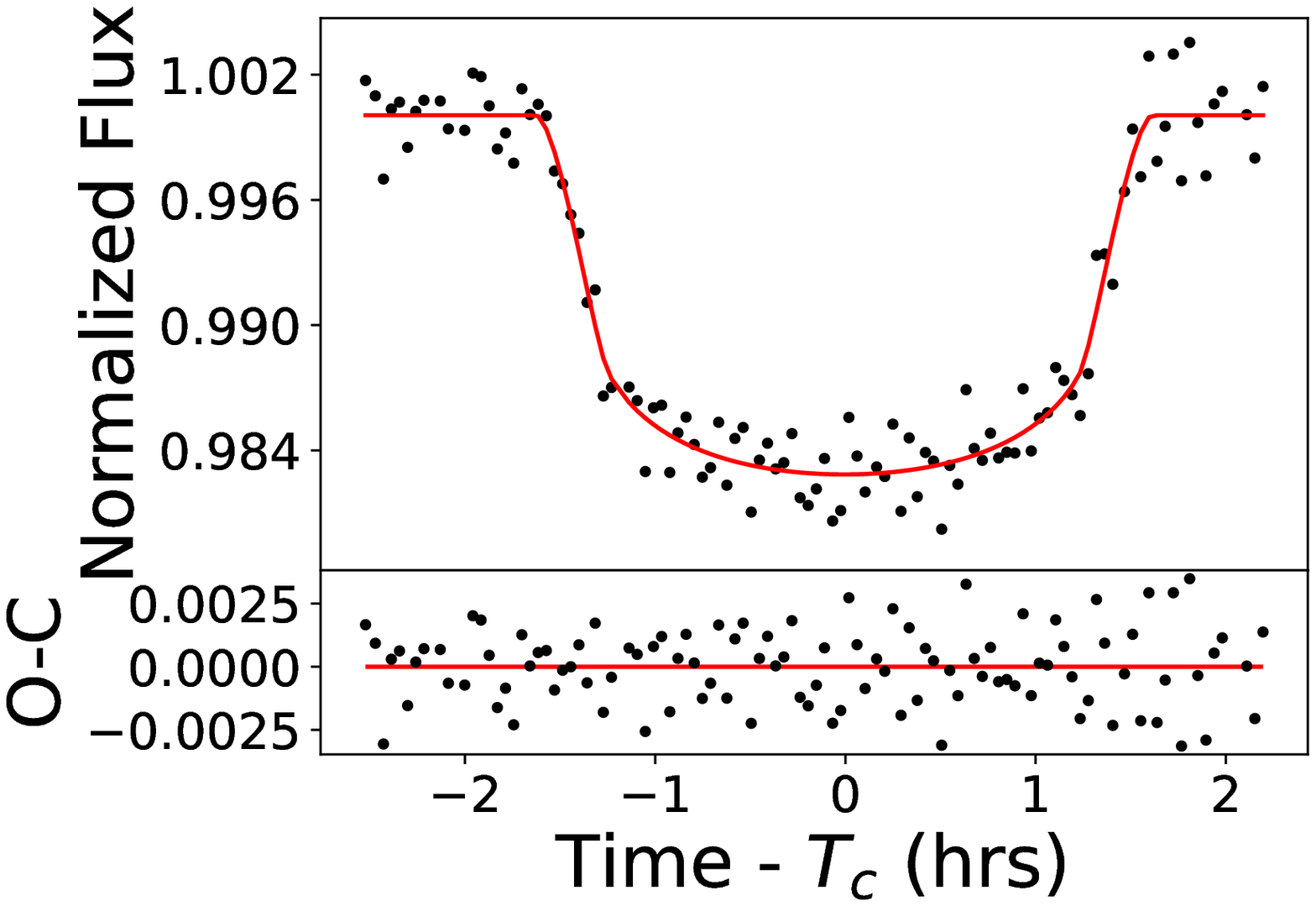}}
  \hfill
  \subfloat[UT50 2020/05/28*]{\includegraphics[width=4cm]{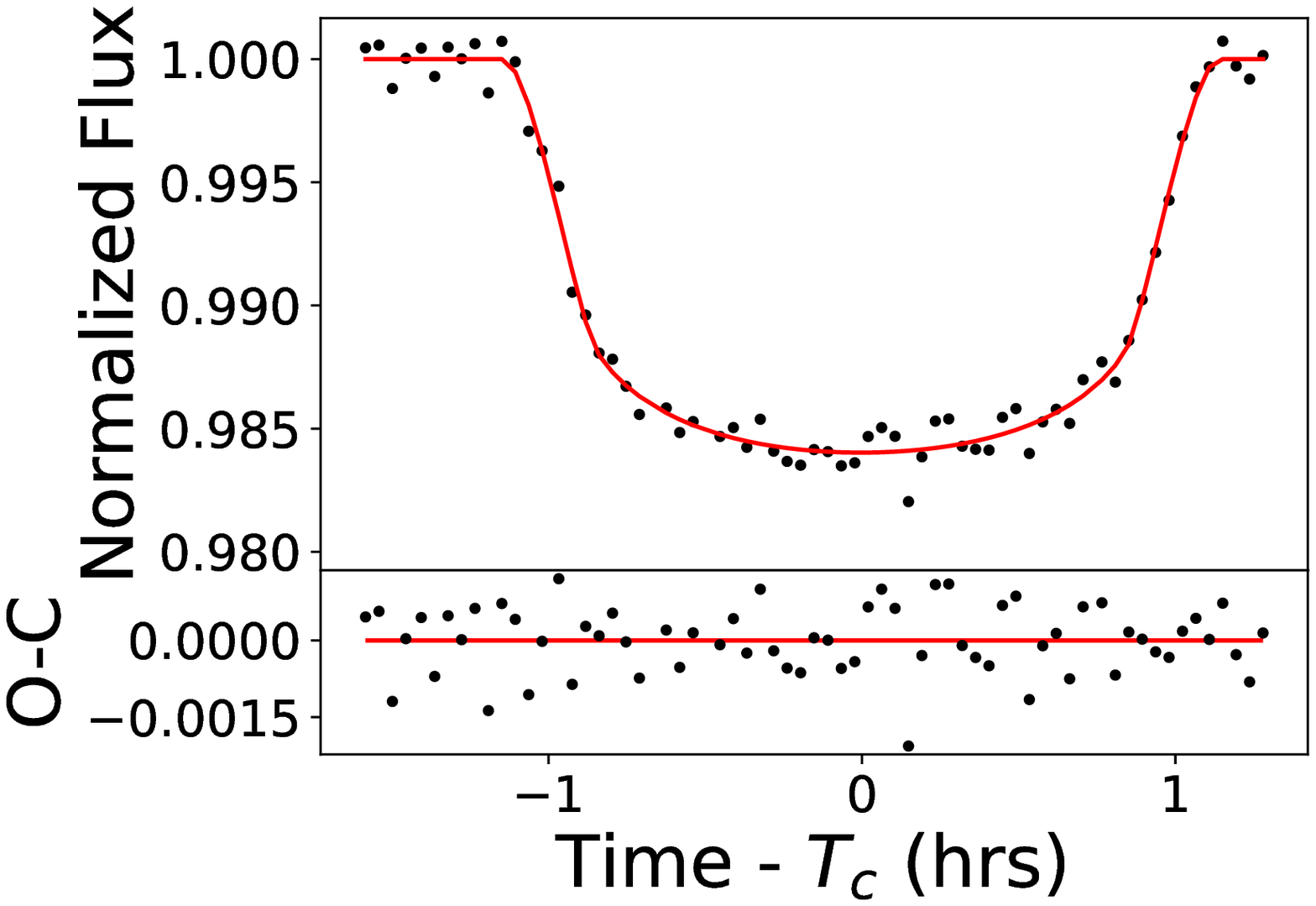}}
  \qquad
  \subfloat[T100 2020/06/22*]{\includegraphics[width=4cm]{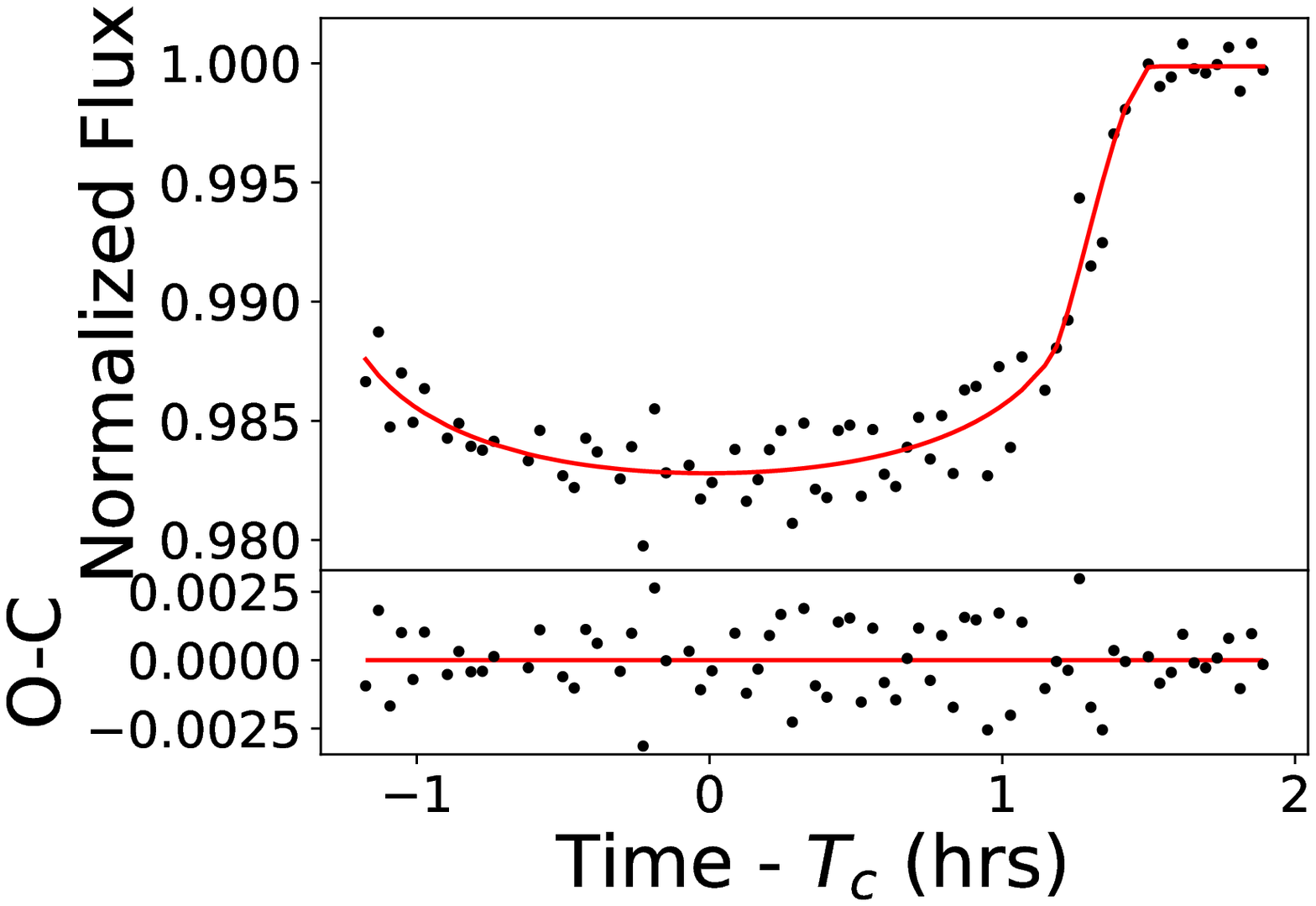}}
    \caption{WASP-37\,b Light Curves}
    \label{fig:wasp37_lightcurves}
\end{figure}
\begin{figure}
  \centering
  \subfloat[CAHA 2015/08/22]{\includegraphics[width=4cm]{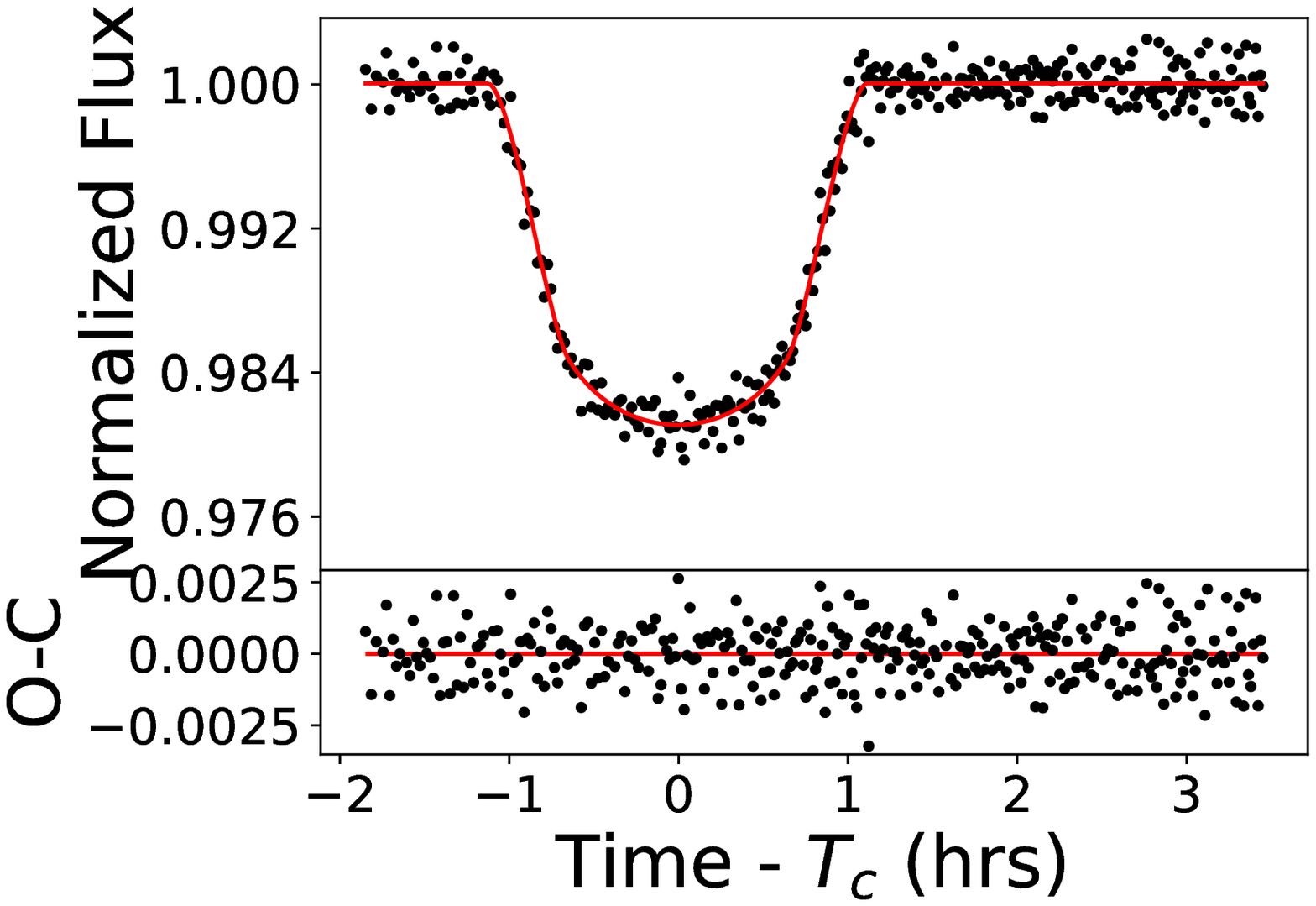}}
  \hfill
  \subfloat[CAHA 2015/09/18]{\includegraphics[width=4cm]{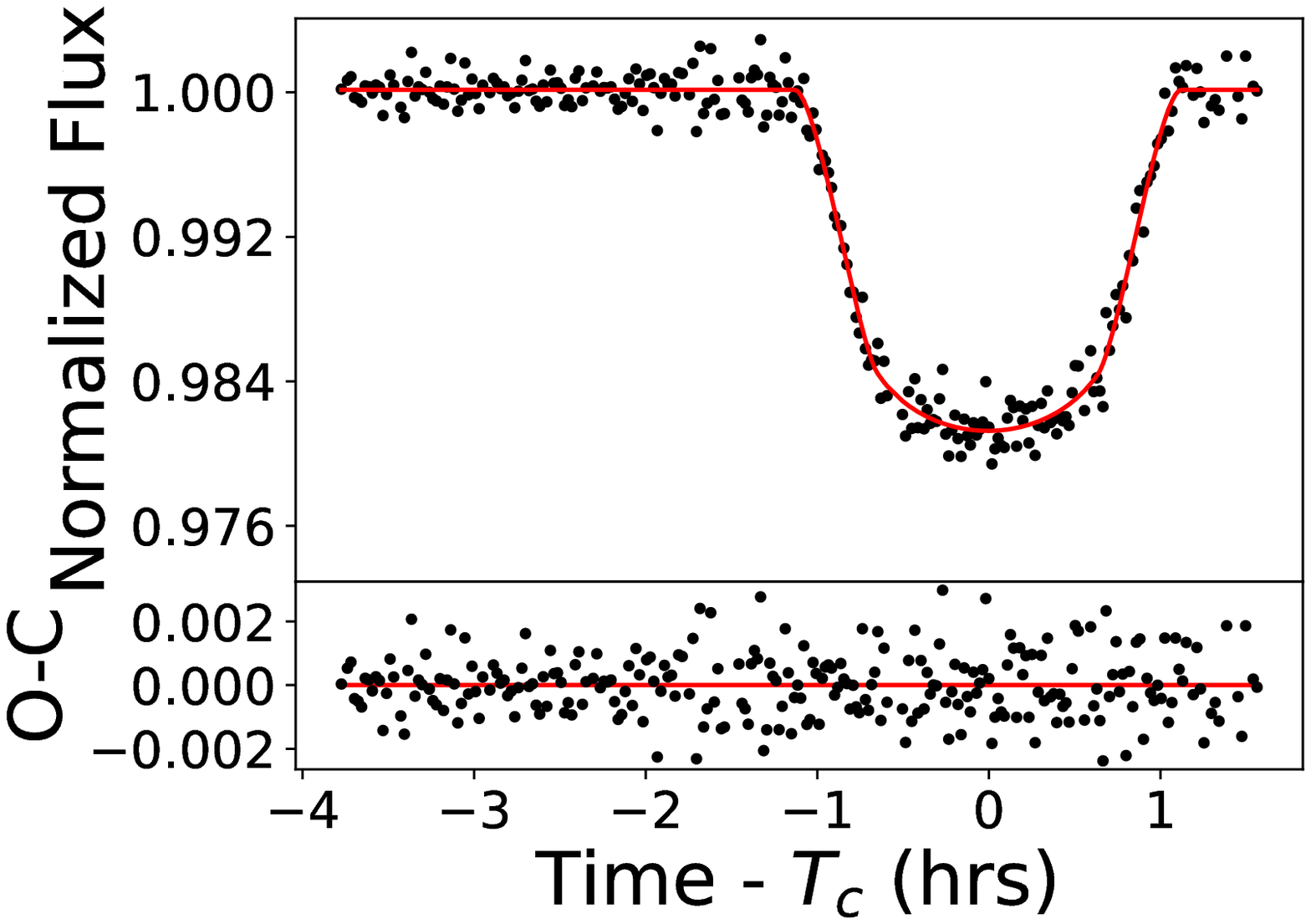}}
  \qquad
  \subfloat[CAHA 2016/09/08]{\includegraphics[width=4cm]{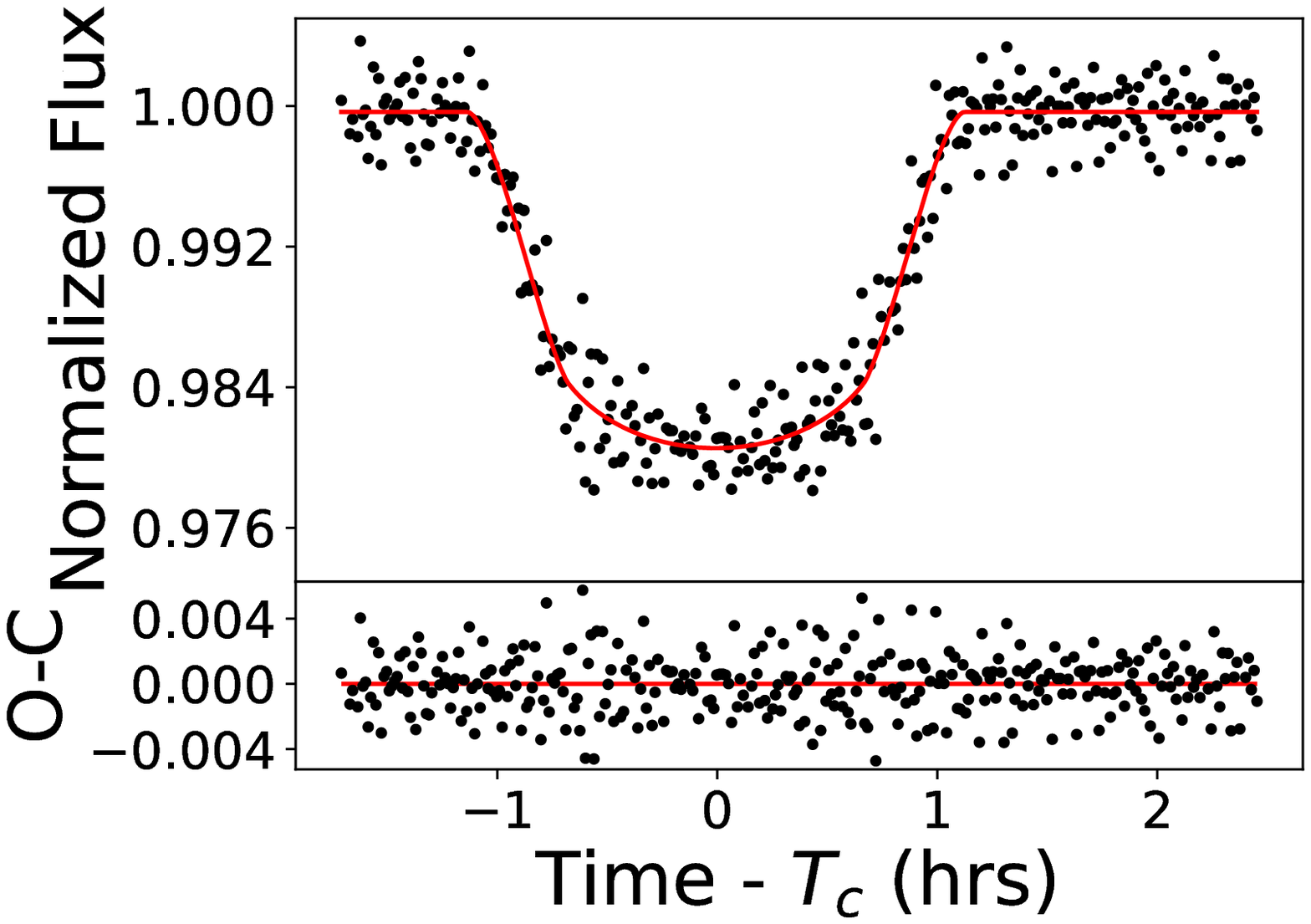}}
  \hfill
  \subfloat[T100 2016/10/09]{\includegraphics[width=4cm]{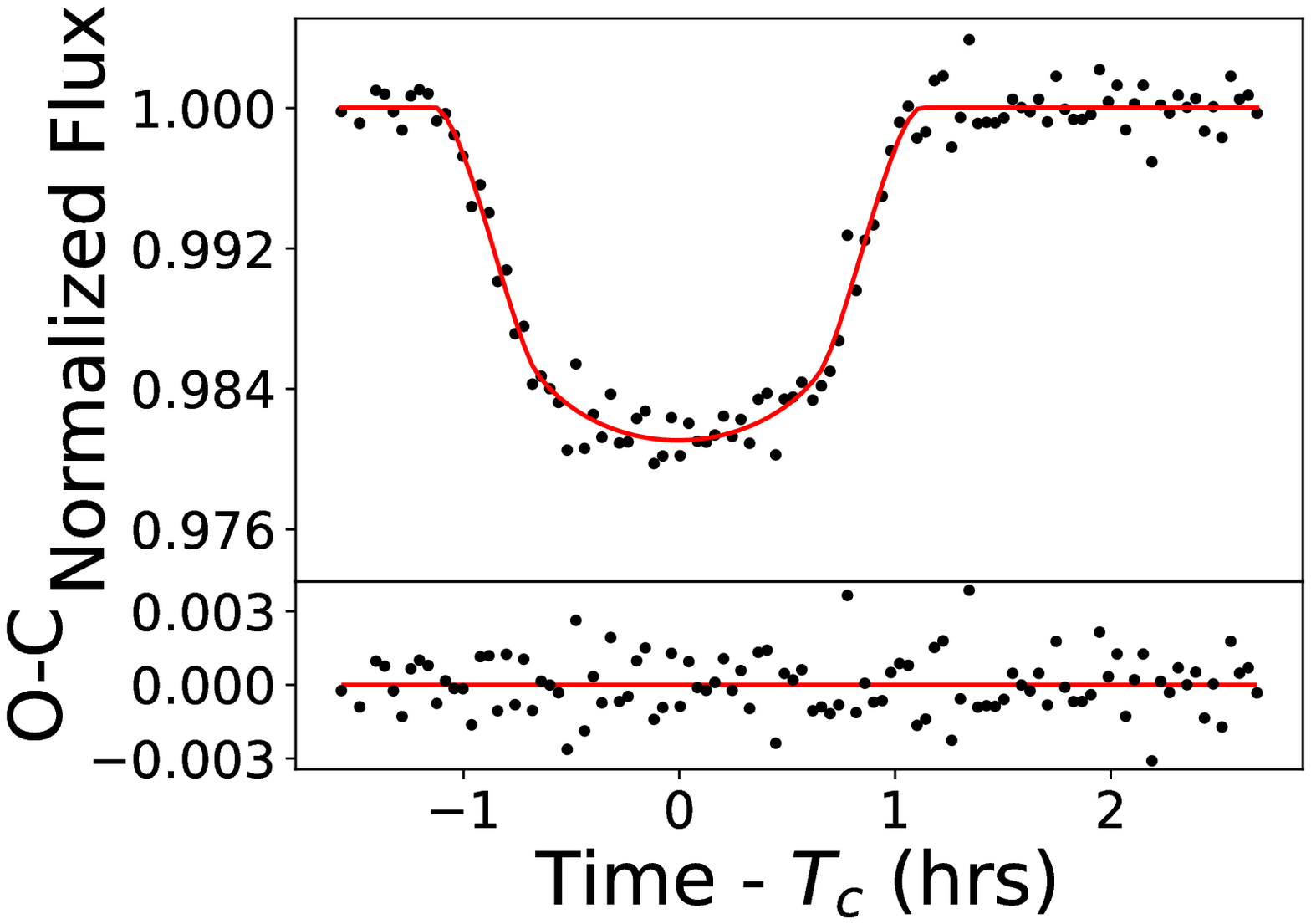}}
  \qquad
  \subfloat[T100 2017/08/26]{\includegraphics[width=4cm]{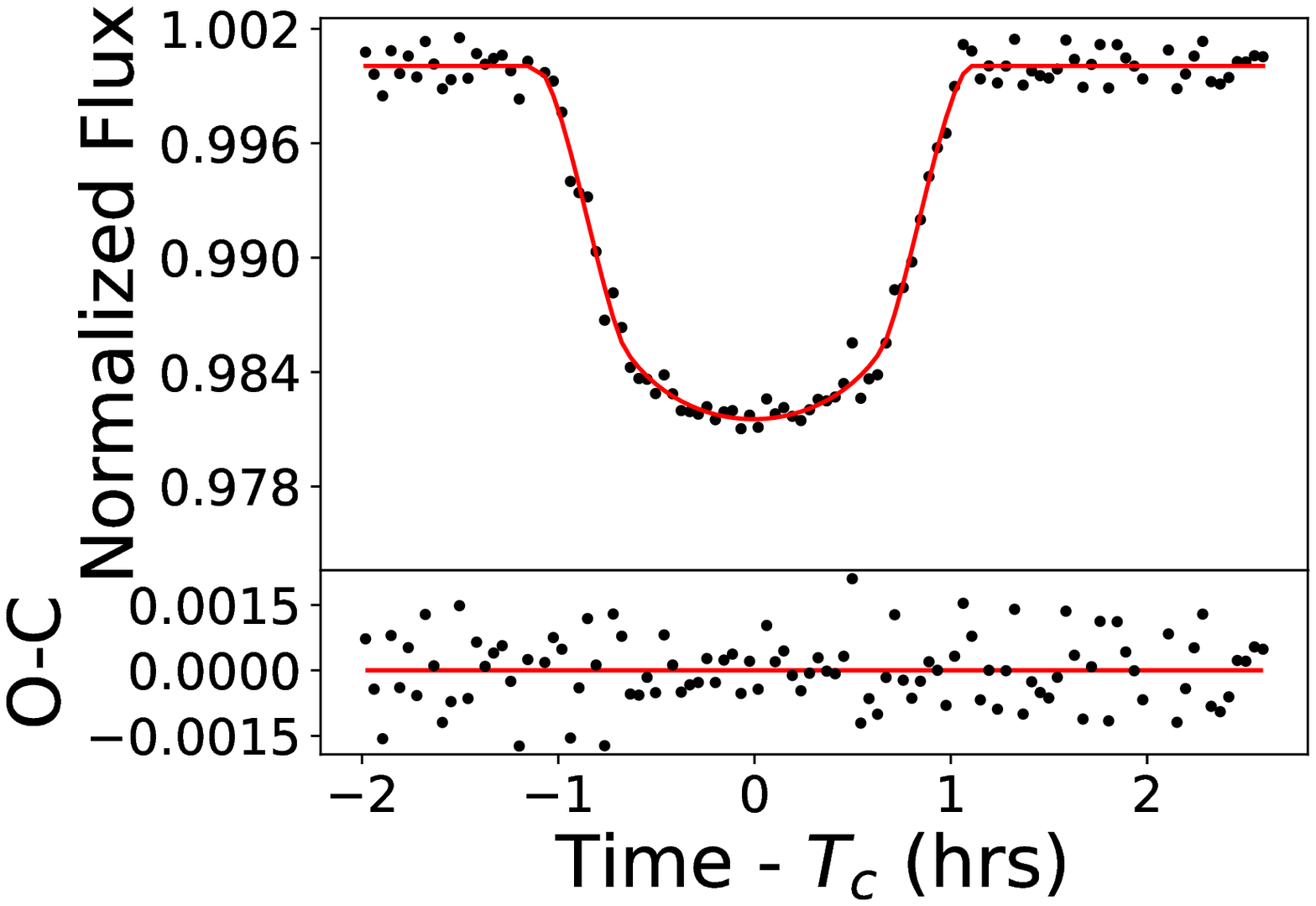}}
  \hfill
  \subfloat[T35 2019/07/27]{\includegraphics[width=4cm]{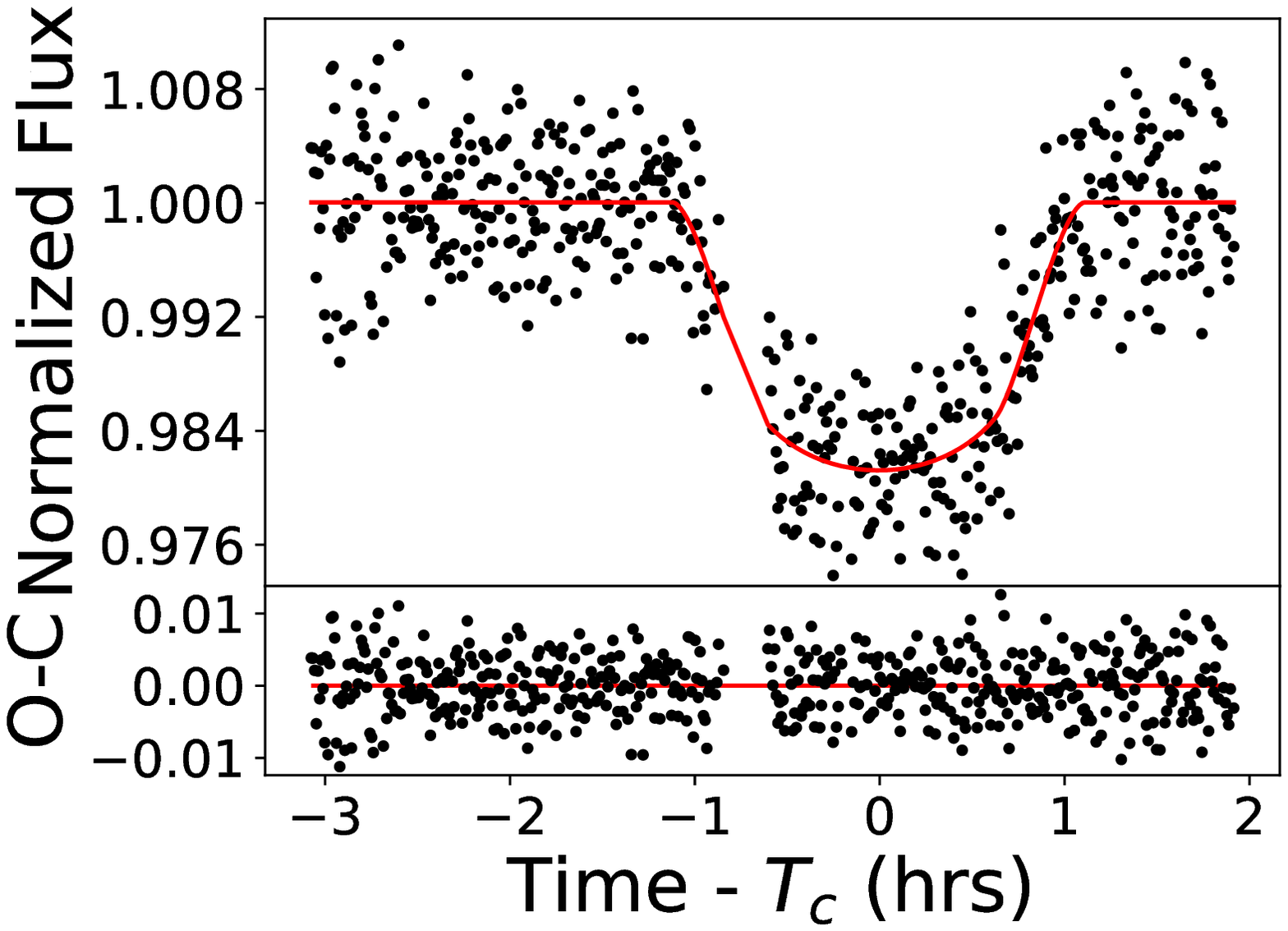}}
  \qquad
    \subfloat[T35 2019/08/27]{\includegraphics[width=4cm]{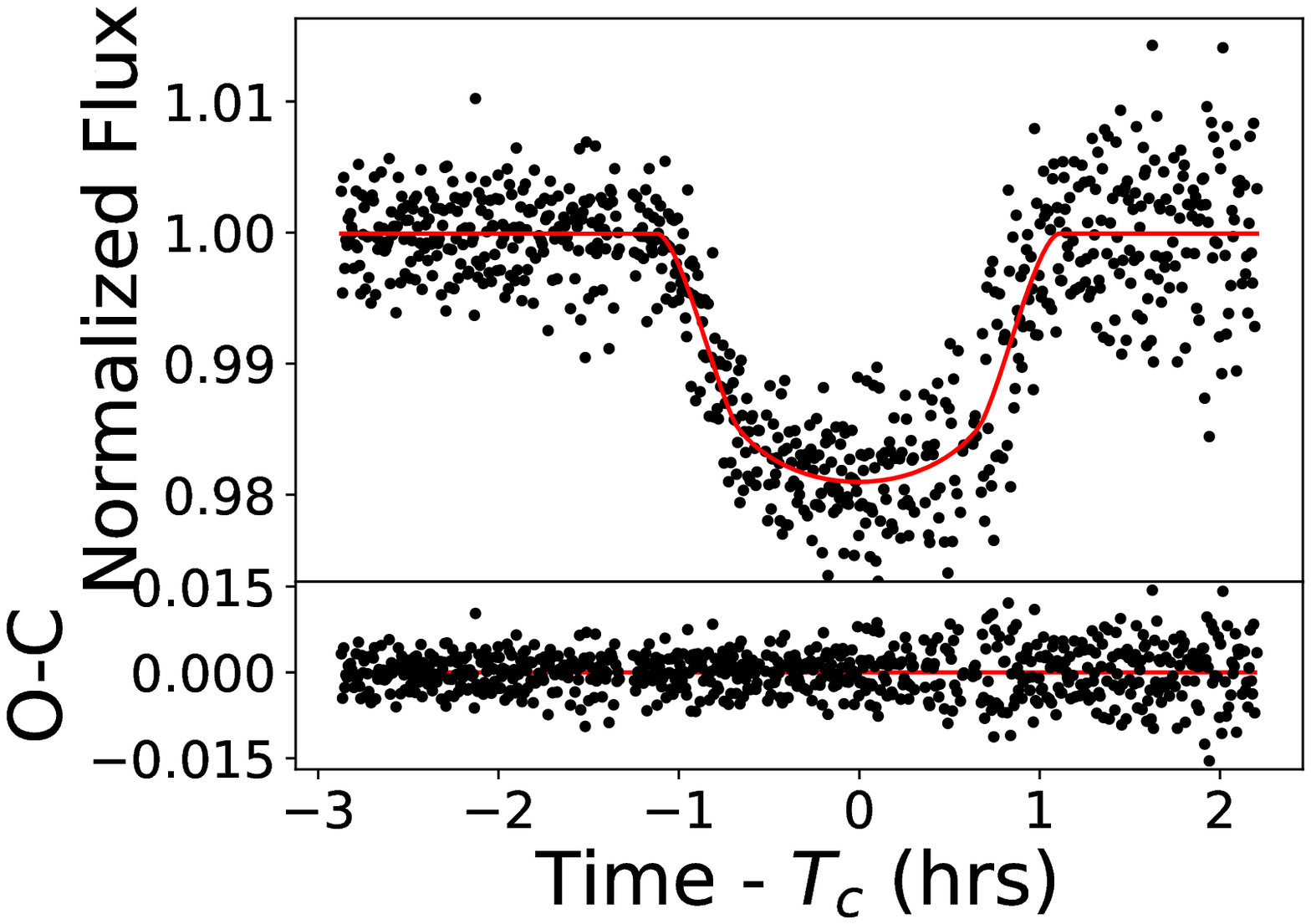}}
    \caption{WASP-69\,b Light Curves}
    \label{fig:wasp69_lightcurves}
\end{figure}

\begin{figure}
  \centering
  \subfloat[T35 2019/07/07*]{\includegraphics[width=4cm]{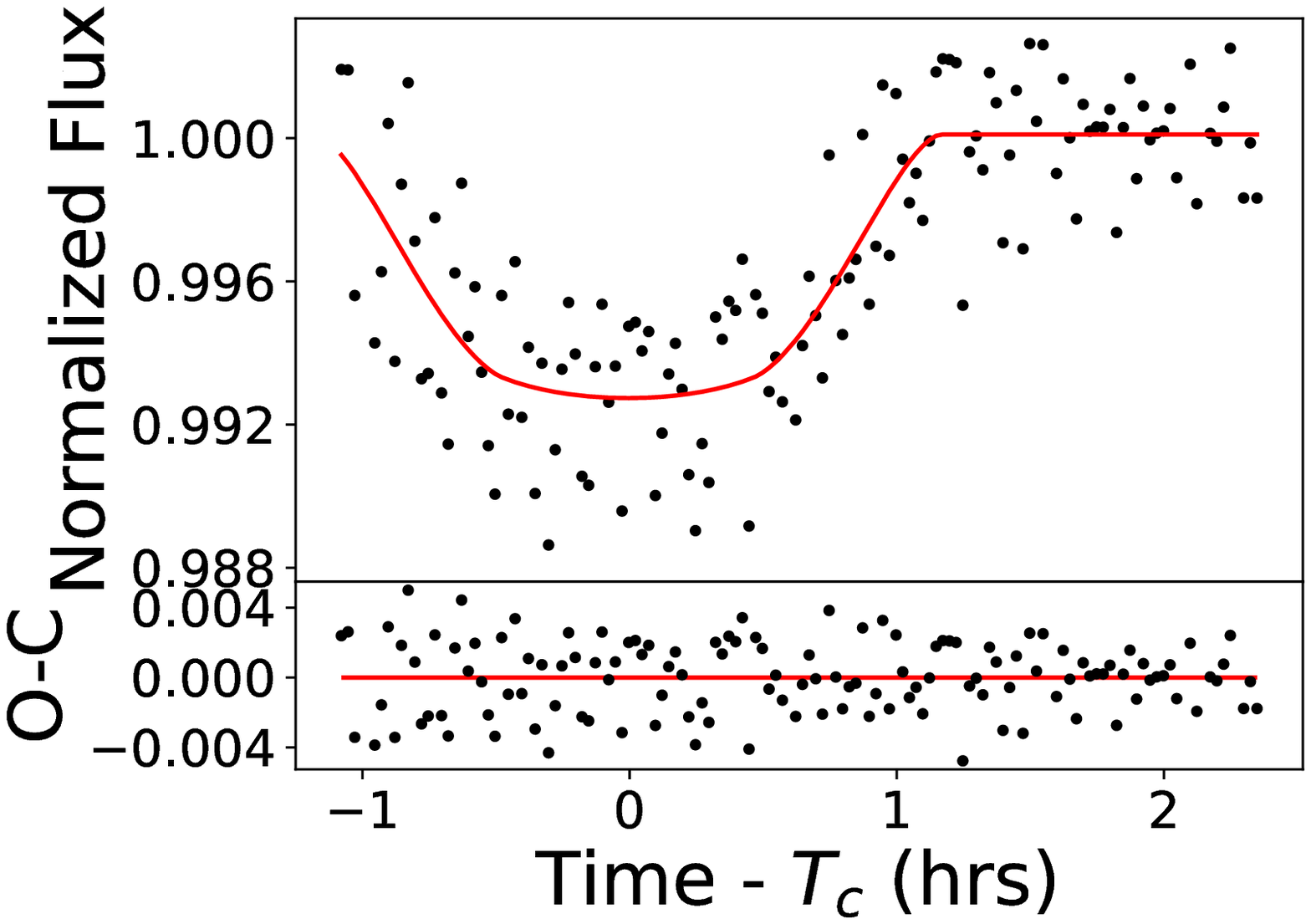}}
  \hfill
  \subfloat[T100 2020/08/12]{\includegraphics[width=4cm]{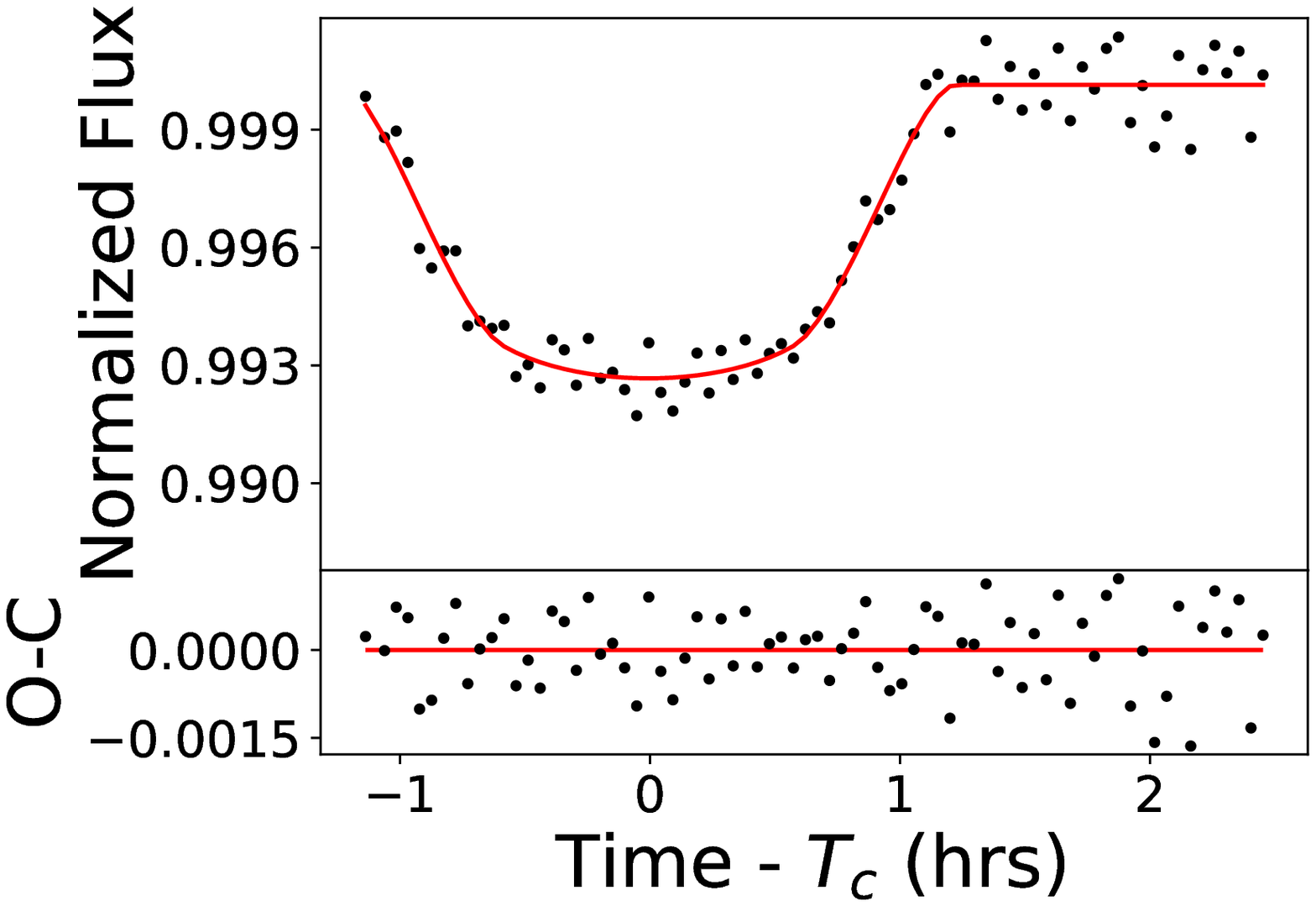}}
  \qquad
  \subfloat[IST60 2020/08/14*]{\includegraphics[width=4cm]{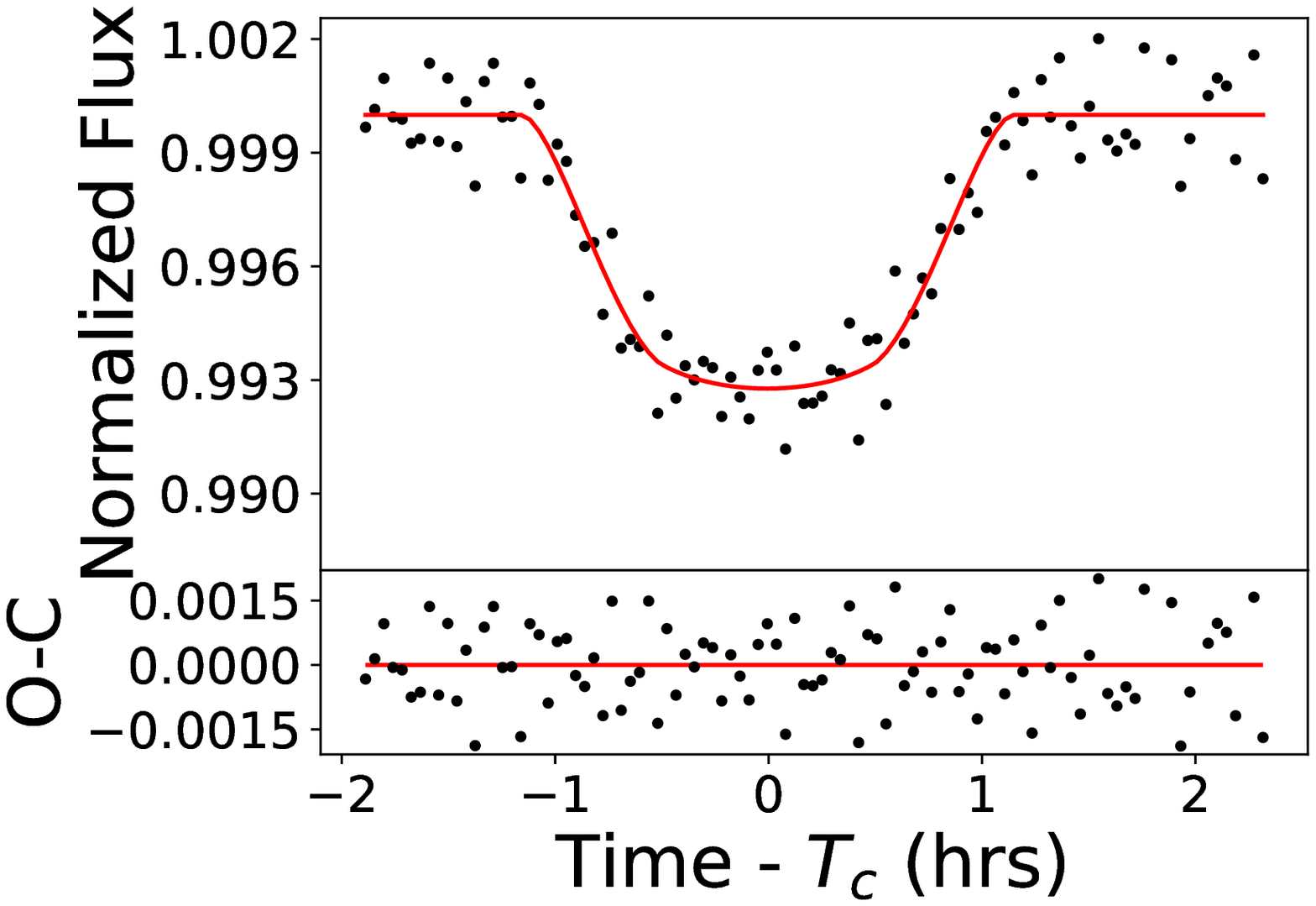}}
  \hfill
  \subfloat[T100 2020/08/27]{\includegraphics[width=4cm]{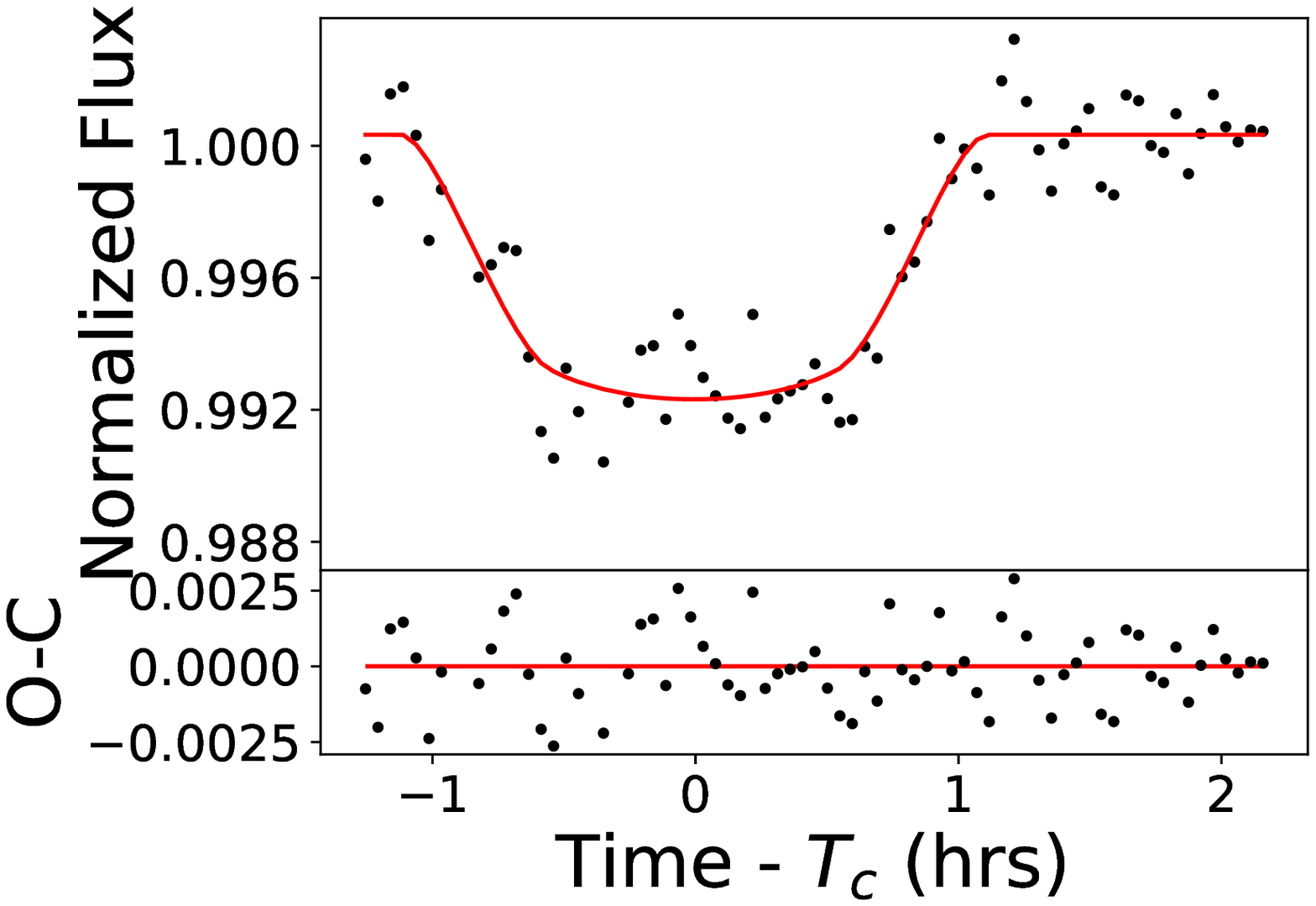}}
    \caption{WASP-74\,b Light Curves}
    \label{fig:wasp74_lightcurves}
\end{figure}

\begin{figure}
  \centering
  \subfloat[T35 2019/02/17]{\includegraphics[width=4cm]{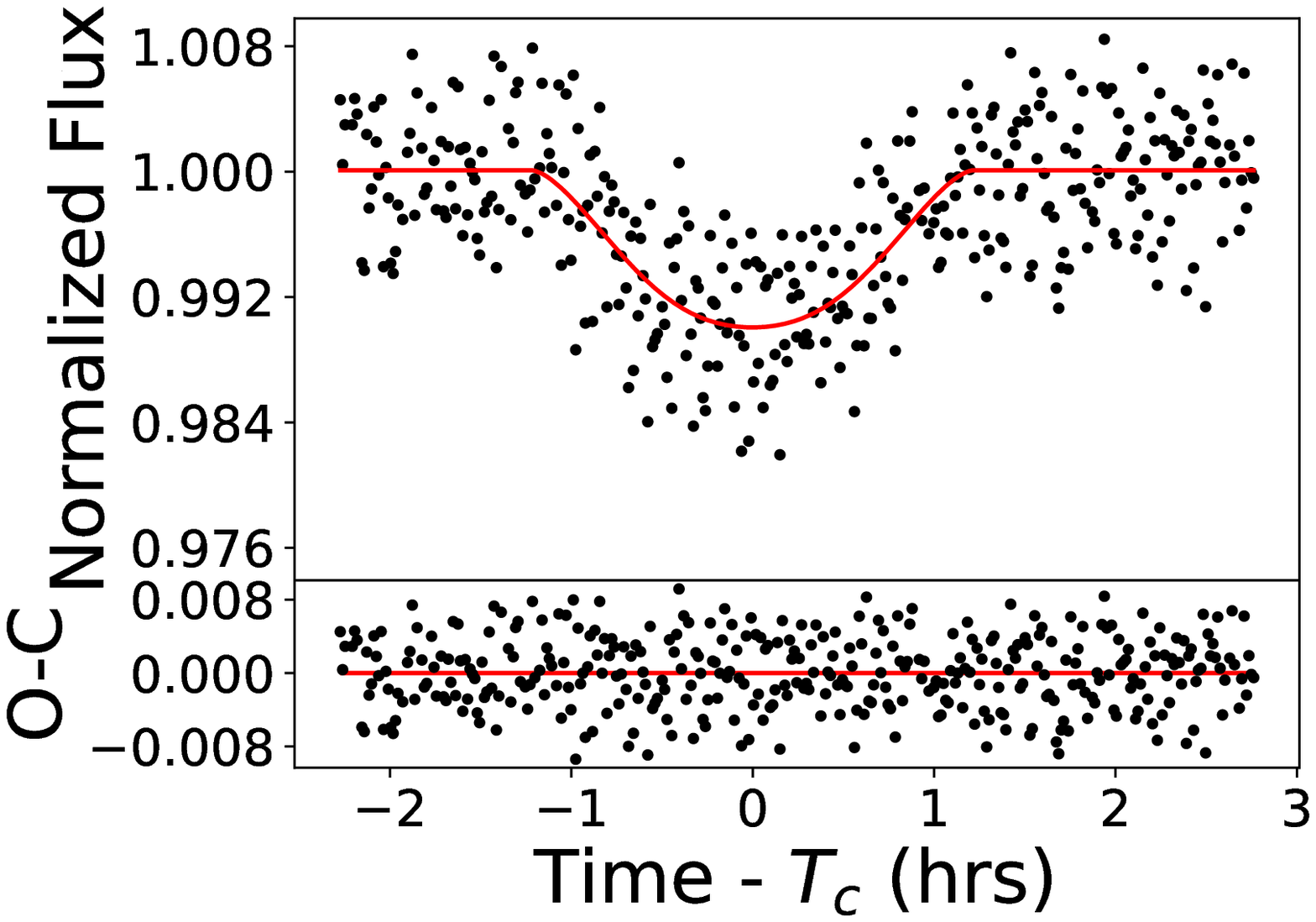}}
  \hfill
  \subfloat[T100 2019/11/09]{\includegraphics[width=4cm]{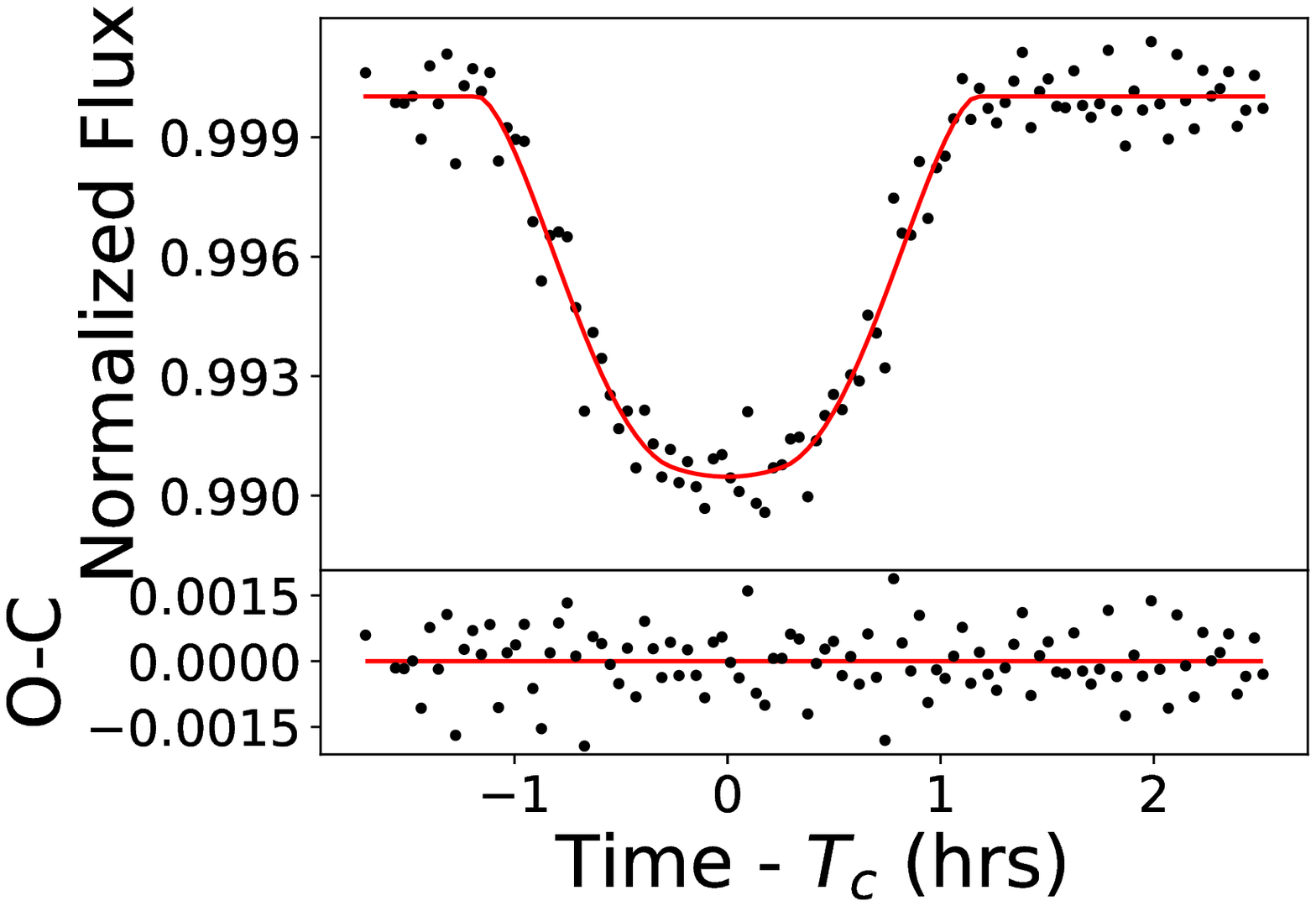}}
    \caption{HAT-P-56\,b Light Curves}
    \label{fig:hatp56_lightcurves}
\end{figure}

\begin{figure}
  \centering
  \subfloat[T35 2019/07/20]{\includegraphics[width=4cm]{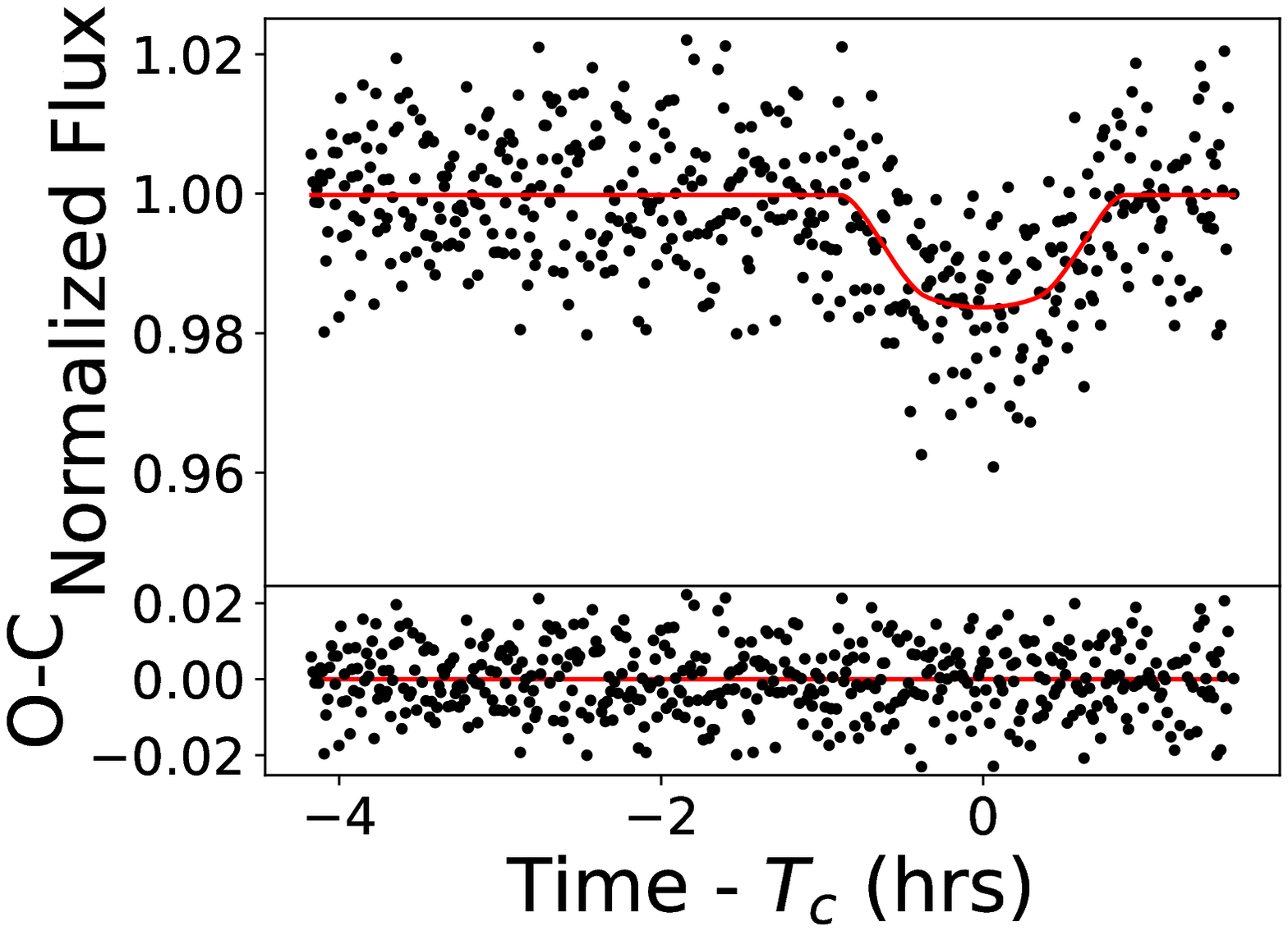}}
  \hfill
  \subfloat[ATA50 2020/10/12]{\includegraphics[width=4cm]{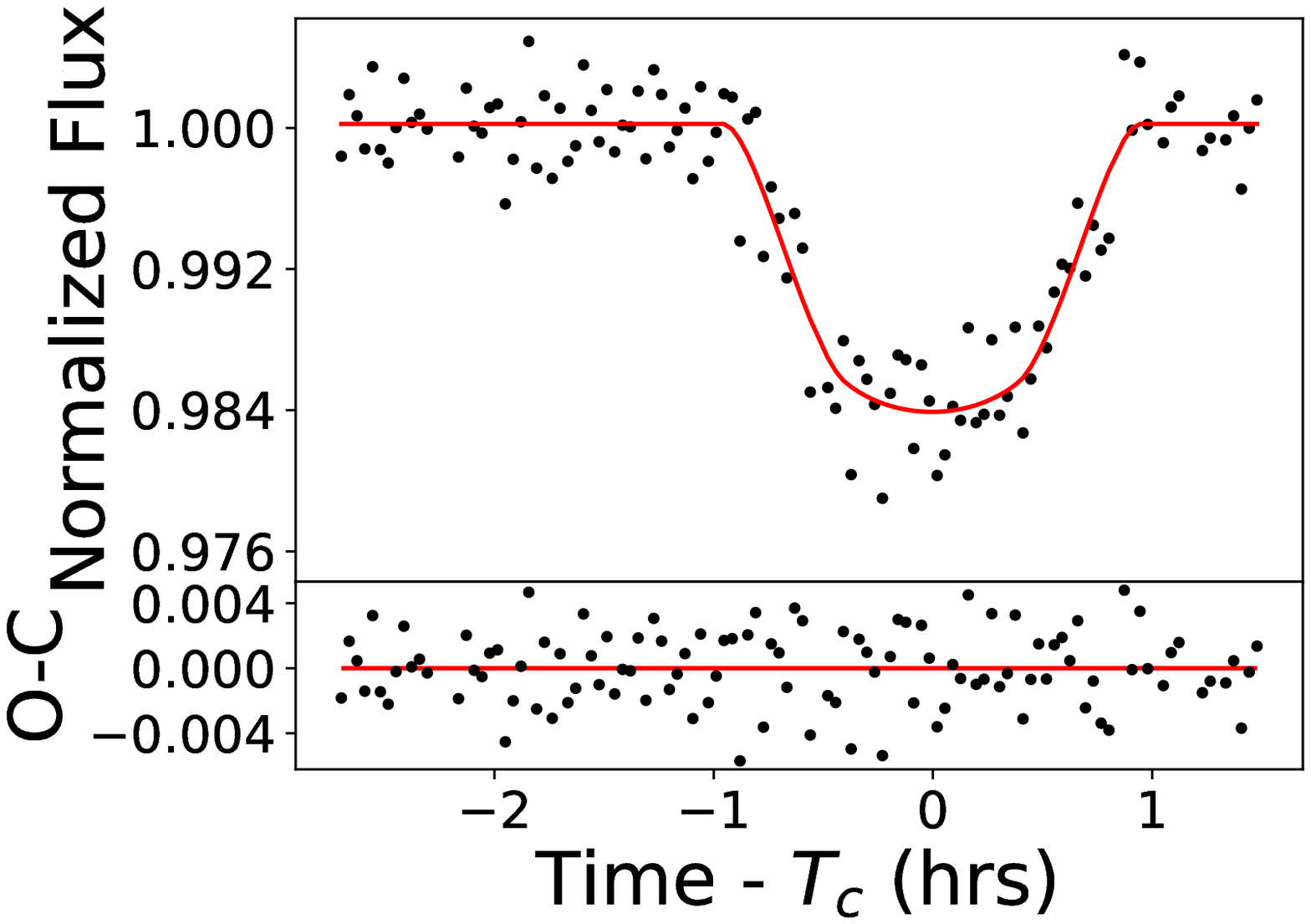}}
  \qquad
  \subfloat[T100 2020/10/25]{\includegraphics[width=4cm]{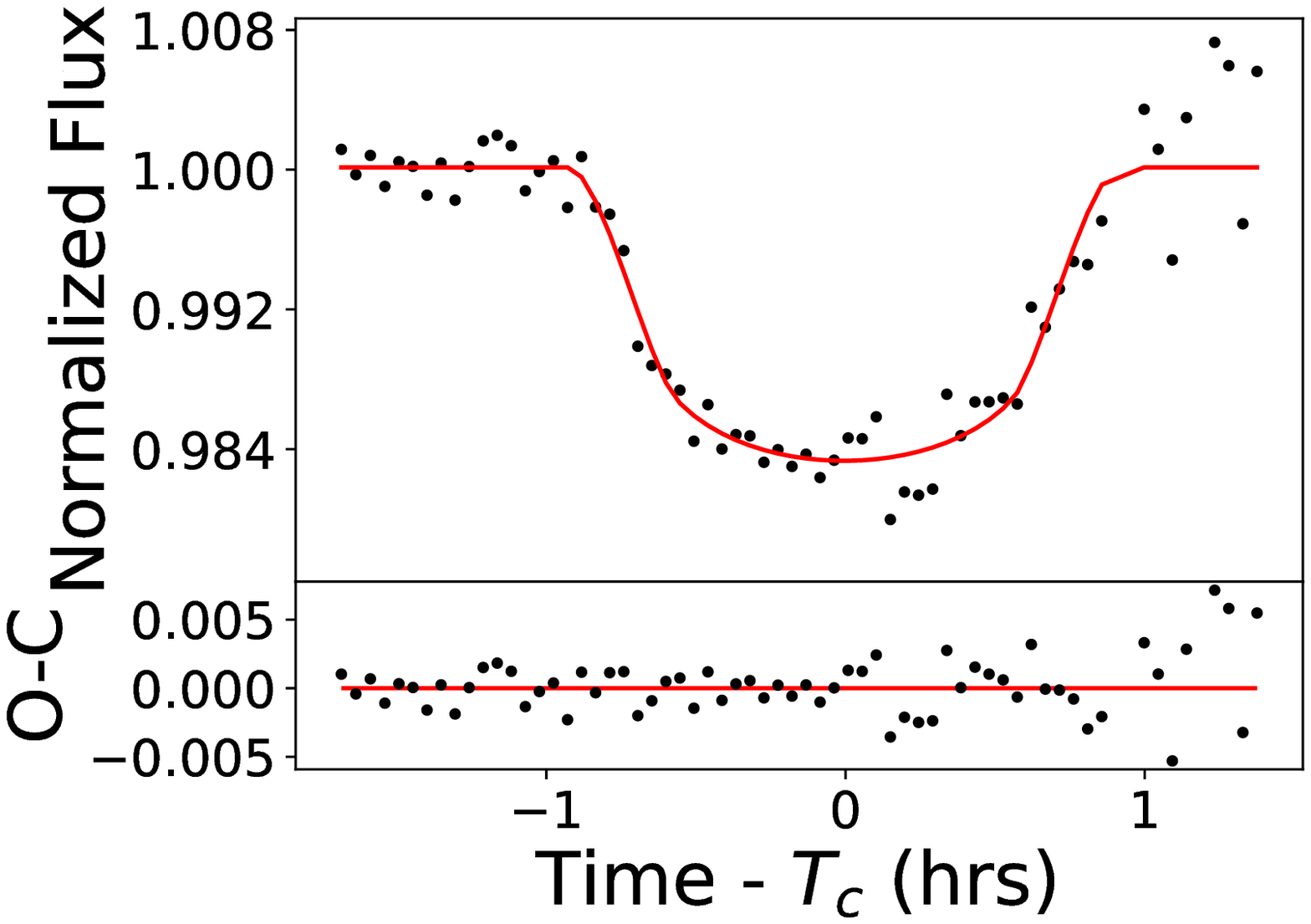}}
    \caption{WASP-2\,b Light Curves}
    \label{fig:wasp2_lightcurves}
\end{figure}

\begin{figure}
  \centering
  \subfloat[T35 2019/11/03]{\includegraphics[width=4cm]{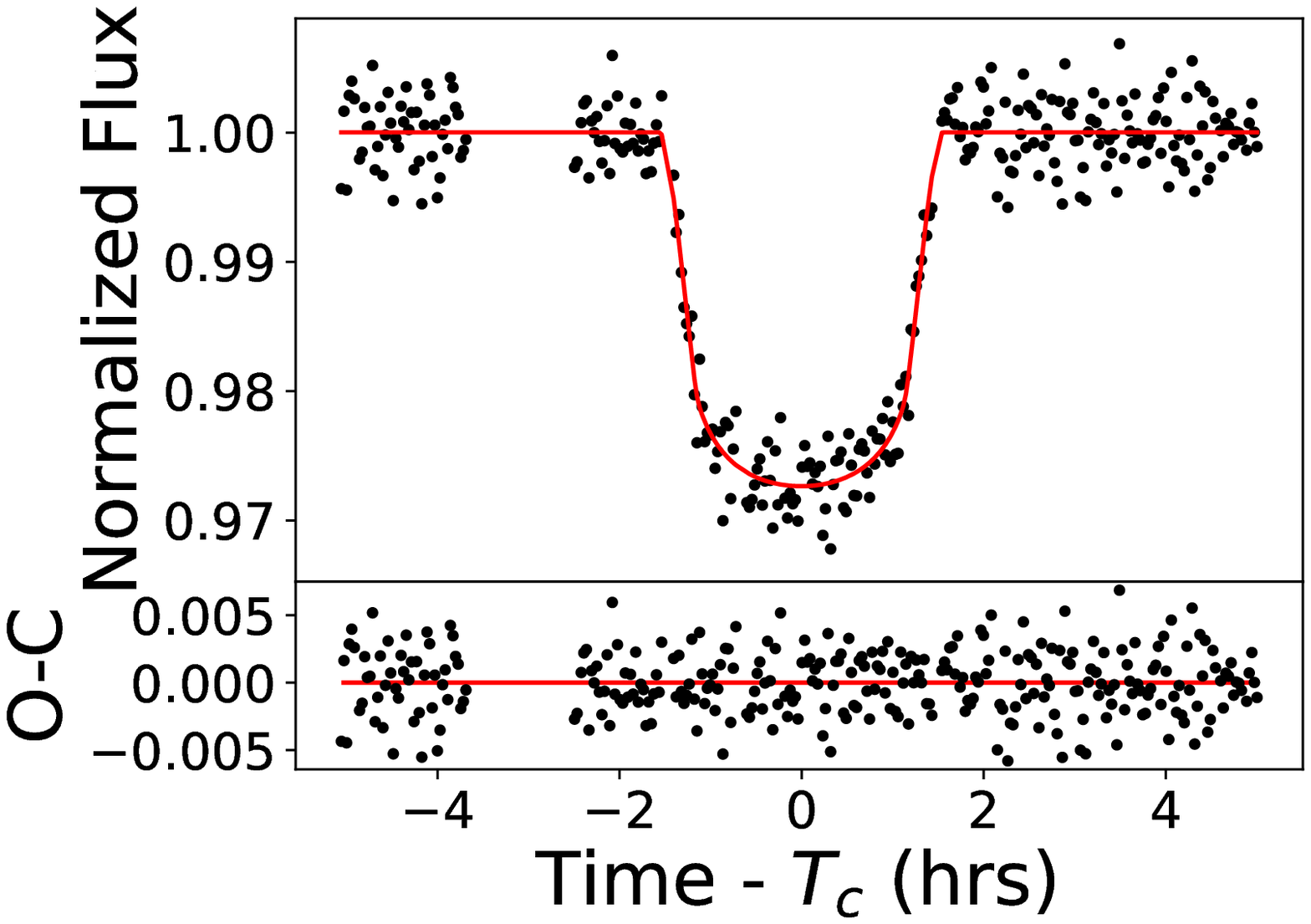}}
    \caption{HAT-P-32\,b Light Curve}
    \label{fig:hatp32_lightcurves}
\end{figure}

\begin{figure}
  \centering
  \subfloat[T35 2019/07/03]{\includegraphics[width=4cm]{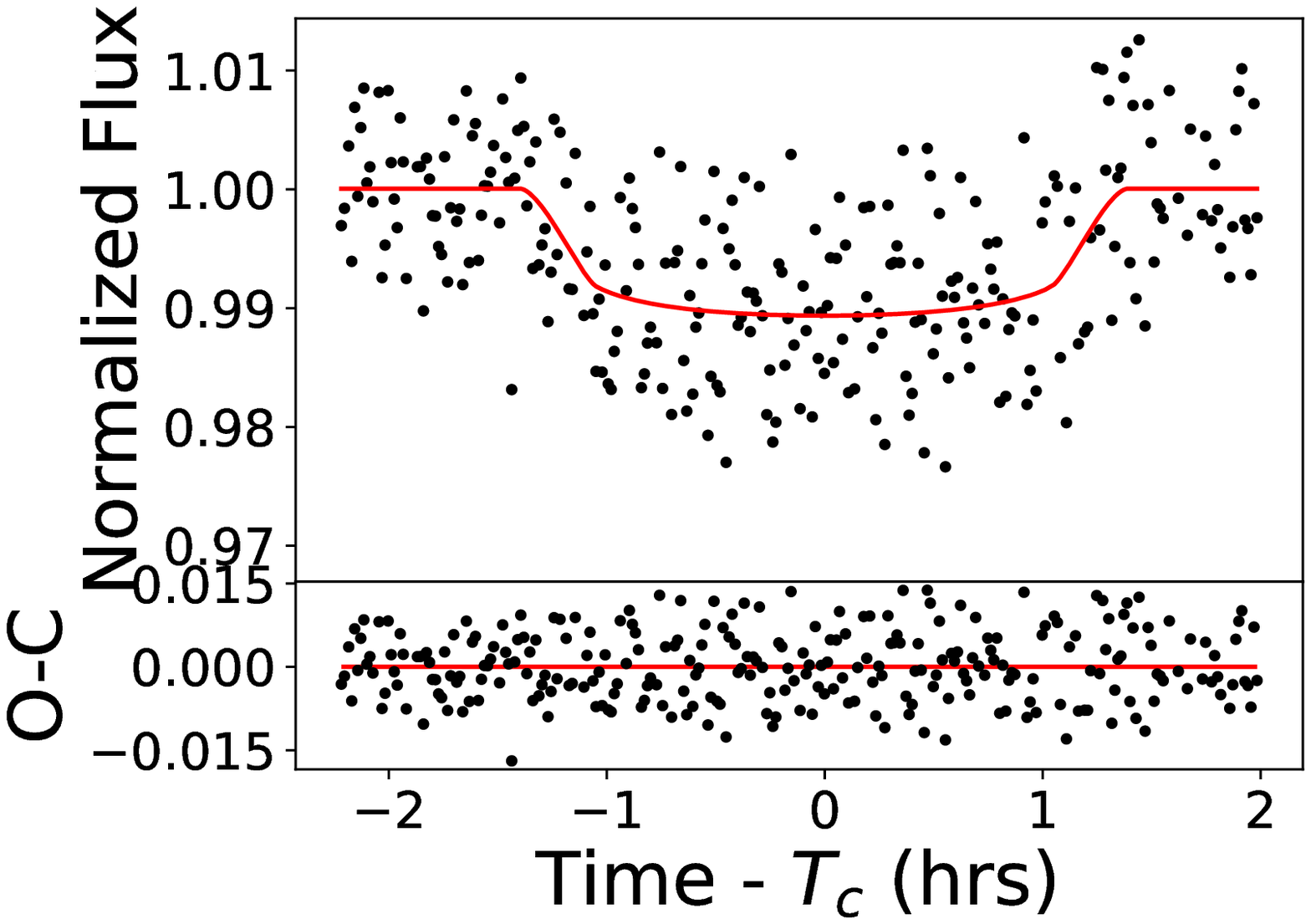}}
  \hfill
  \subfloat[ATA50 2020/05/14*]{\includegraphics[width=4cm]{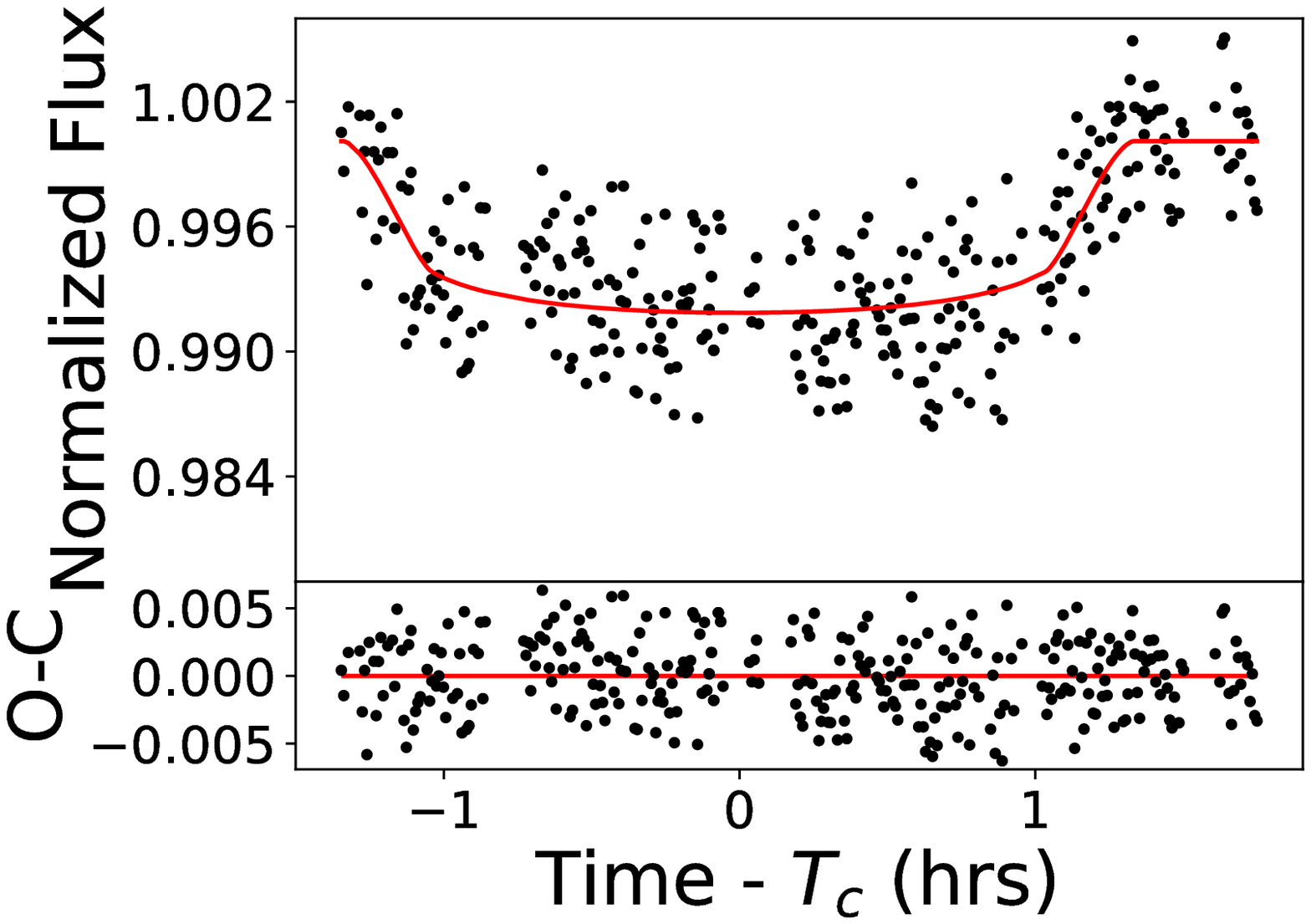}}
    \caption{WASP-14\,b Light Curves}
    \label{fig:wasp14_lightcurves}
\end{figure}

\begin{figure}
  \centering
  \subfloat[T100 2014/05/30]{\includegraphics[width=4cm]{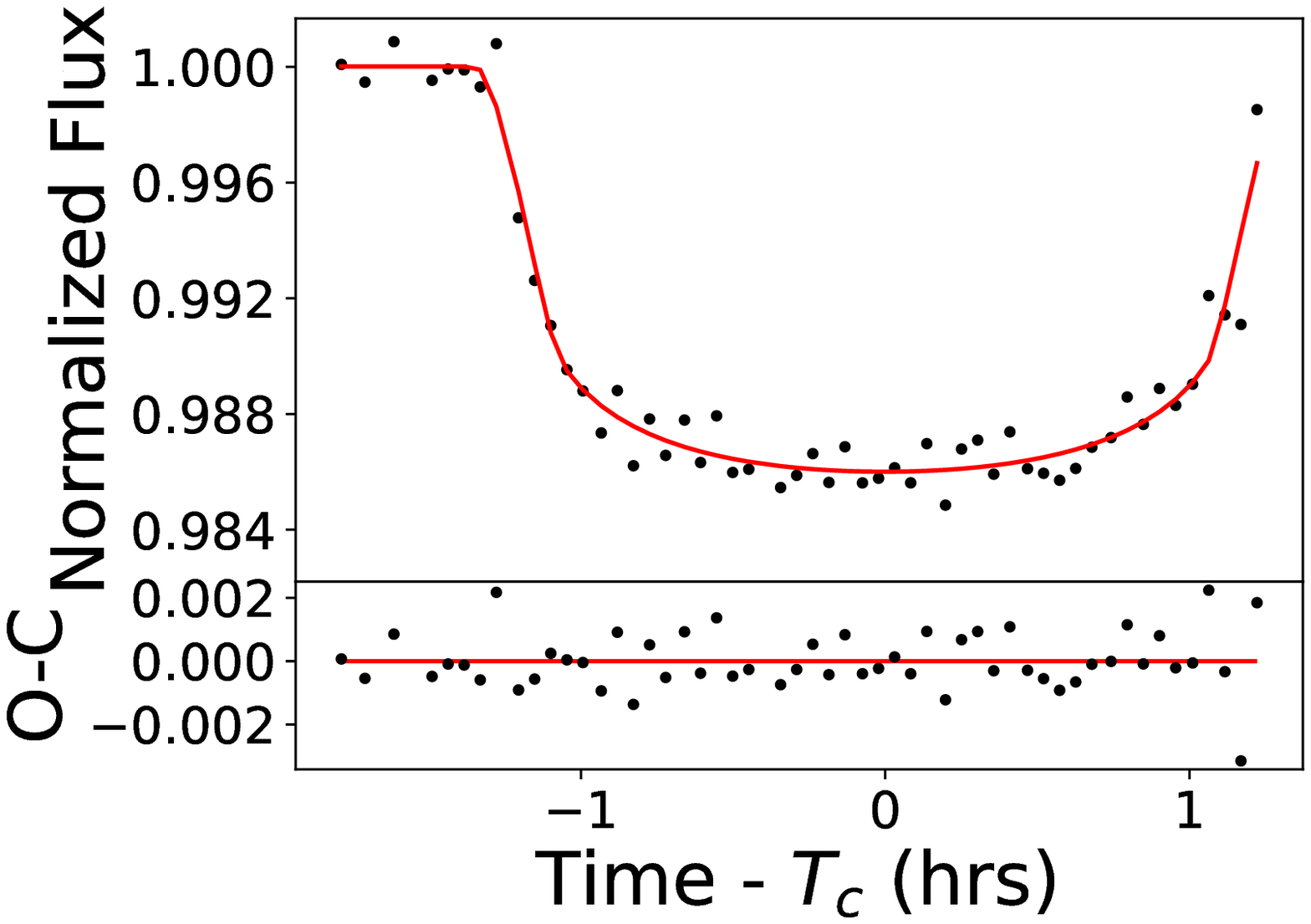}}
  \hfill
  \subfloat[CAHA 2015/04/30]{\includegraphics[width=4cm]{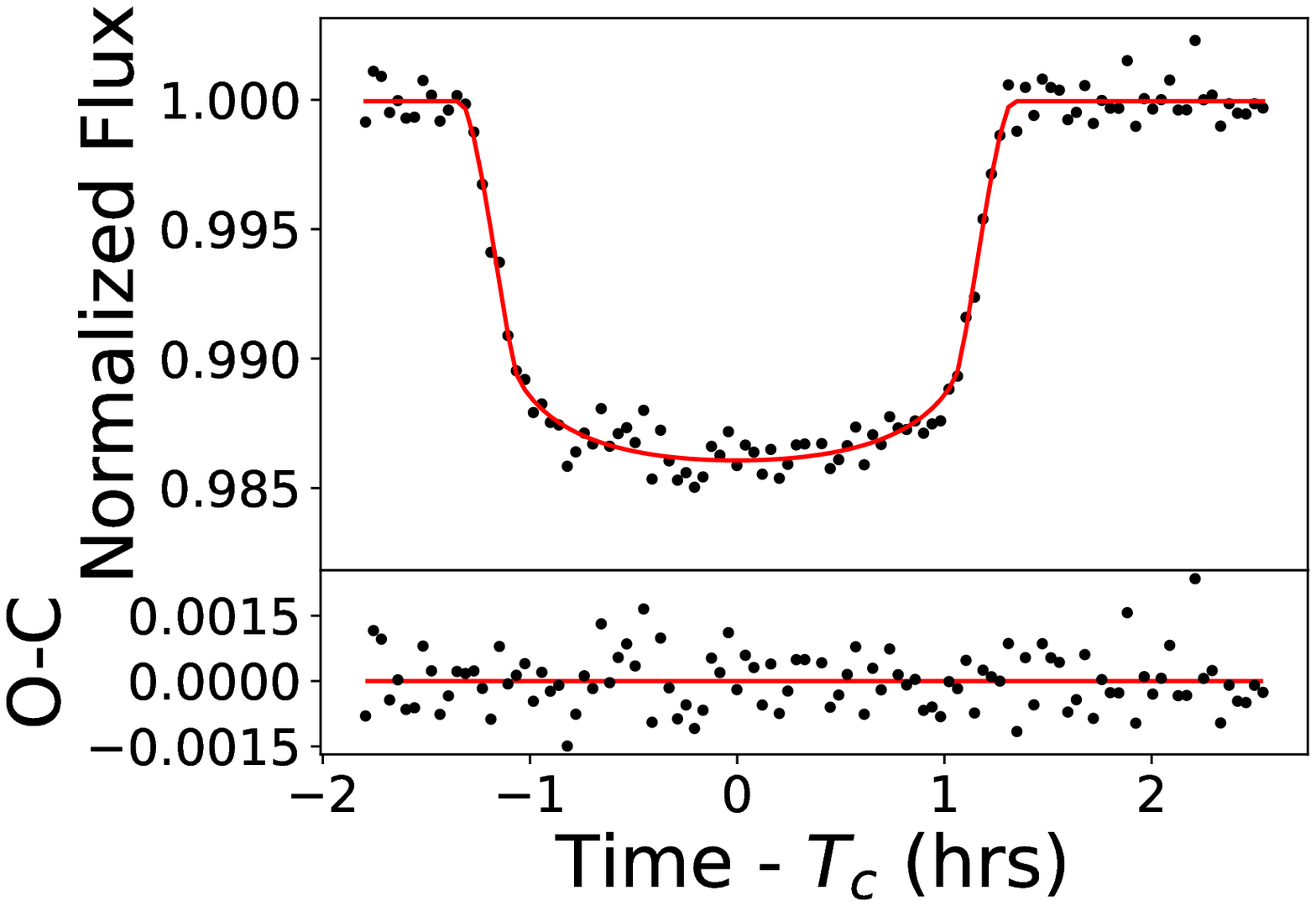}}
  \qquad
  \subfloat[CAHA 2015/06/07]{\includegraphics[width=4cm]{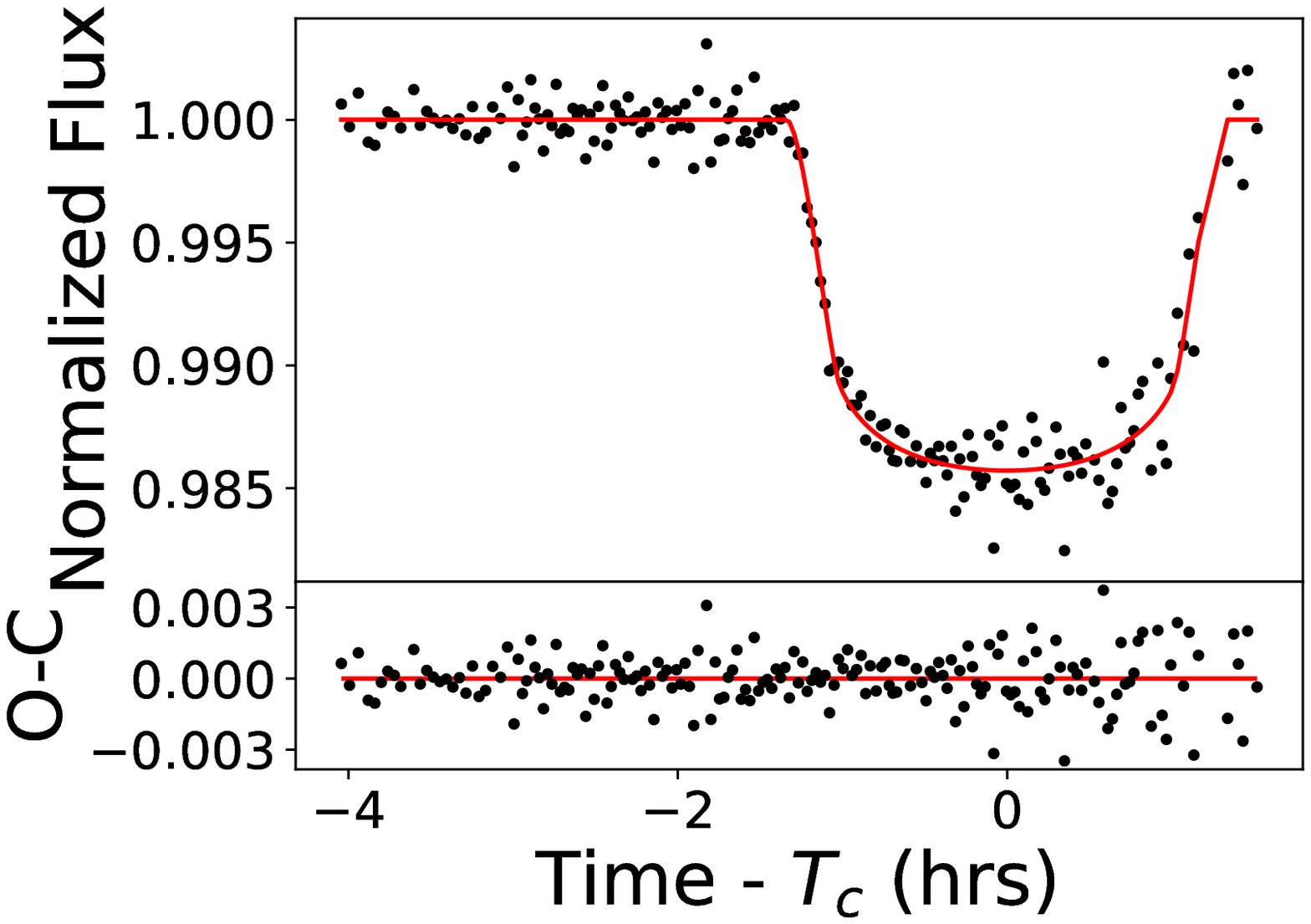}}
  \hfill
  \subfloat[T100 2017/06/11]{\includegraphics[width=4cm]{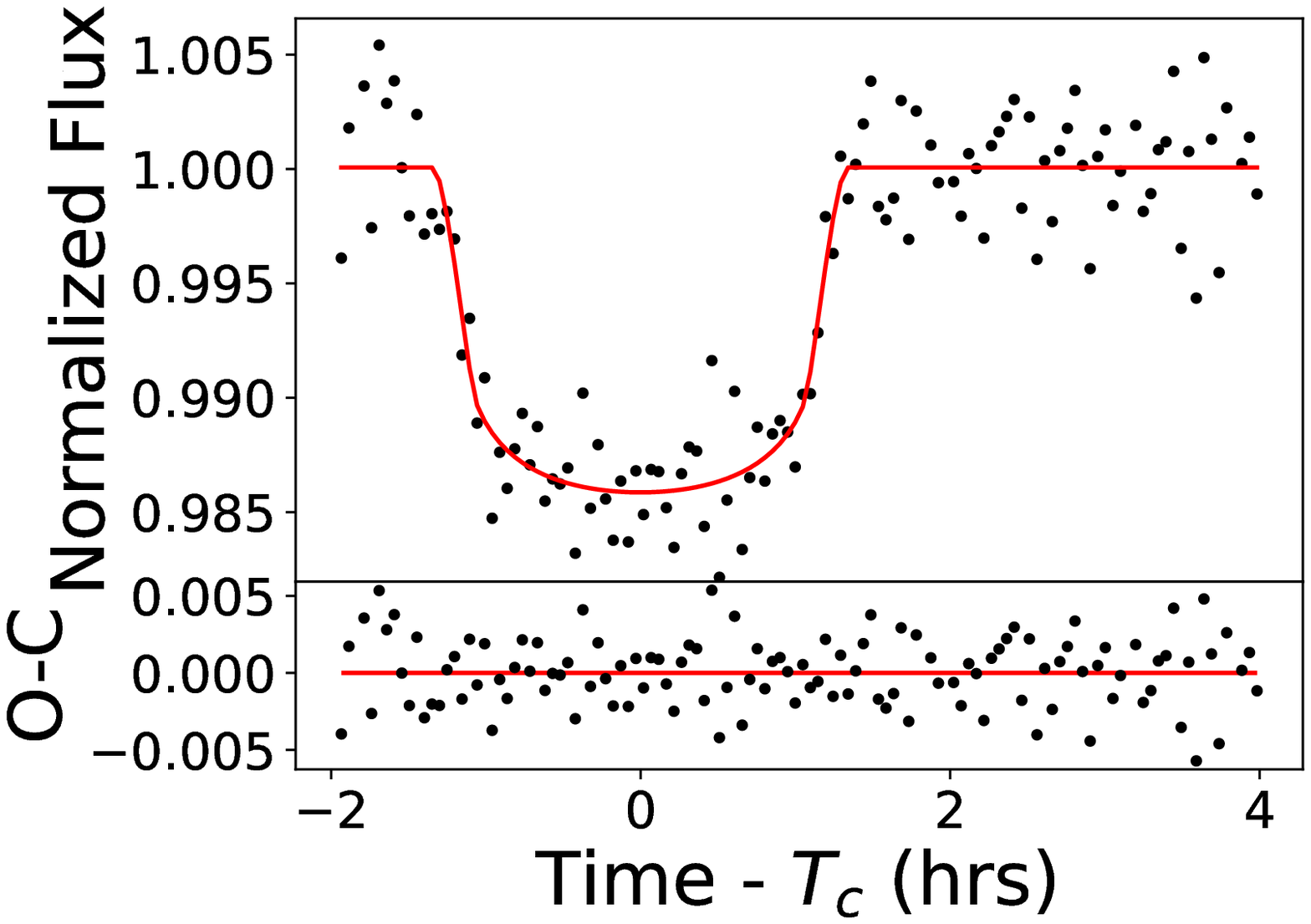}}
  \qquad
  \subfloat[T35 2018/07/01]{\includegraphics[width=4cm]{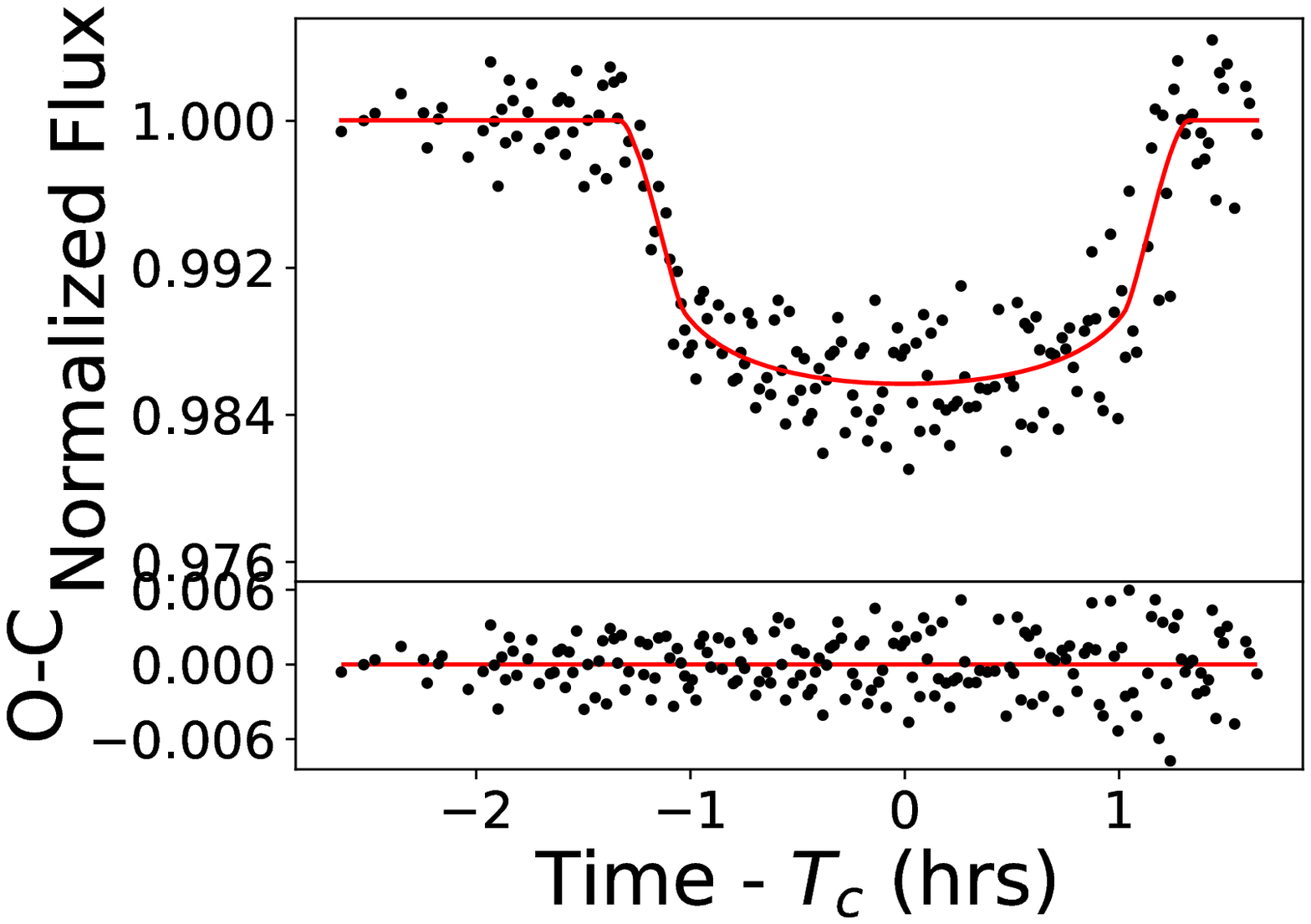}}
  \hfill
  \subfloat[T100 2020/07/29]{\includegraphics[width=4cm]{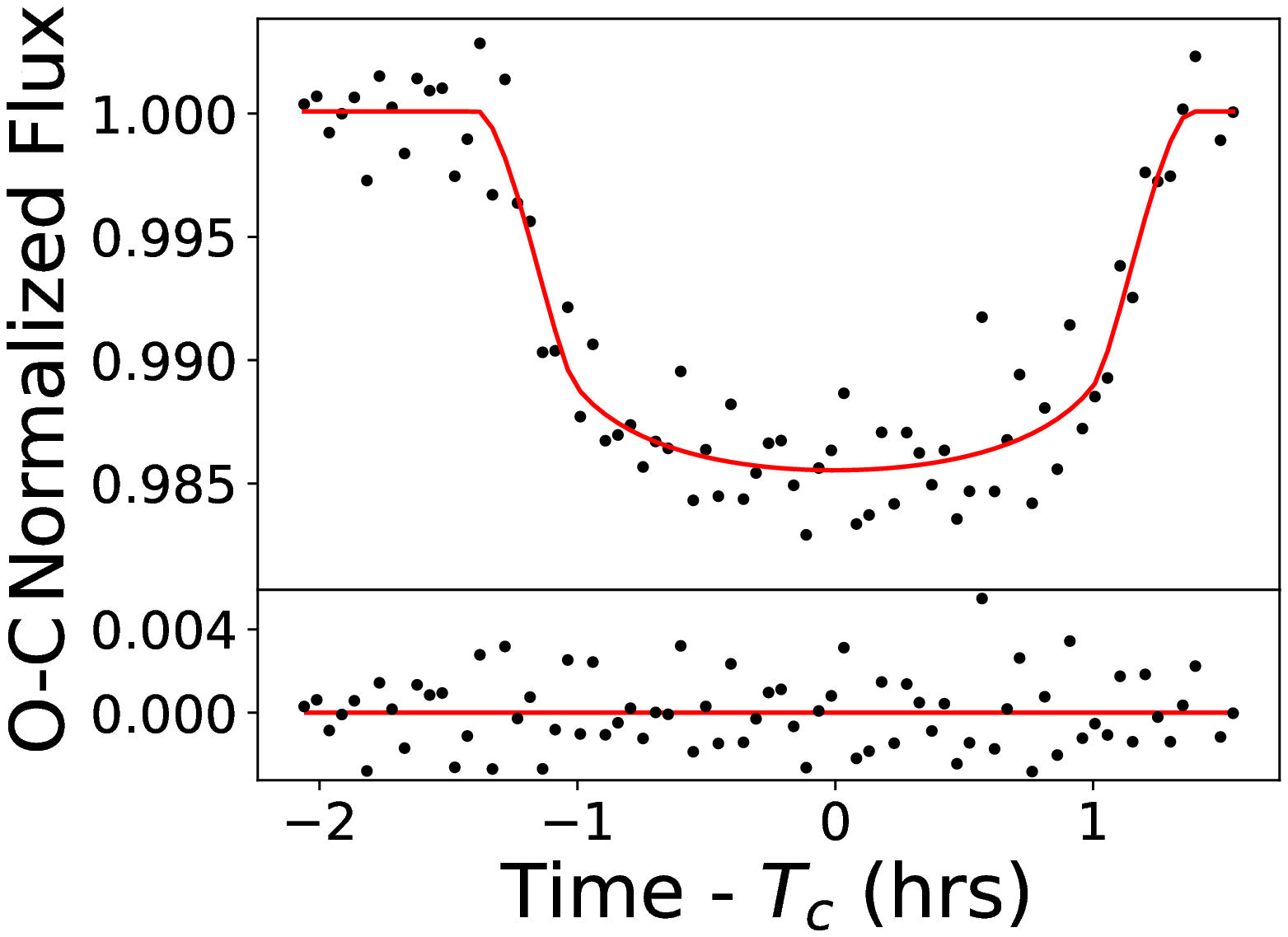}}
    \caption{WASP-103\,b Light Curves}
    \label{fig:wasp103_lightcurves}
\end{figure}

\begin{figure}
  \centering
  \subfloat[T100 2015/08/04]{\includegraphics[width=4cm]{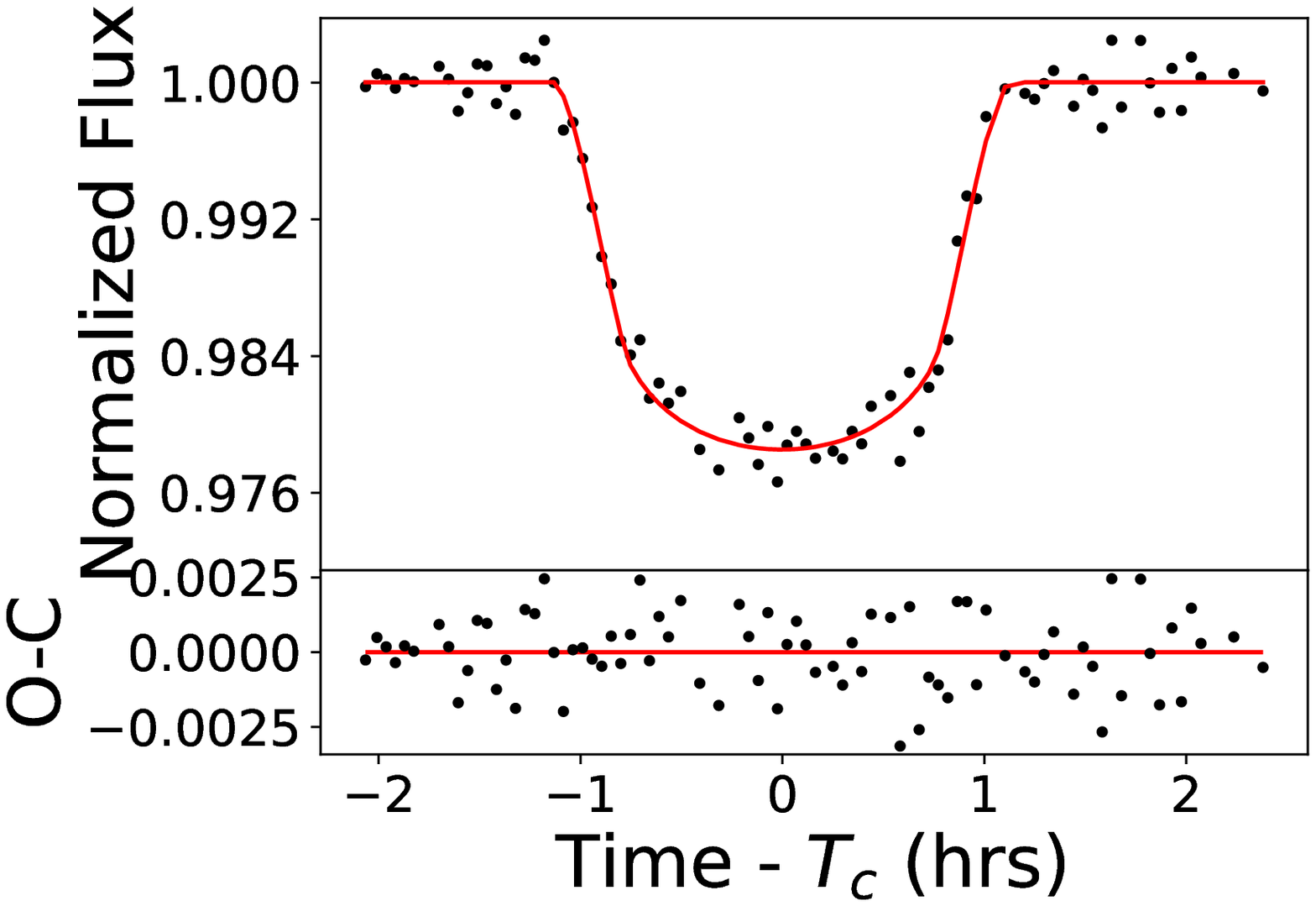}}
  \hfill
  \subfloat[T100 2019/05/17]{\includegraphics[width=4cm]{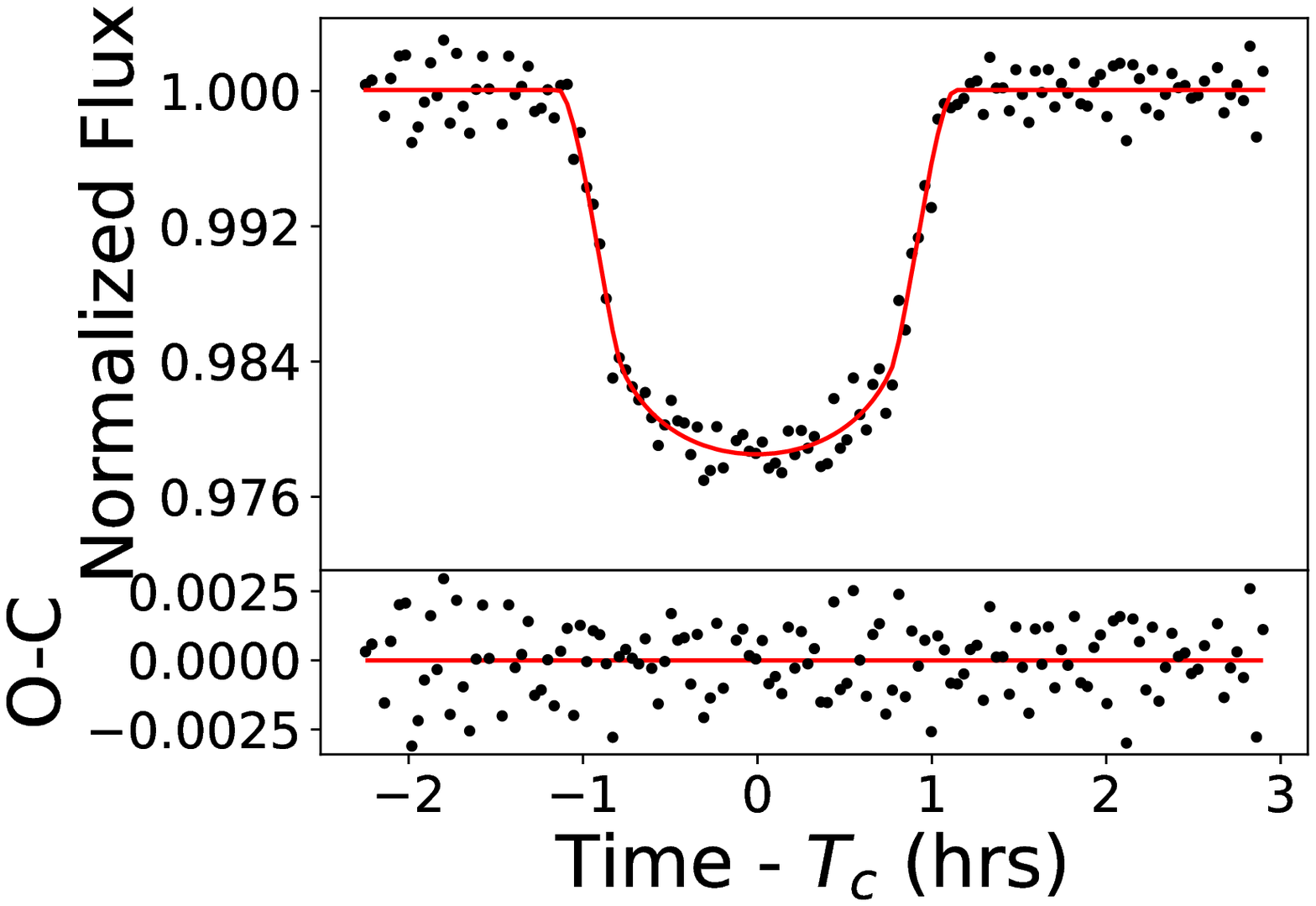}}
    \caption{HAT-P-37\,b Light Curves}
    \label{fig:hatp37_lightcurves}
\end{figure}

\section{Corner Plots}
\label{sec:appendixb}

We provide the posterior probability distributions of the parameters of the linear fits to TTV data to illustrate the distributions as well as the correlations between the fit parameters.

\begin{figure}
  \centering
  \includegraphics[width=\columnwidth]{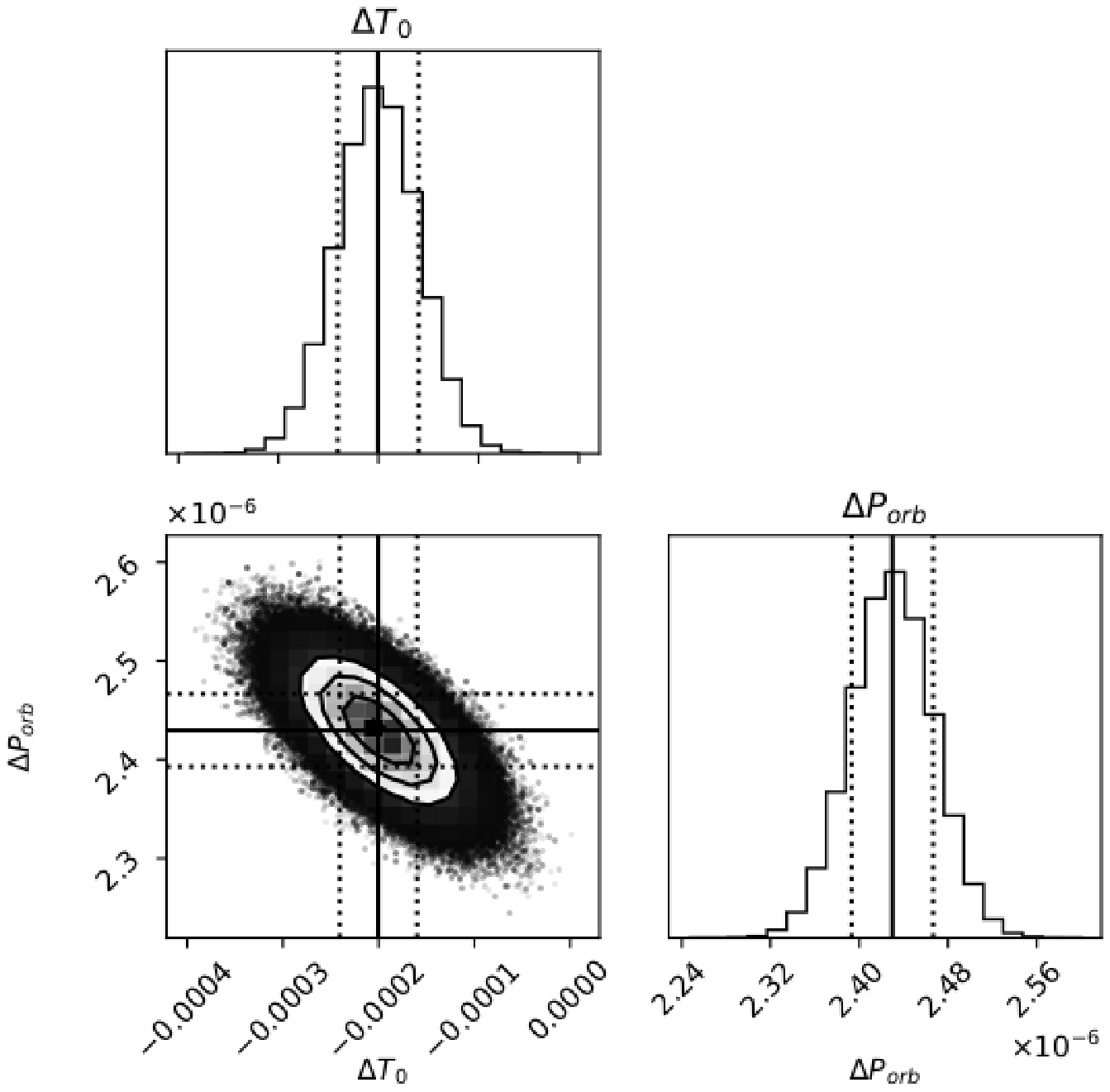}
  \caption{Corner plot showing the posterior probability distributions of the change in the reference mid-transit time (dT) (upper left) and the change in the orbital period (dP) (lower right), and the correlation between these two fit parameters (lower left) for HAT-P-23\,b.}
  \label{fig:hatp23_corner_plot}
\end{figure}

\begin{figure}
  \centering
  \includegraphics[width=\columnwidth]{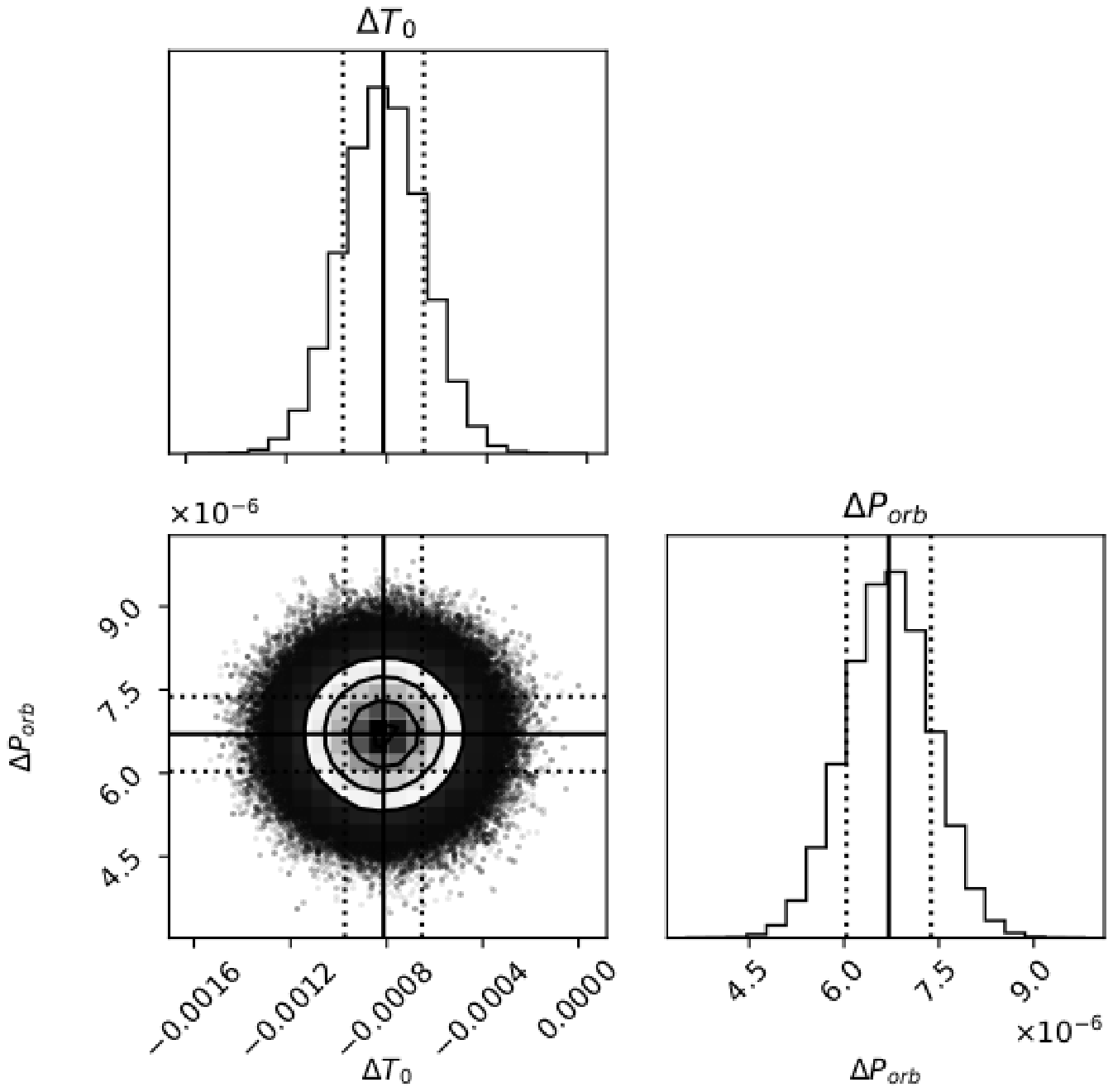}
  \caption{The same as Fig. \ref{fig:hatp23_corner_plot} only for WASP-37\,b.}
  \label{fig:wasp37_corner_plot}
\end{figure}

\begin{figure}
  \centering
  \includegraphics[width=\columnwidth]{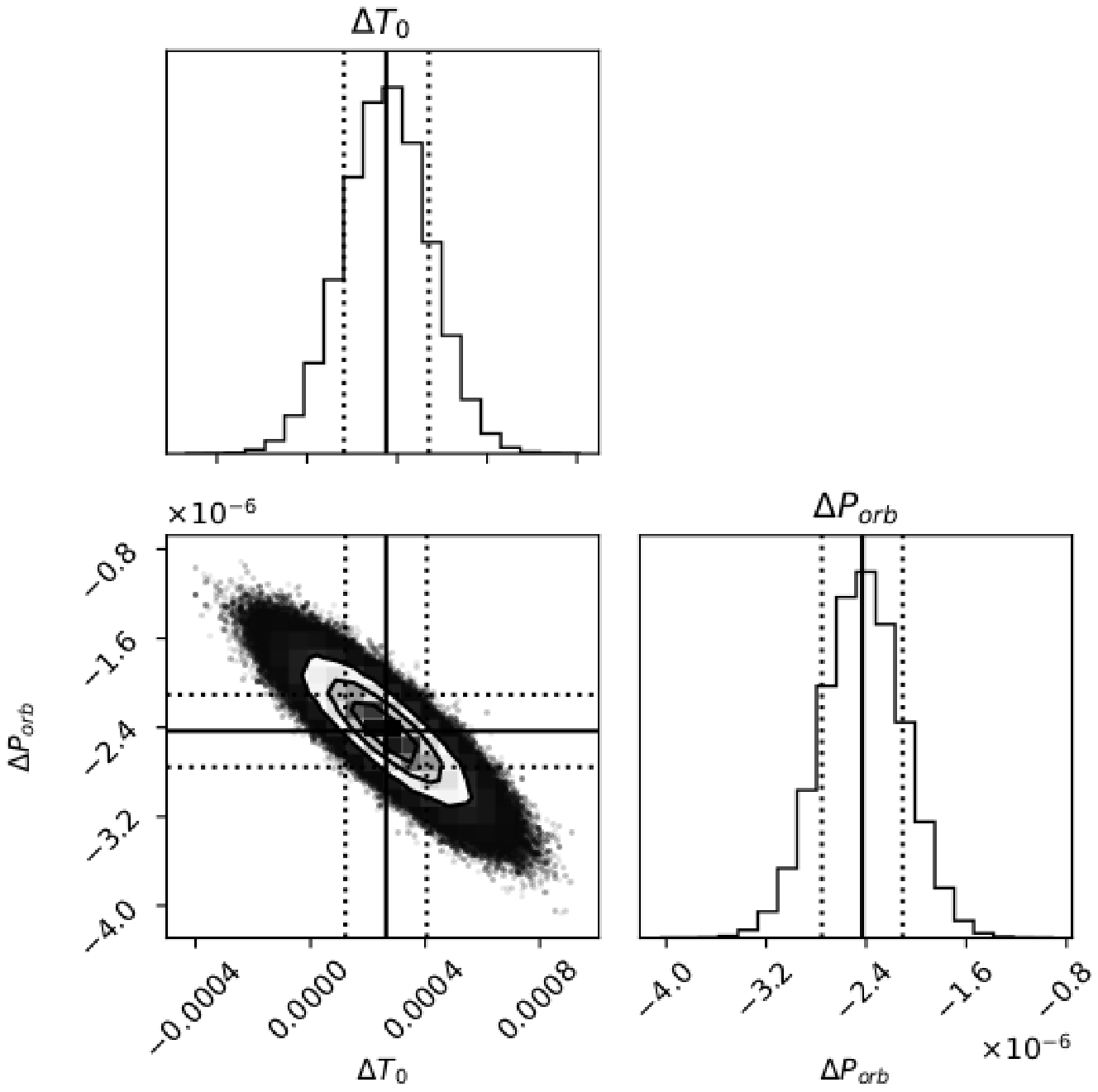}
  \caption{The same as Fig. \ref{fig:hatp23_corner_plot} only for WASP-69\,b.}
  \label{fig:wasp69_corner_plot}
\end{figure}

\begin{figure}
  \centering
  \includegraphics[width=\columnwidth]{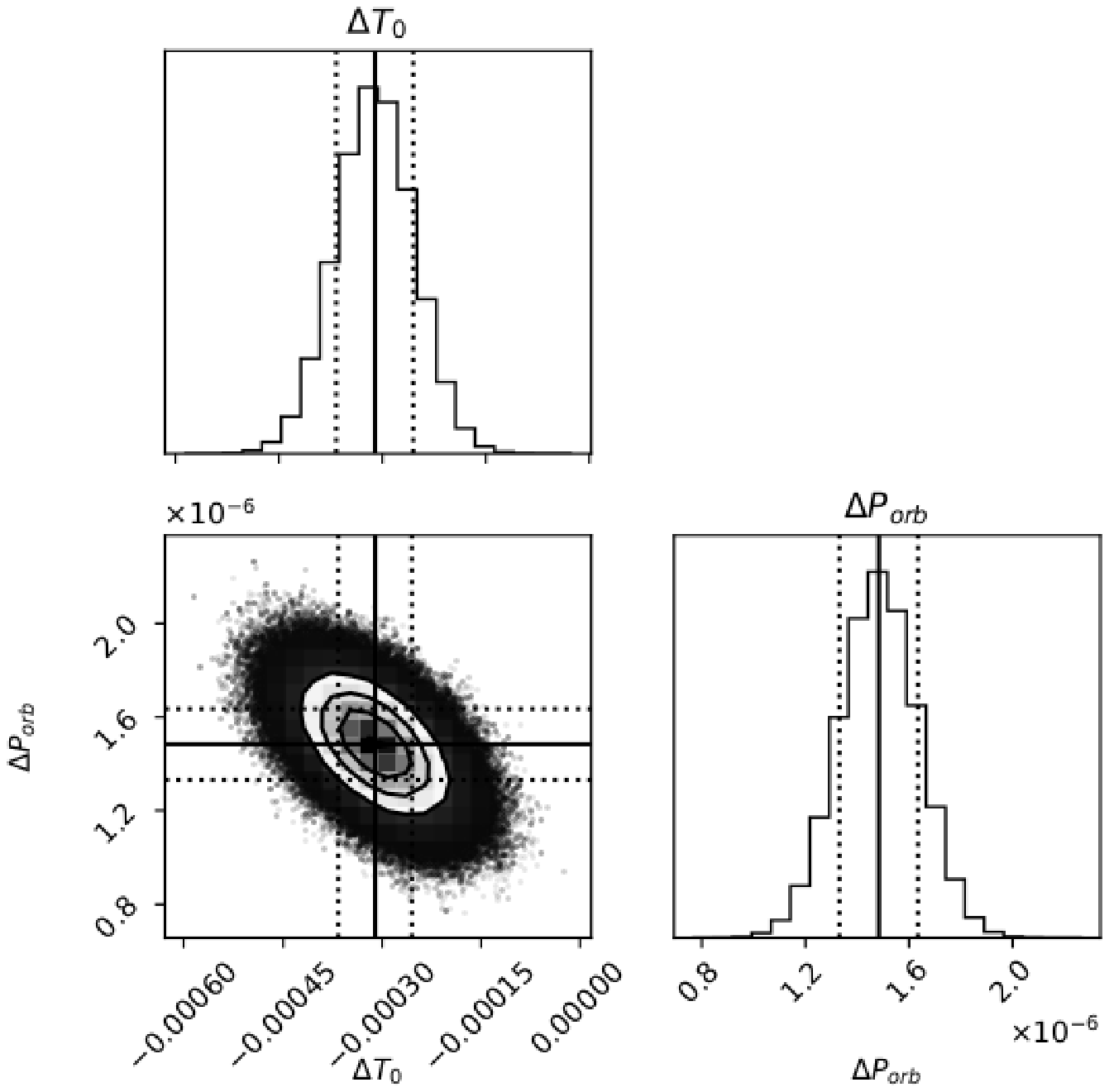}
  \caption{The same as Fig. \ref{fig:hatp23_corner_plot}  only for WASP-74\,b.}
  \label{fig:wasp74_corner_plot}
\end{figure}

\begin{figure}
  \centering
  \includegraphics[width=\columnwidth]{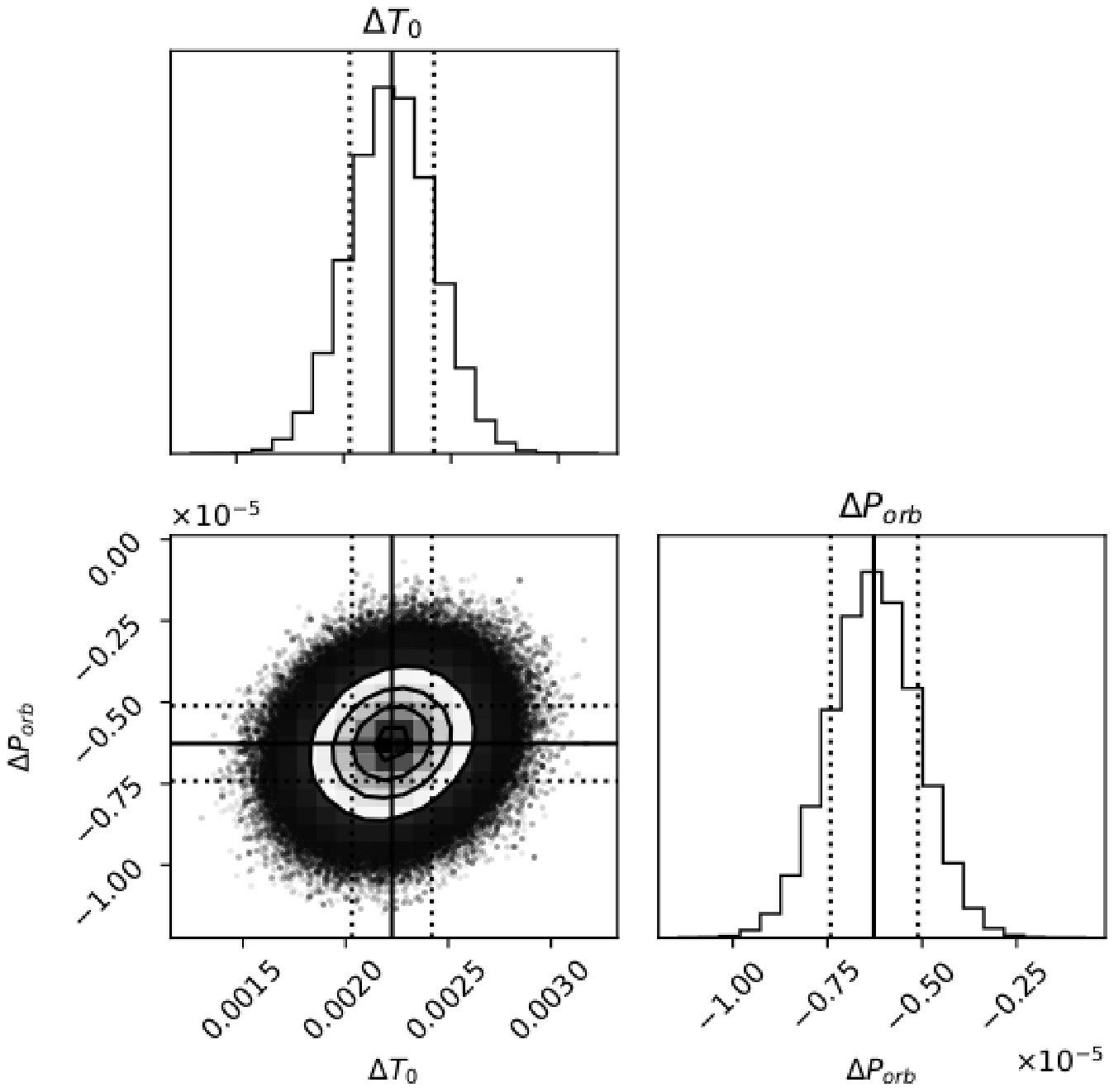}
  \caption{The same as Fig. \ref{fig:hatp23_corner_plot} only for HAT-P-56\,b.}
  \label{fig:hatp56_corner_plot}
\end{figure}

\begin{figure}
  \centering
  \includegraphics[width=\columnwidth]{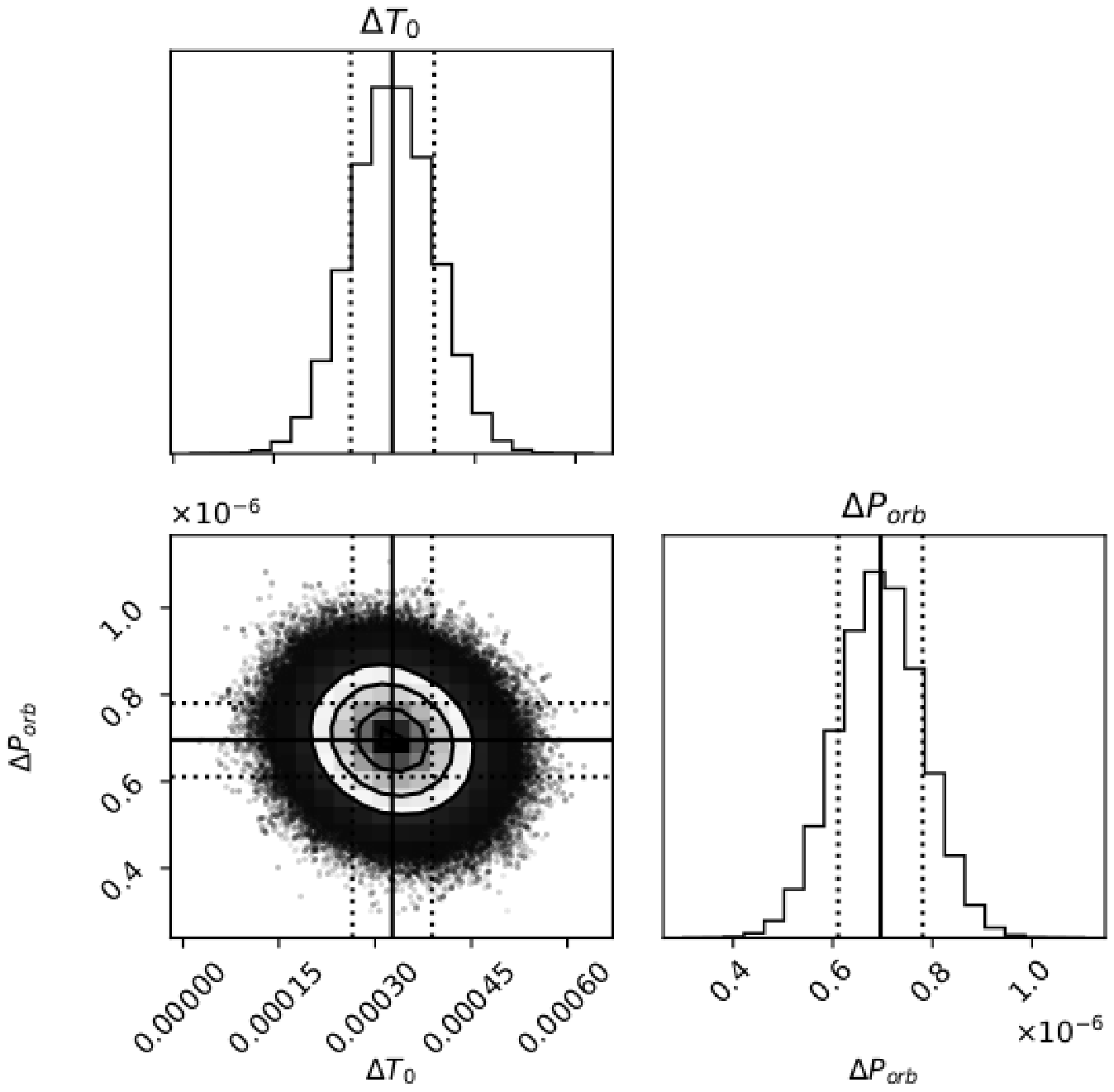}
  \caption{The same as Fig. \ref{fig:hatp23_corner_plot} only for WASP-2\,b.}
  \label{fig:wasp2_corner_plot}
\end{figure}

\begin{figure}
  \centering
  \includegraphics[width=\columnwidth]{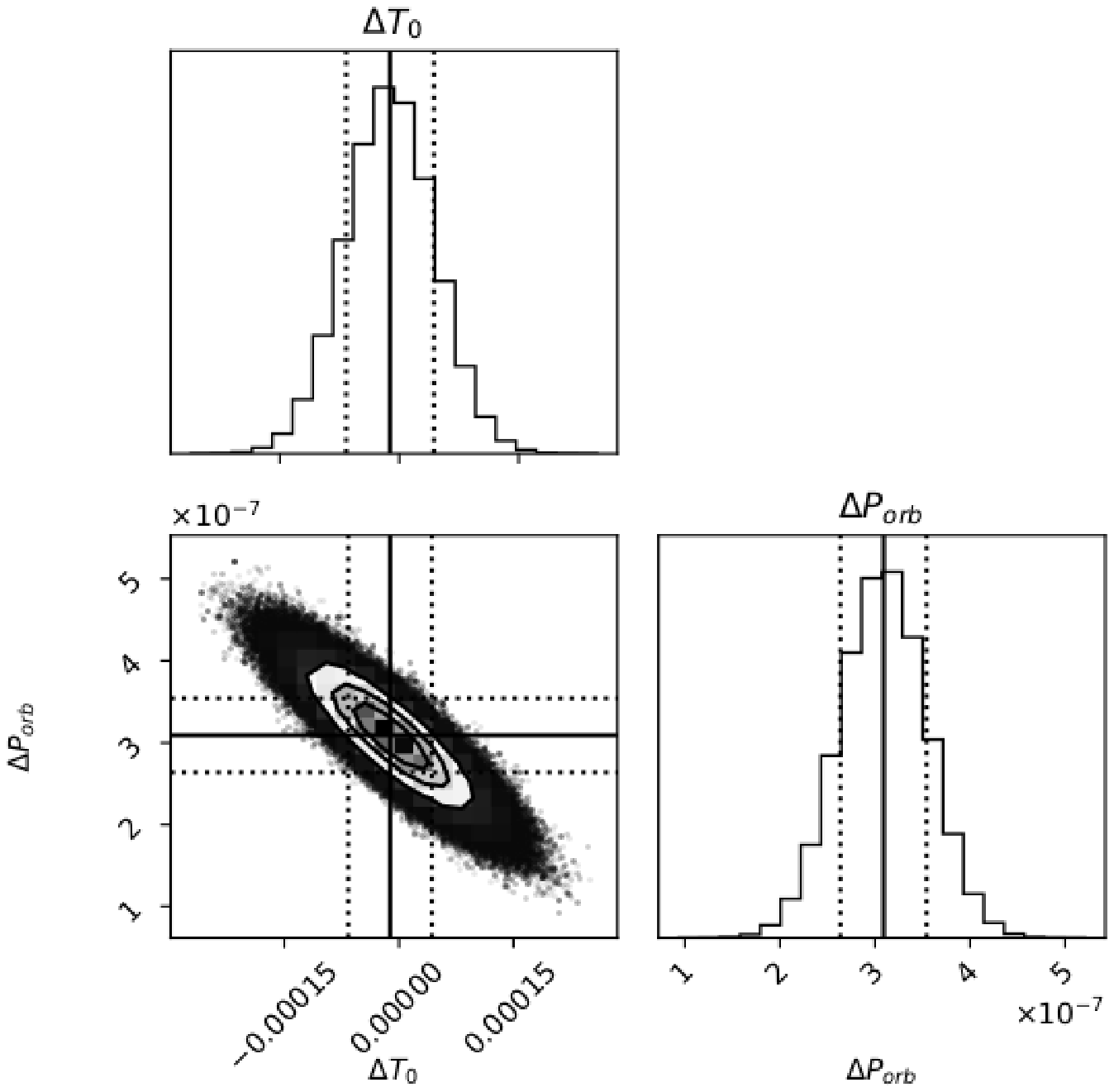}
  \caption{The same as Fig. \ref{fig:hatp23_corner_plot} only for HAT-P-32\,b.}
  \label{fig:hatp32_corner_plot}
\end{figure}

\begin{figure}
  \centering
  \includegraphics[width=\columnwidth]{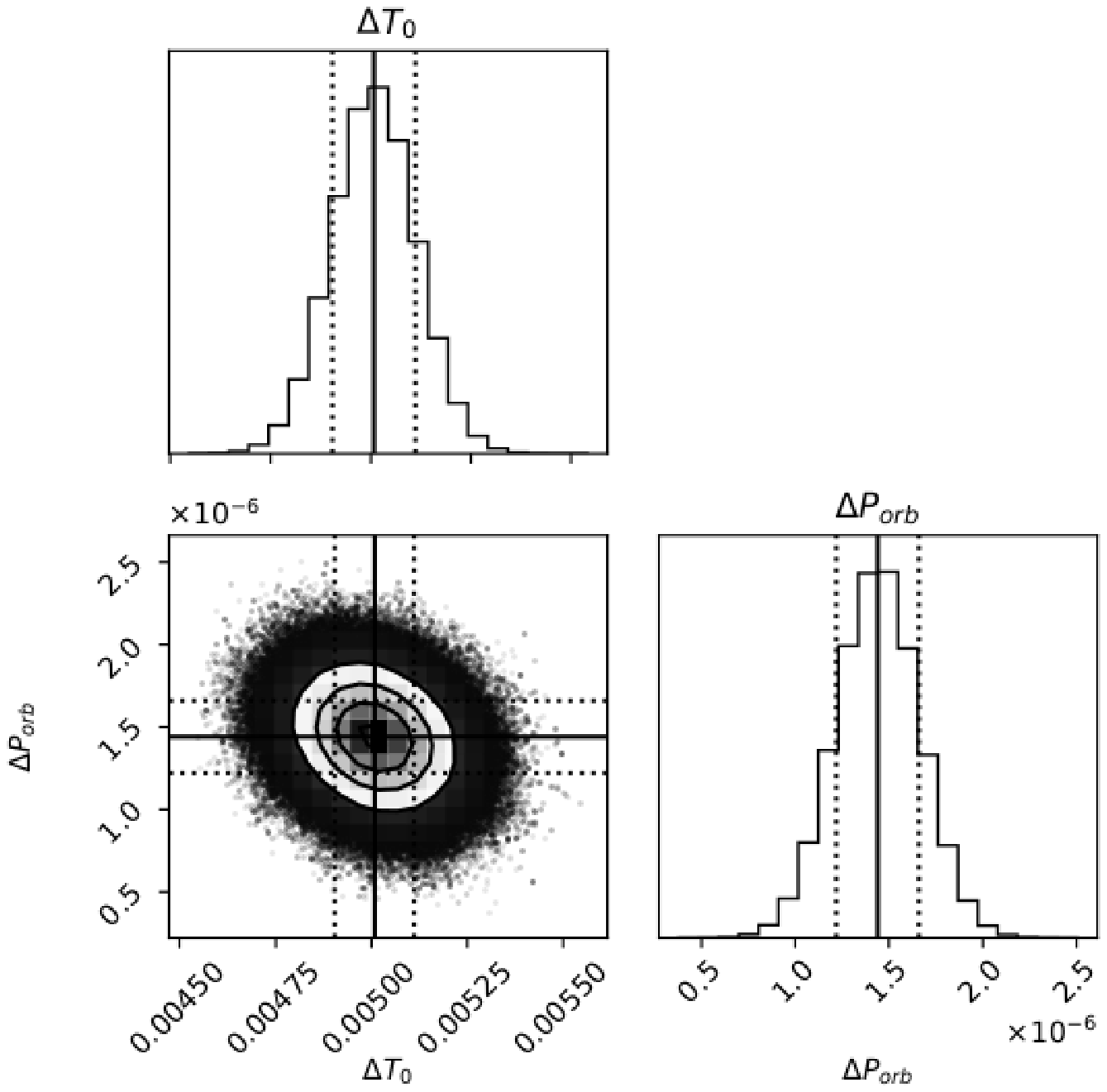}
  \caption{The same as Fig. \ref{fig:hatp23_corner_plot} only for WASP-14\,b.}
  \label{fig:wasp14_corner_plot}
\end{figure}

\begin{figure}
  \centering
  \includegraphics[width=\columnwidth]{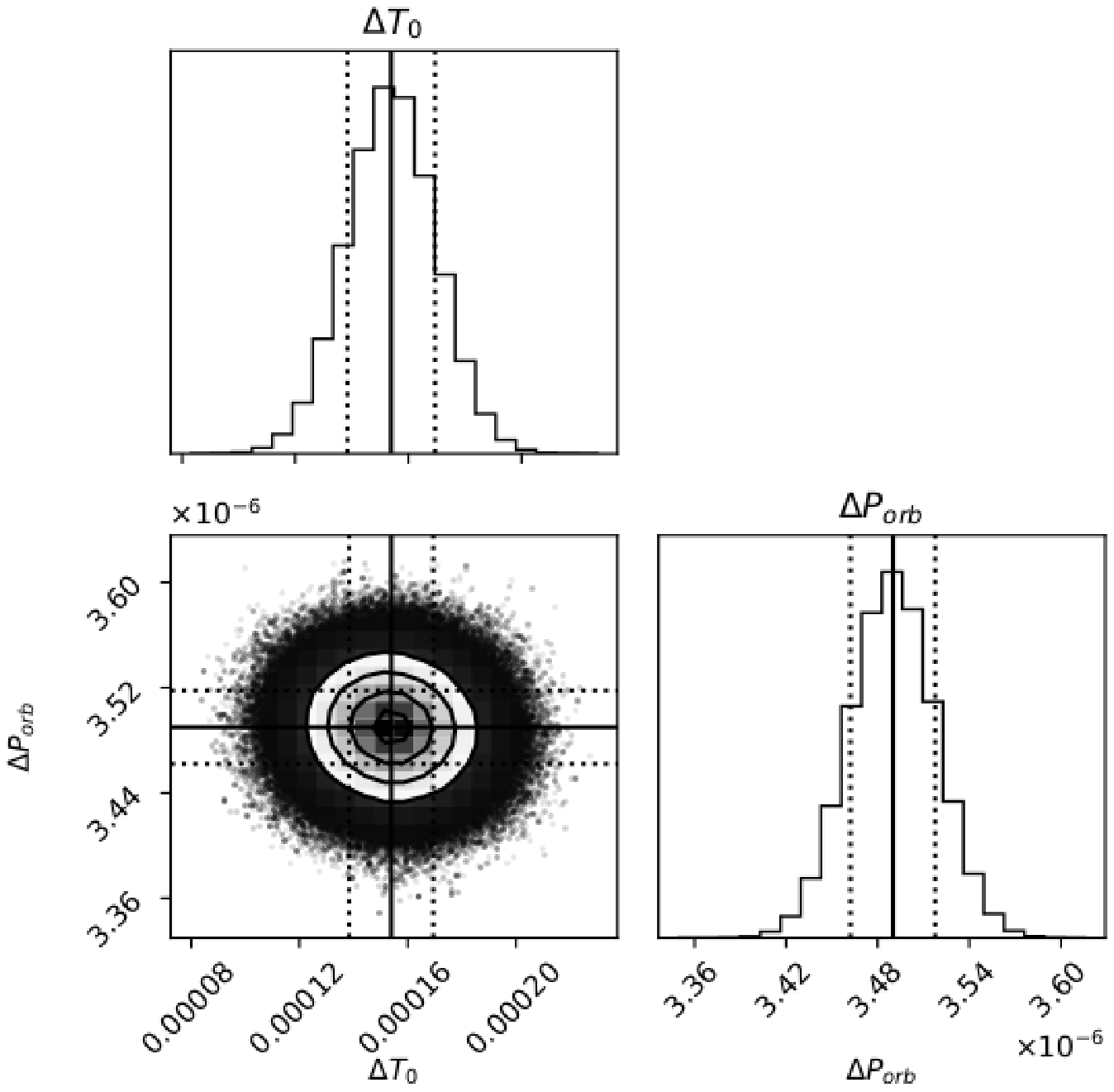}
  \caption{The same as Fig. \ref{fig:hatp23_corner_plot} only for WASP-103\,b.}
  \label{fig:wasp103_corner_plot}
\end{figure}

\begin{figure}
  \centering
  \includegraphics[width=\columnwidth]{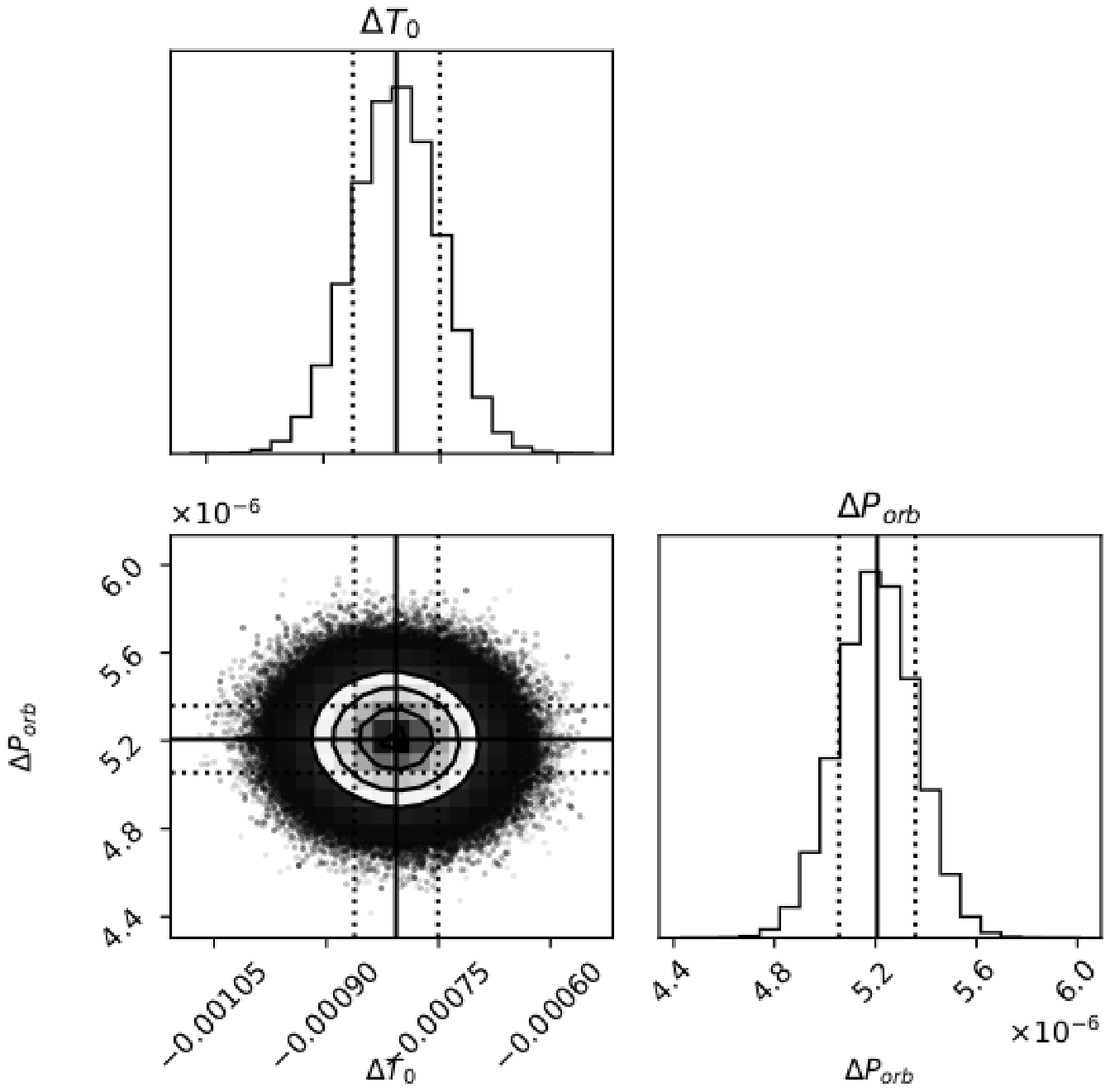}
  \caption{The same as Fig. \ref{fig:hatp23_corner_plot}  only for HAT-P-37.}
  \label{fig:hatp37_corner_plot}
\end{figure}



\bsp	
\label{lastpage}
\end{document}